\documentclass{aa}
\usepackage{natbib}
\usepackage{graphicx}
\usepackage{times}
\usepackage{amsmath}
\usepackage{txfonts}
\newcommand{\etal}{et al.,~}
\newcommand{\msun}{{\,\rm M}_{\odot}}
\newcommand{\lsun}{{\,\rm L}_{\odot}}

\newcommand{\kms}{\,{\rm km.s}^{-1}}

\newcommand{\nht}{\ifmmode {{\rm NH}_3} \else {NH{\bas 3}} \fi}
\newcommand{\tco}{\ifmmode {^{13}{\rm CO}} \else {$^{13}{\rm CO}$}\fi}
\newcommand{\dco}{\ifmmode {^{12}{\rm CO}} \else {$^{12}{\rm CO}$}\fi}
\newcommand{\cdo}{\ifmmode {{\rm C}^{18}{\rm O}} \else {${\rm C}^{18}{\rm O}$}\fi}

\newcommand{\htco}{\ifmmode {{\rm H}^{13}{\rm CO}^{+} } \else {${\rm H}^{13}
{\rm CO}^{+}$ }\fi}
\newcommand{\hco}{\ifmmode {{\rm H}^{12}{\rm CO}^{+} } \else {${\rm H}^{12}
{\rm CO}^{+}$ }\fi}
\newcommand{\juz}{\ifmmode {{\rm J}=1\rightarrow 0} \else
{J=1$\rightarrow$0}\fi}
\newcommand{\jdu}{\ifmmode {{\rm J}=2\rightarrow 1} \else
{J=2$\rightarrow$1}\fi}
\newcommand{\jtd}{\ifmmode {{\rm J}=3\rightarrow 2} \else
{J=3$\rightarrow$2} \fi}
\newcommand{\jcq}{\ifmmode {{\rm J}=5\!\rightarrow\!4} \else
{${\rm J}=5\!\rightarrow\!4$} \fi}
\newcommand{\as}{\ifmmode {^{\scriptscriptstyle\prime\prime}}
        \else $^{\scriptscriptstyle\prime\prime}$\fi}
\newcommand{\am}{\ifmmode {^{\scriptscriptstyle\prime}}
        \else $^{\scriptscriptstyle\prime}$\fi}
\newcommand{\app}{$a_+$}
\newcommand{\amm}{$a_-$}

\newcommand{\lya}{Ly\,${\alpha}$}

\newcommand{\msol}{${\,\rm M}_{\odot}$}

\renewcommand{\textbf}{\textrm}

\begin{document}
\title{Chemistry in Disks. V:
CN and HCN in proto-planetary disks \thanks{Based on observations carried out with the IRAM Plateau de Bure Interferometer. IRAM is
supported by INSU/CNRS (France), MPG (Germany), and IGN (Spain).}}
\author{Edwige Chapillon\inst{1,2,3,4}, St\'ephane Guilloteau \inst{2,3}, Anne Dutrey\inst{2,3}, Vincent Pi\'etu\inst{4} and Michel Gu\'elin\inst{4}}
\institute{MPIfR, Auf dem H\"ugel 69, 53121 Bonn, Germany\\
\email{echapill@mpifr-bonn.mpg.de}
 \and{}Universit\'e de Bordeaux, Observatoire Aquitain des Sciences de l'Univers, 2 rue de l'Observatoire BP 89, F-33271 Floirac, France
  \and{}
  CNRS/INSU - UMR5804, Laboratoire d'Astrophysique de Bordeaux;  2 rue de l'Observatoire BP 89, F-33271 Floirac, France \\
  \email{guilloteau@obs.u-bordeaux1.fr,dutrey@obs.u-bordeaux1.fr}
  \and{}
  IRAM, 300 rue de la Piscine, F-38406 Saint Martin d'H\`eres, France.\\
  \email{pietu@iram.fr,guelin@iram.fr}
  }
\date{Received xx-xxx-xxxx, Accepted xx-xxx-xxxx}
\authorrunning{Chapillon \etal}
\titlerunning{CN and HCN in proto-planetary disks}


\abstract
{The chemistry of proto-planetary disks is thought to be dominated by two major processes: photodissociation near the disk surface, and depletion on dust grains in the disk mid-plane, resulting in a layered
structure with molecules located in a warm layer above the disk mid-plane.}
{We attempt here to confront this warm molecular layer model prediction with the distribution of two key molecules for dissociation processes: CN and HCN}
{Using the IRAM Plateau de Bure interferometer, we obtained high spatial and spectral resolution images of the CN J=2-1 and HCN J=1-0 lines in the disks surrounding the two T-Tauri DM\,Tau and LkCa\,15 and
the Herbig Ae MWC\,480. Disk properties are derived assuming power law distributions. The hyperfine structure of the observed transitions allows us to constrain the line opacities and excitation temperatures. We compare the observational results with predictions from existing chemical models, and use a simple PDR model
(without freeze-out of molecules on grains and surface chemistry)
to illustrate dependencies on UV field strength, grain size and gas-to-dust ratio. We also evaluate the impact of \lya\  radiation.}
{The temperature ordering follows the trend found from CO lines, with DM Tau being the coldest object and MWC 480 the warmest. Although CN indicates somewhat higher excitation temperatures than HCN, the derived values in
the T Tauri disks are very low (8-10 K). They agree with results obtained from C$_2$H, and are in contradiction with thermal and chemical model predictions. These very low temperatures, as well as geometrical constraints, suggest that substantial amounts of CN and HCN remain in the gas phase close to the disk mid-plane, and that this mid-plane is quite cold.  The observed CN/HCN ratio ($\simeq 5-10$) is in better agreement with the existence of large grains, and possibly also a substantial contribution of \lya\ radiation.
}
{}


\maketitle{}


\section{Introduction}

Studying the physical and chemical structure of proto-planetary disks is a prerequisite for
our understanding of planet formation. While much theoretical work has been made on the
structure and evolution of these disks \citep[][and references therein]{Bergin+etal_2007},
key parameters such as gas density and temperature, or molecular content and distribution
remain poorly constrained. Since H$_2$, which represents 80 \% of the gas mass, cannot be
observed with a good spatial resolution, our knowledge of the molecular component
relies on low abundance tracers such as CO \citep{Guilloteau+Dutrey_1998}. 
Interferometric maps of the CO emission, at mm and sub-mm wavelengths, resolve the gaseous
disks, revealing their radial and vertical structure. Studies of the rare CO isotopologues,
$^{13}$CO and C$^{18}$O, confirm the existence of a vertical temperature gradient
\citep{Dartois+etal_2003,Pietu+etal_2007}, allowing first face to face confrontation between
model predictions and observations in the outer parts of the disks.

This led to a global picture of a flared disk consisting of 3 strata, or layers. The upper layer,
close to the disk surface, is directly illuminated by the stellar UV and dominated by
photo-dissociation reactions. The lowest layer, near the disk plane, is very cold, so that molecules
are expected to be depleted on grains. Between those, the third layer, where the gas is lukewarm and
the chemistry dominated by gas-phase reactions. Such a simple picture is already challenged by new
observations which reveal that large amounts of cold CO are still in the gas phase
\citep{Dartois+etal_2003,Pietu+etal_2007}, contrary to predictions. Vertical mixing was proposed
to refurbish CO in the cold zone
\cite[][see also Willacy et al. 2006]{Aikawa+Nomura_2006,Aikawa_2007,Semenov+etal_2006}.

In this scheme, molecules are displaced from the upper layers to the cold mid-plane and have
the time to cool down before sticking onto grains; the results, however, are strongly dependent on the
grain surface reaction efficiencies and on the disk structure. \citet{Hersant+etal_2009} have also shown
that photo-desorption may increase the amount of cold CO in the gas.

A deeper understanding of the disk structure relies on the study of
molecules probing different physical and chemical conditions. Apart from CO, only few
molecules show strong enough mm/sub-mm emission to be detected: HCO$^+$, N$_2$H$^+$, HCN,
CN, HNC, CCH, CS and H$_2$CO
\citep{Dutrey+etal_1997,Kastner+etal_1997,Thi+etal_2004,vanZadelhoff+etal_2001}. Even fewer can
be mapped \citep{Aikawa+etal_2003,Qi+etal_2004,Dutrey+etal_2007,Schreyer+etal_2008}. Among
those, HCN and CN are particularly interesting because they trace different conditions and
their abundances are closely related: HCN may be photo-dissociated (e.g., by \lya\ radiation
from the star) to yield CN \citep{Bergin+etal_2003}. One thus expects HCN to trace the
UV-shielded lukewarm and cold layers of the disk and CN the uppermost layer.

In this article, we present high angular resolution (\textbf{3$''$} to 1.5$''$) observations of HCN\,J=1-0 and
CN\,J=2-1 emission in 3 protoplanetary disks. The sources belong to the nearby
Taurus-Aurigae star forming region, but have moved outside their parent cloud, so that they
can be observed without interference. Table \ref{tabcoord} summarizes their stellar properties.

- \object{DM\,Tau} is a classical T-Tauri star of mass
0.5\,\msol\ surrounded by a disk in Keplerian rotation \citep{Guilloteau+Dutrey_1998,
Pietu+etal_2007}. \citet{Dartois+etal_2003} have shown the existence of a vertical
temperature gradient from high angular resolution observations of several transitions of
\dco\, \tco\ and \cdo . 
Eight different molecular species have been detected in this source \citep{Dutrey+etal_1997}.

- \object{LkCa\,15} is a 1 \msol\ T-Tauri star surrounded by a disk with a central cavity \citep{Pietu+etal_2006}. \citet{Pietu+etal_2007} found no evidence of a vertical temperature gradient from their CO multi-isotopologue data. The cavity radius derived from dust emission is $\sim 45$\,AU, but some CO gas remains down
to 20 AU or so \citep{Pietu+etal_2007}.

- Finally, \object{MWC\,480} is a Herbig Ae star of 1.8\,\msol\ surrounded by a Keplerian disk
\citep{Mannings+etal_1997,Simon+etal_2000,Pietu+etal_2007}. CO isotopologues indicate a
vertical temperature gradient. The disk continuum emission was resolved by \citet{Pietu+etal_2006},
suggesting even lower temperatures for the dust than from $^{13}$CO, but \citet{Guilloteau+etal_2010}
show that alternative solutions with radially varying dust properties are possible.

Section 2 presents the observations. The method of analysis and the results are shown
in Section 3 and Section 4 is dedicated to the chemical modeling. We discuss the
implications of our study in Section 5 and then we conclude.

\begin{table*}[ht]
\caption{Star properties}\label{tabcoord}
\begin{center}
\begin{tabular}{lllllll}
\hline \hline
Source   & Right ascension      & Declination & Spec.~type & Effective temp.(K) & Stellar lum.($\lsun$) & CO paper \\
\hline
MWC\,480  & 04:58:46.264  & 29:50:36.86 & A4   &  8460 & 11.5 & 1,2\\
LkCa\,15  & 04:39:17.790  & 22:21:03.34 & K5   &  4350 & 0.74 & 1,2\\
DM\,Tau   & 04:33:48.733  & 18:10:09.89 & M1   &  3720 & 0.25 & 1,3  \\
 \hline
\end{tabular}
\end{center}
\tablefoot{Columns 2\&3: J2000 coordinates deduced from the fit of the PdBI 1.3\,mm continuum map.
Errors on the astrometry are of order $\leq 0.05''$. Columns 4, 5, and 6, the spectral
type, effective temperature and the stellar luminosity are those given in \citet{Simon+etal_2000}.
Column 6, CO interferometric papers are: \textbf{1=}\citet{Pietu+etal_2007}, 2 =\citet{Simon+etal_2000}, 3 = \citet{Dartois+etal_2003}.}
\end{table*}

\section{Observations}
\label{sec:observations}

All the observations reported here were made with the IRAM Plateau de Bure interferometer
(PdBI).

The CN \jdu\ observations of DM\,Tau were performed in Feb., Apr., Sep. and Oct.
1997 with baselines ranging from 15 m to 180 m. The spatial resolution (HPBW) was $1.7
\times 1.2\arcsec$ (PA $122^\circ$) for ``natural weighting''. The HCN \juz\ data were
obtained in Nov. 1999, Jan. 2001 and Jan. 2002 with baselines from 15 to 330 m. The HPBW
was $3.7 \times 2.4 \arcsec$ (PA $39^\circ$) using natural weighting, and $3.0 \times
1.8\arcsec$ at PA $45^\circ$ for ``robust'' weighting.

The LkCa15 and MWC\,480 data were observed in HCN \juz\ and CN \jdu\ data in a track
sharing mode, to obtain an homogeneous calibration for both sources. Baselines ranged from 15 to
330 meters, with an emphasis on short baselines: natural weighting yielded an HPBW of $2.2
\times 1.3\arcsec$ (PA $56^\circ$) and robust weighting an HPBW of $1.2 \times 0.7\arcsec$
(PA $40^\circ$) for CN, and proportionally larger by a factor 2.5 for HCN.

\begin{figure} 
  {\large \textbf{LkCa 15}} \\ 
  \includegraphics[width=0.9\columnwidth]{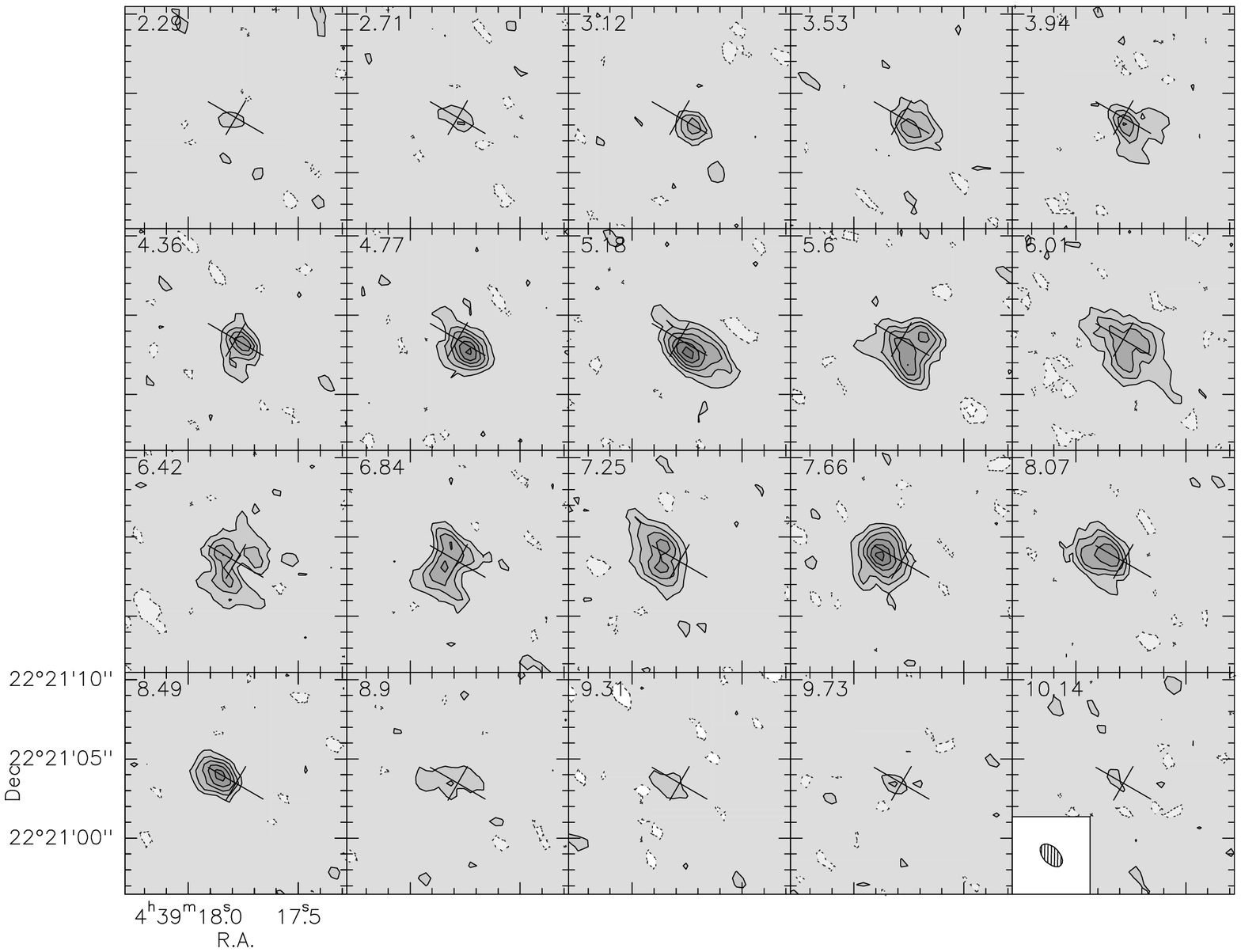}\\
  {\large \textbf{DM Tau}}\\ 
  \includegraphics[width=0.9\columnwidth]{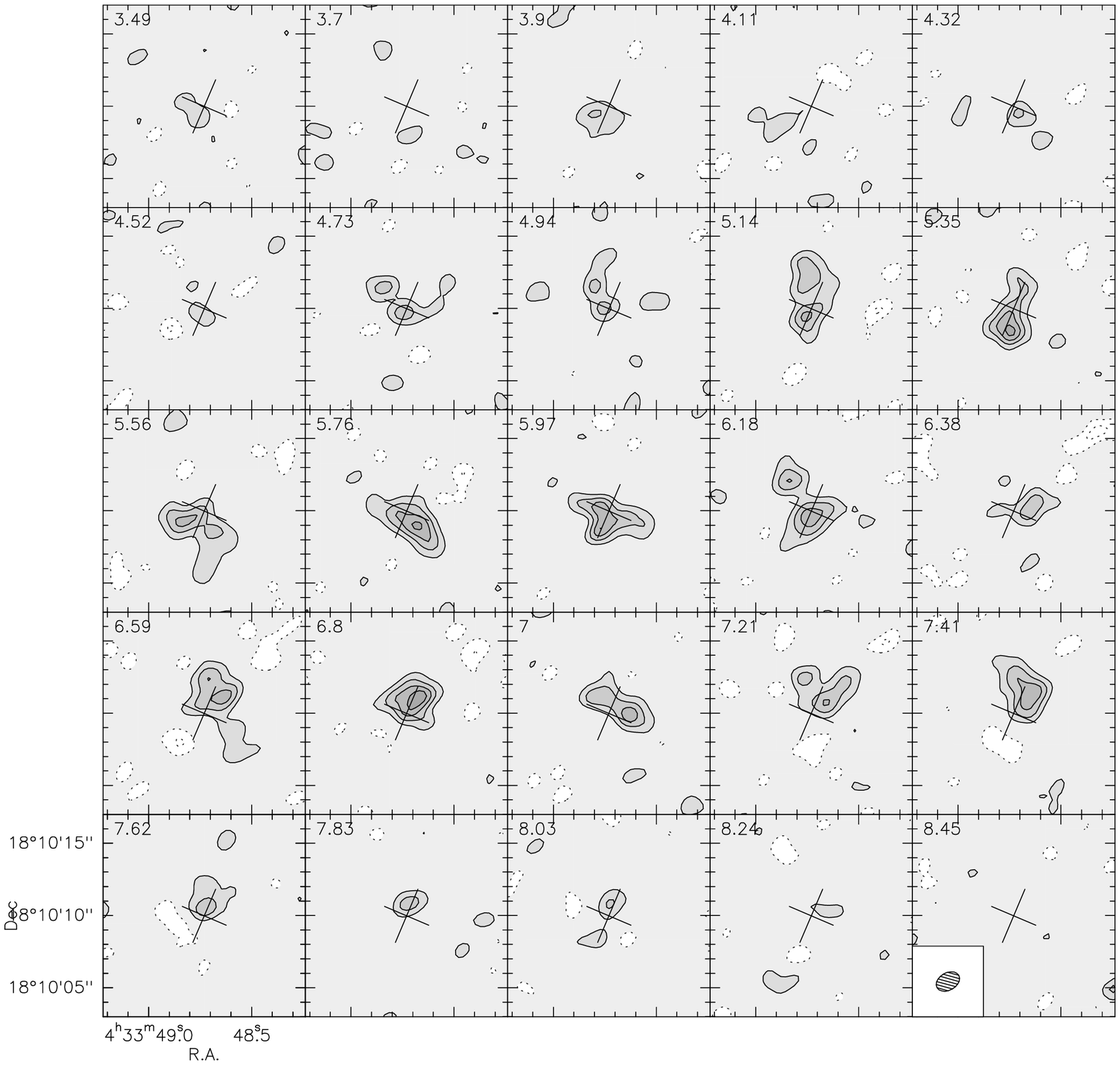}
  {\large\textbf{MWC 480}}\\
  \includegraphics[width=0.9\columnwidth]{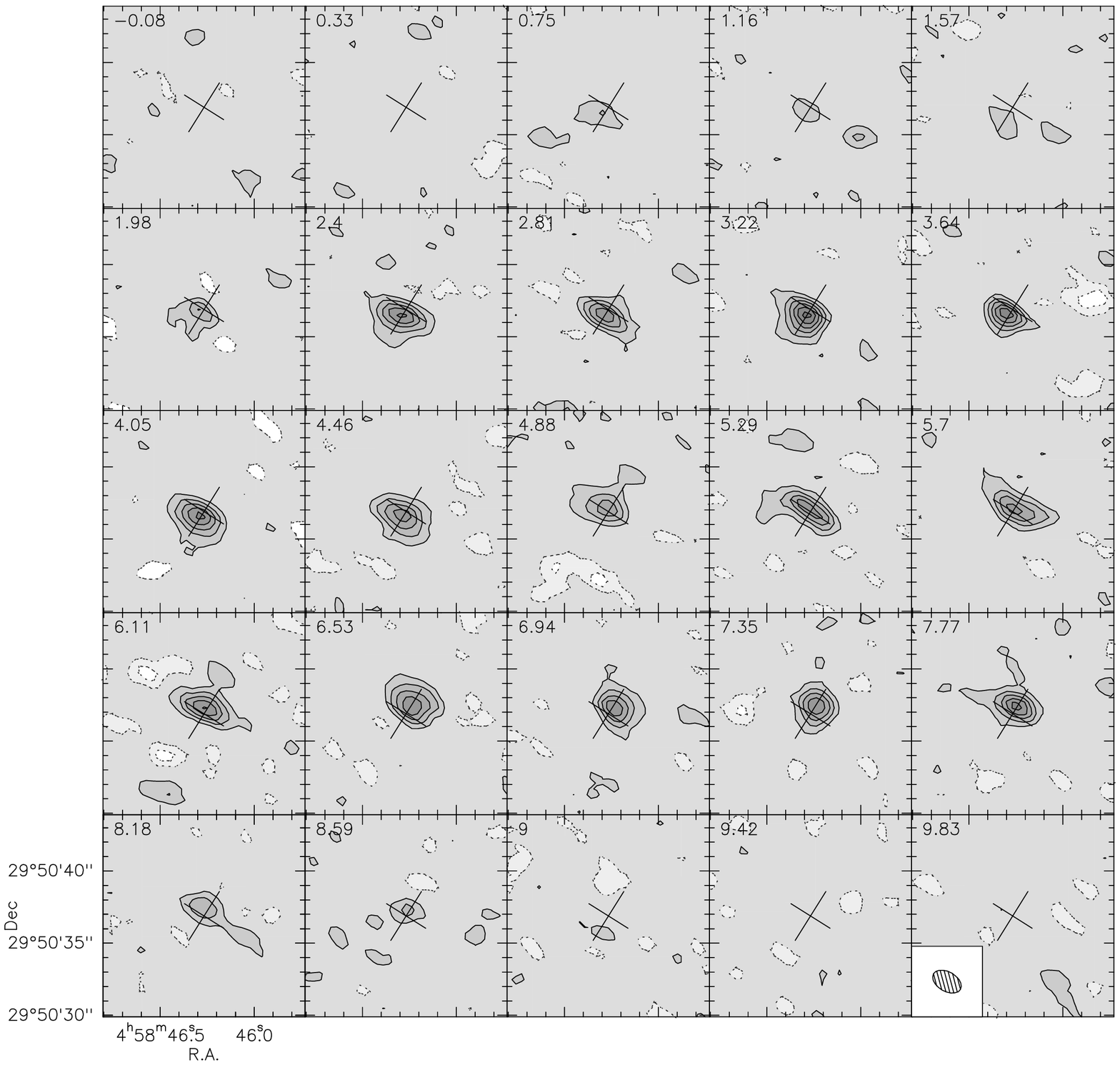}
  \caption{Channel maps of the CN 2-1 emission for the strongest group
  of three blended hyperfine components.
   Top: LkCa\,15; the spatial resolution is
   $1.7\times 1.0''$ at PA $44^\circ$, contour spacing is 60 mJy/beam, or 0.81 K and $2.1\sigma$.
   Middle: DM\,Tau; the spatial resolution is
   $1.7\times 1.2''$ at PA $120^\circ$, contour spacing is 53 mJy/beam, or 0.61 K, $2.0 \sigma$.
   Bottom: MWC\,480; the spatial resolution is
  $2.2\times1.35''$ at PA $60^\circ$, contour spacing is 55 mJy/beam, or 0.45 K and $2 \sigma$.
   }\label{fig:map:cn}
\end{figure}

\begin{figure} 
  {\large \textbf{LkCa 15}} \\ 
  \includegraphics[width=\columnwidth]{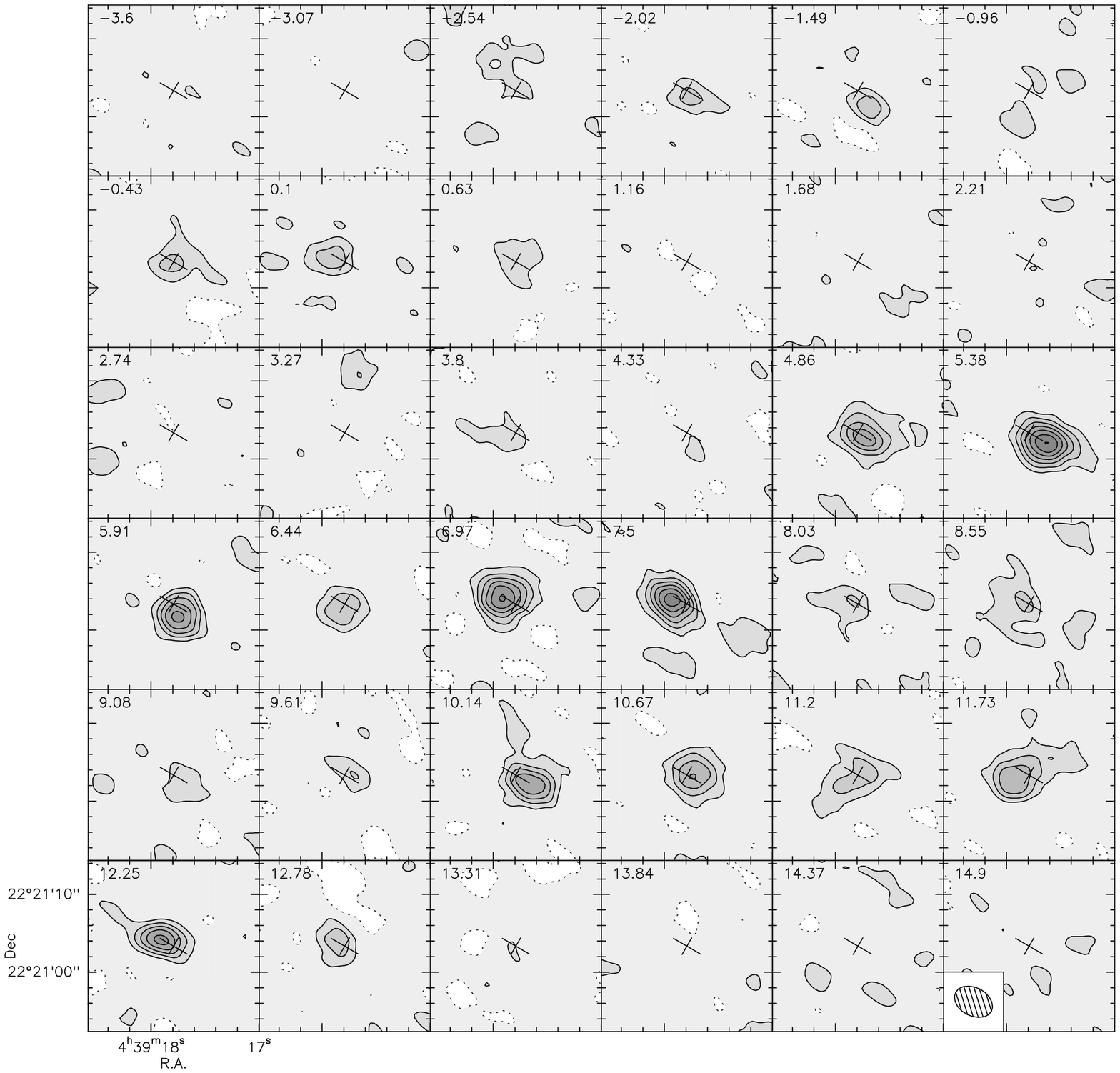}\\
  {\large \textbf{DM Tau}}\\ 
  \includegraphics[width=\columnwidth]{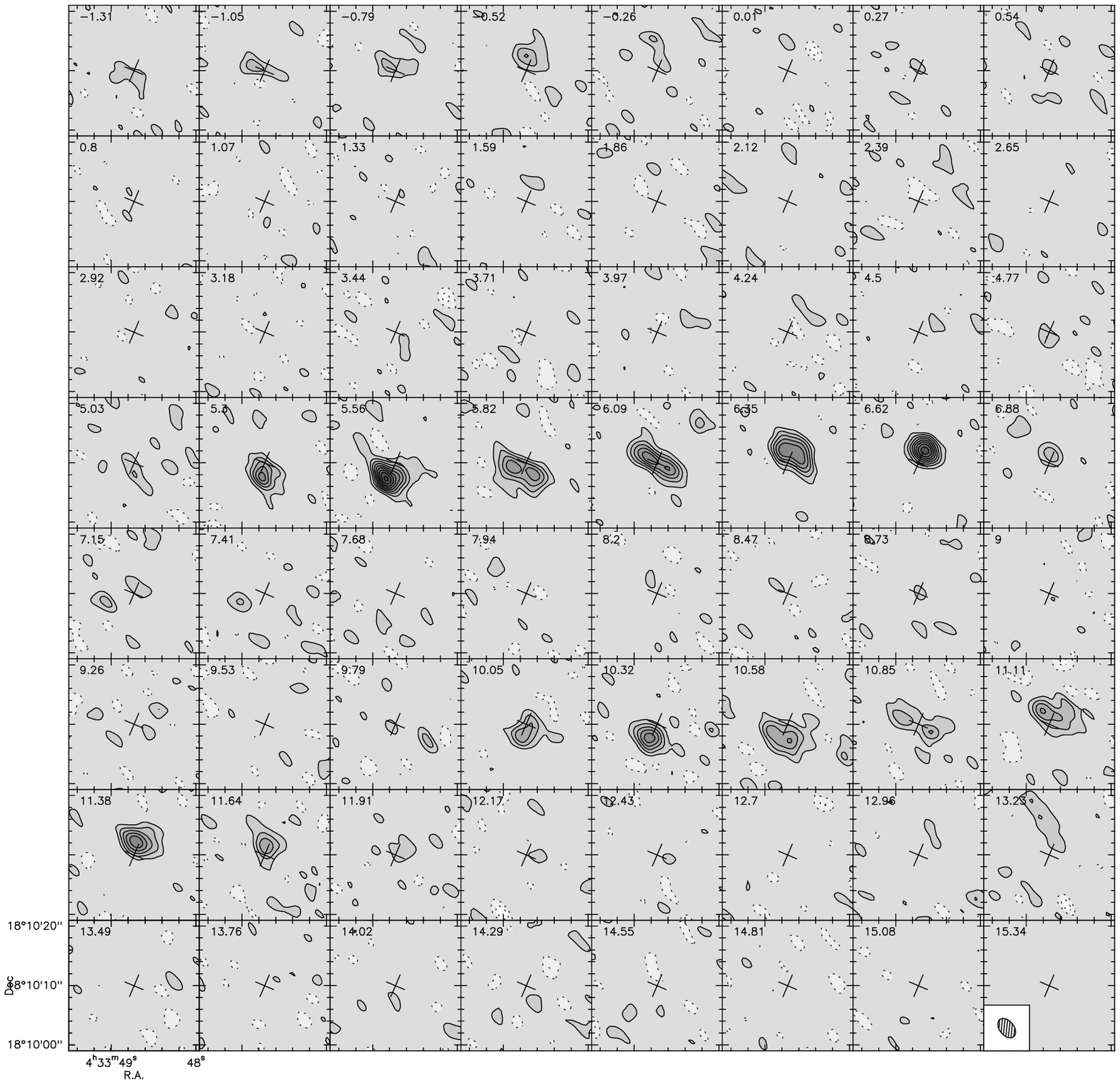}
  \caption{
\textbf{Channel maps} of the HCN 1-0 emission.
   Top: LkCa\,15; the spatial resolution is $4.1\times 2.9''$ at PA $51^\circ$, contour spacing is
   16 mJy/beam, or 0.21 K and $2.0 \sigma$. Bottom: DM\,Tau; the spatial resolution is
   $3.7\times 2.4''$ at PA $39^\circ$, contour spacing is 12 mJy/beam, or 0.21 K, $2.0 \sigma$.}
   \label{fig:map:hcn}
\end{figure}

\begin{figure} 
  \includegraphics[angle=270,width=\columnwidth]{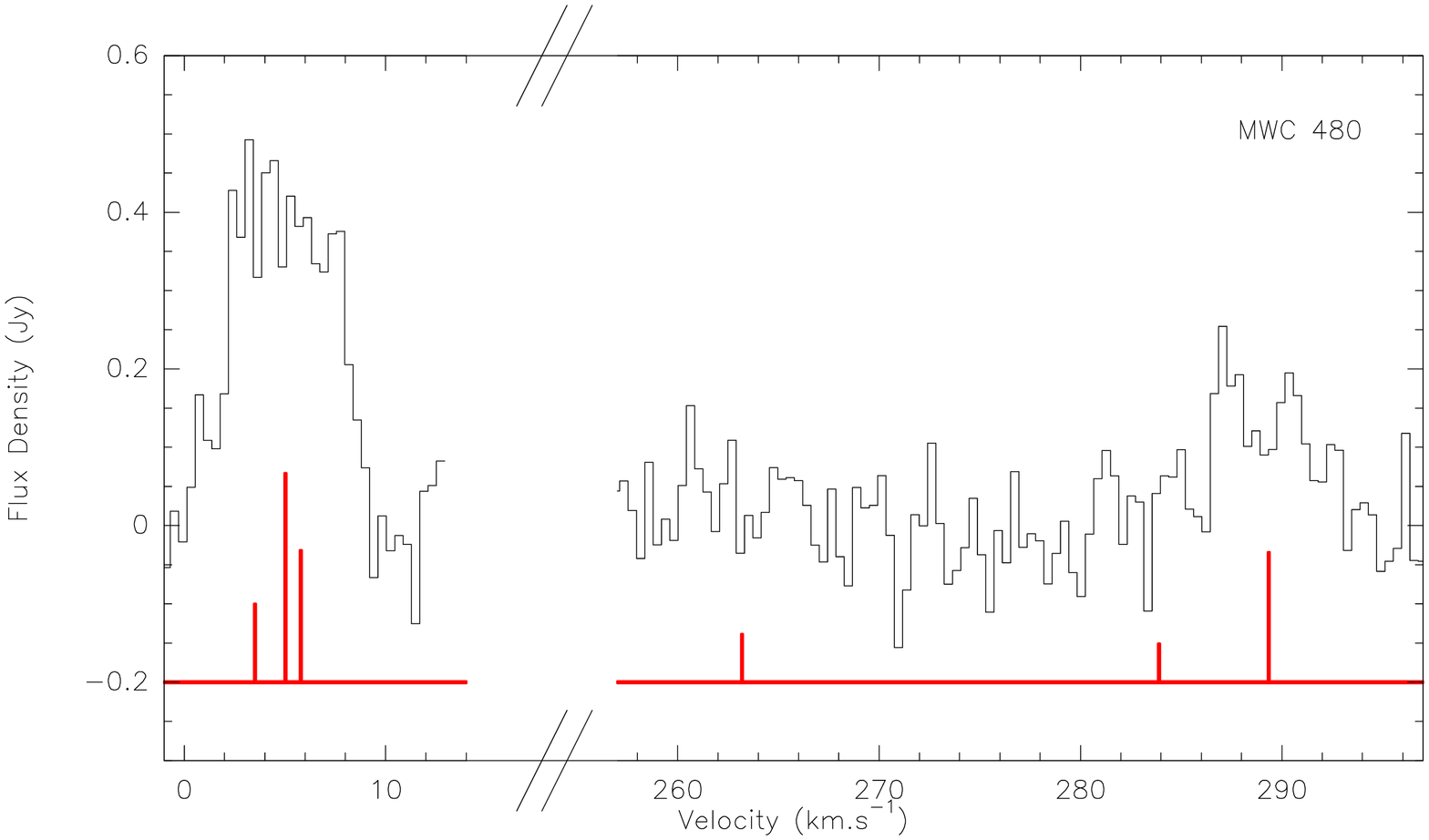}\\
  \includegraphics[angle=270,width=\columnwidth]{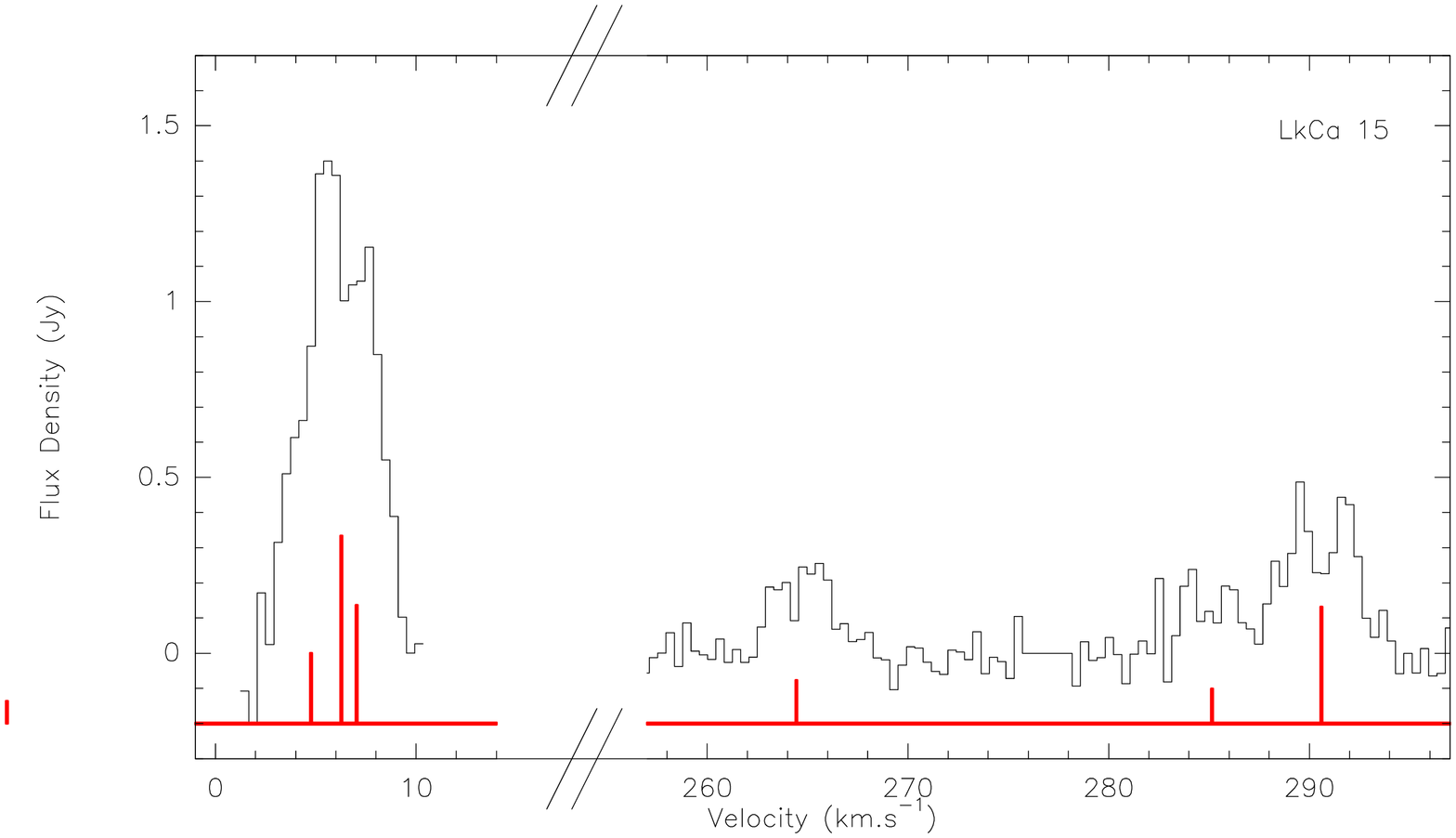}\\
  \includegraphics[angle=270,width=\columnwidth]{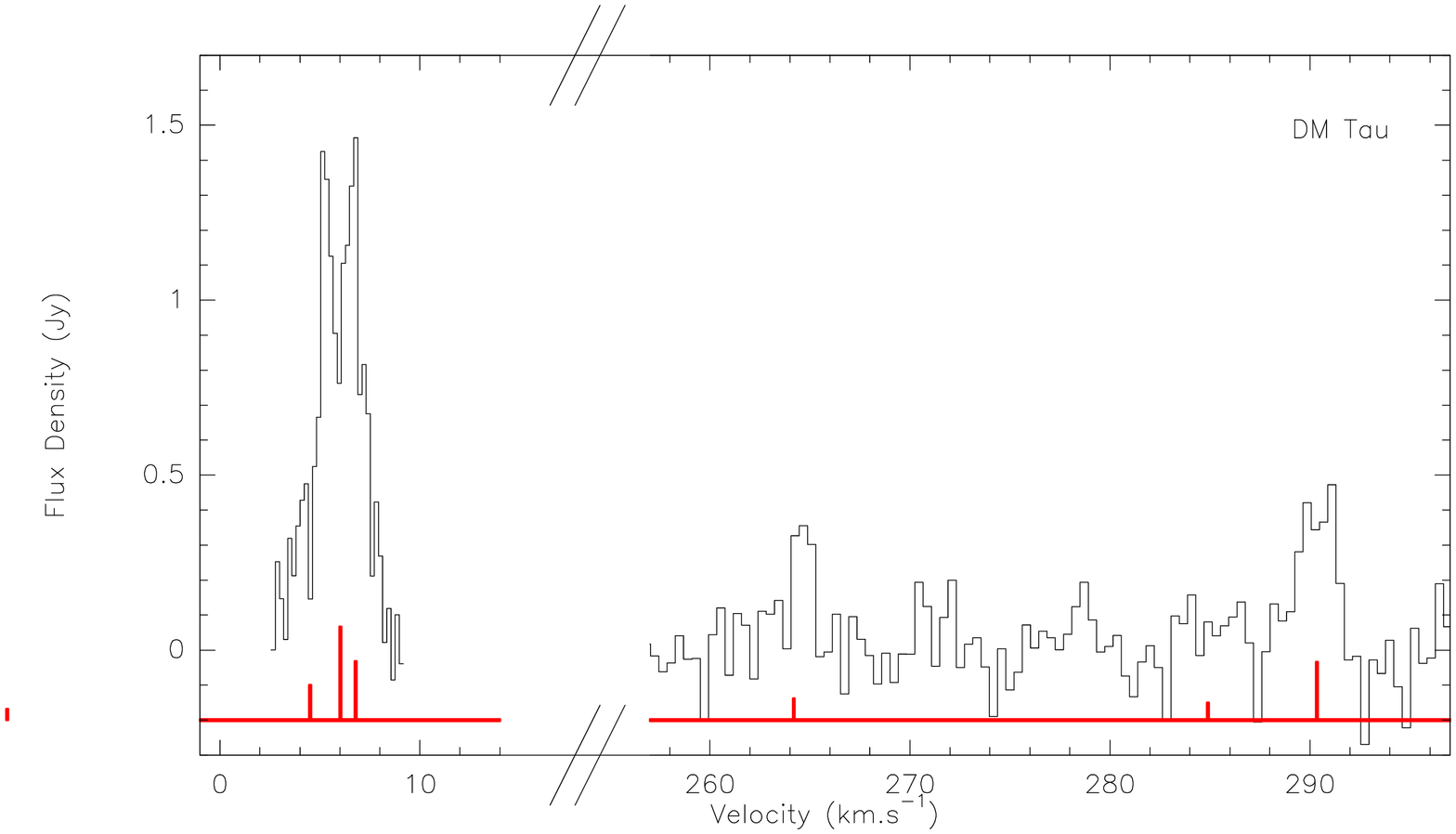}
  \caption{Integrated spectra for CN 2-1. The relative positions and intensities of
   the hyperfine components are indicated in red. Top: MWC\,480; middle: LkCa\,15; bottom: DM\,Tau}\label{fig:spectra:cn}
\end{figure}

\begin{figure} 
  \includegraphics[width=\columnwidth]{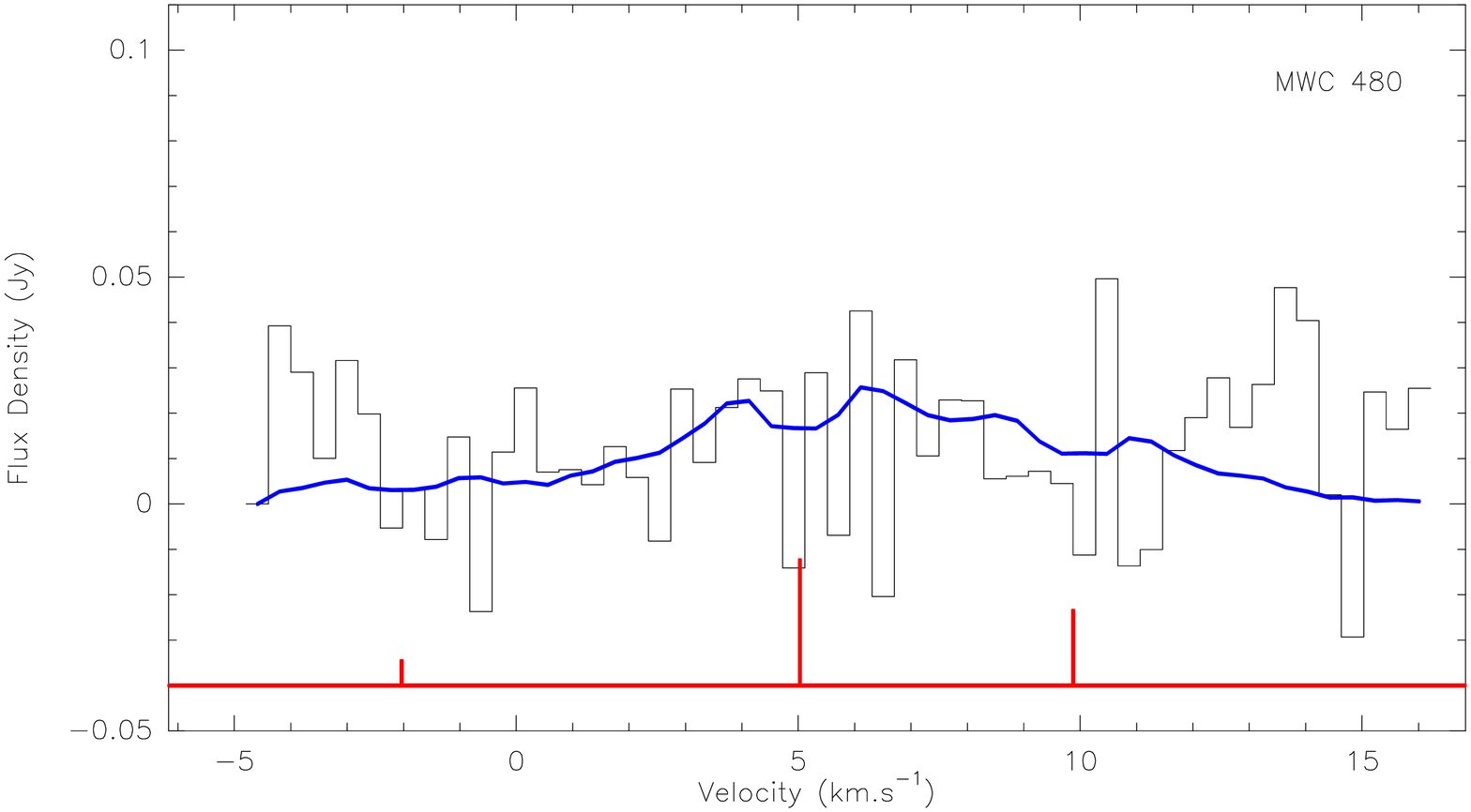}\\
  \includegraphics[width=\columnwidth]{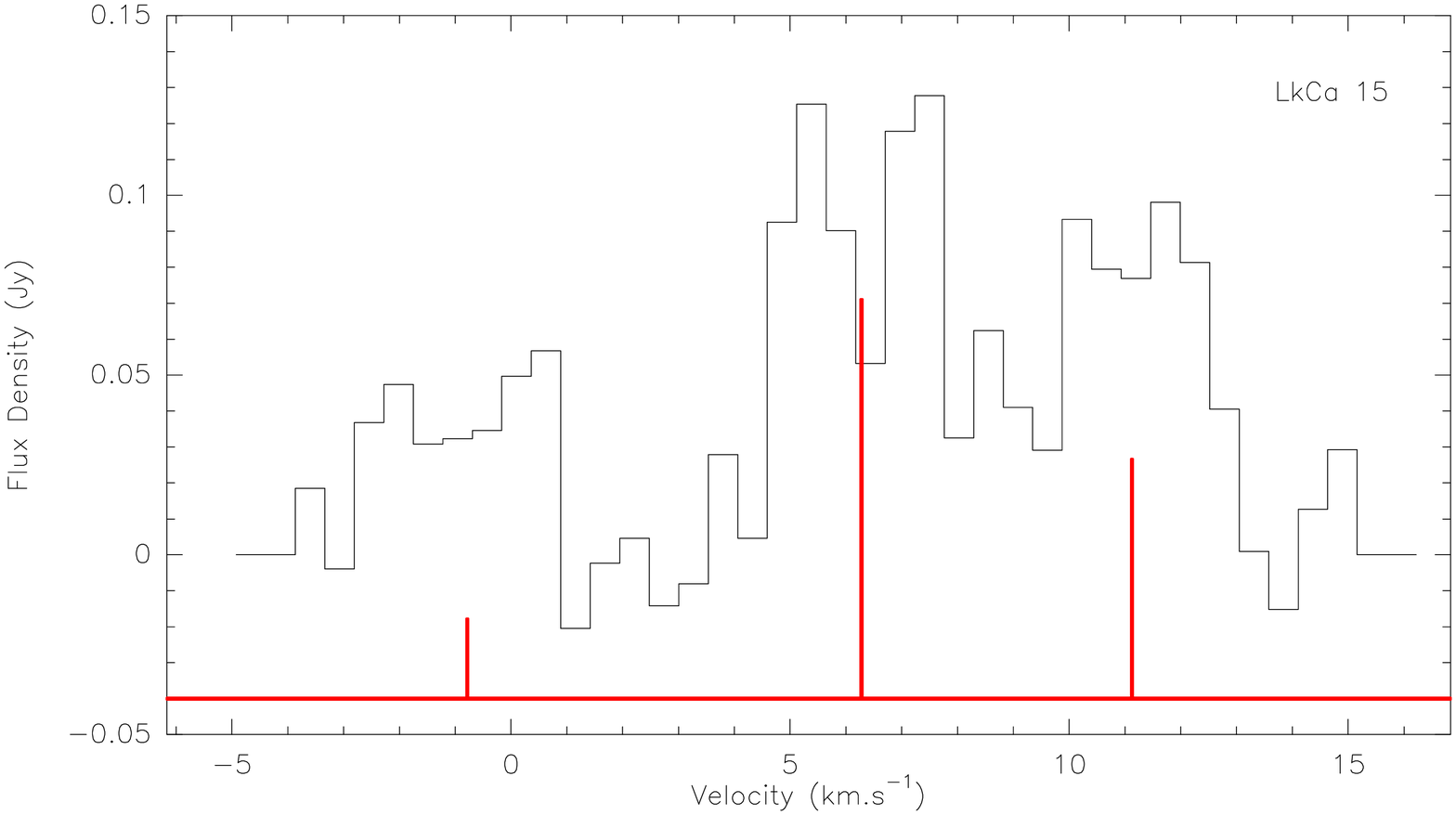}\\
  \includegraphics[width=\columnwidth]{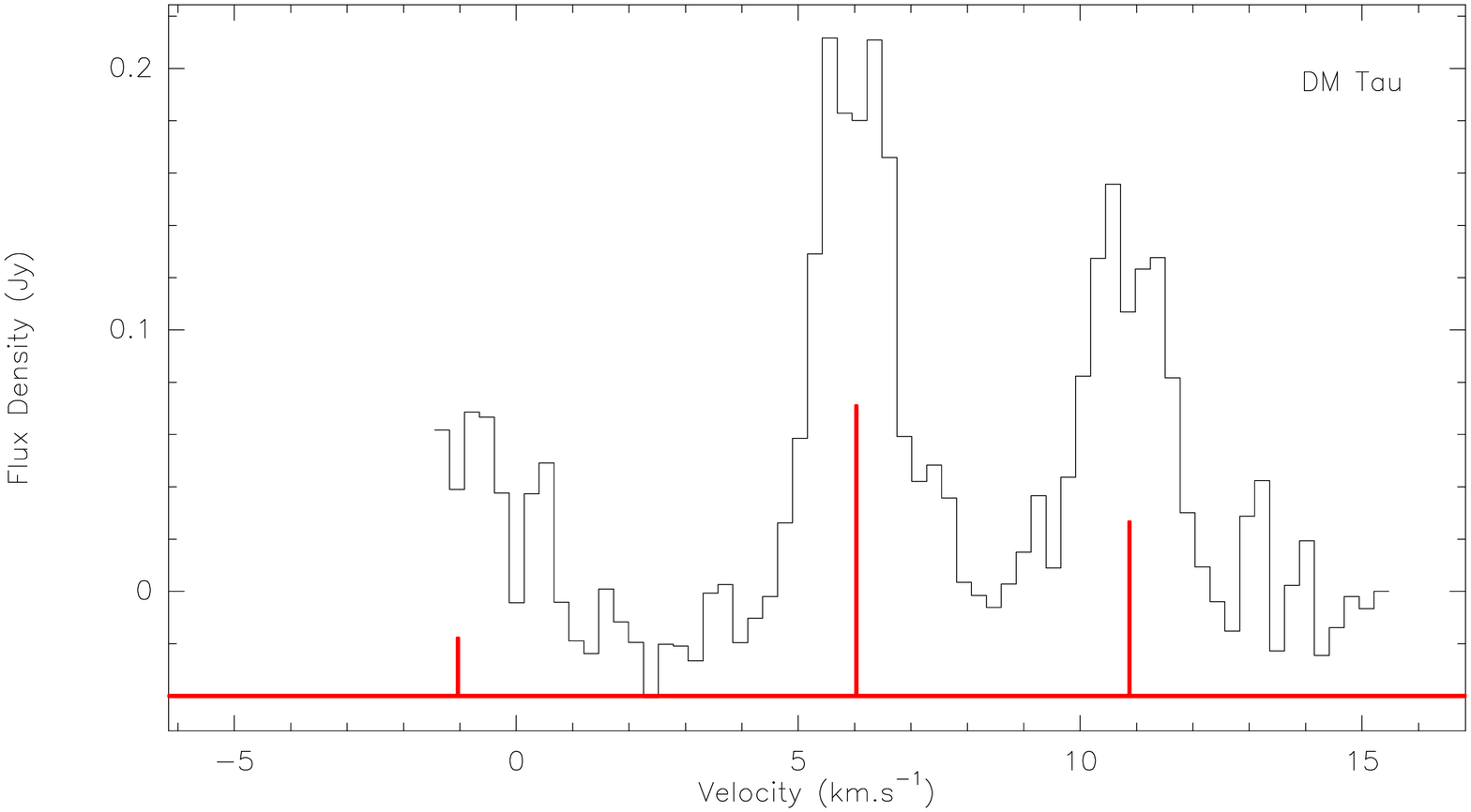}
  \caption{Integrated spectra for HCN 1-0. The relative positions and intensities of
   the hyperfine components are indicated in red. Top: MWC\,480, with the best fit model (see Tab.\ref{tab:results}) superimposed to
   illustrate the weak detection; middle: LkCa\,15; bottom: DM\,Tau}\label{fig:spectra:hcn}
\end{figure}

We used the GILDAS\footnote{See \texttt{http://www.iram.fr/IRAMFR/GILDAS} for more
information about the GILDAS software.}  software package \citep{pety05} for the data
calibration and deconvolution, and as framework for the implementation our minimization
method. The resulting velocity-channel maps in the HCN~\juz\ and CN~\jdu\ lines are shown
in Fig.\,\ref{fig:map:hcn} and \ref{fig:map:cn}) and the spectra integrated over the
disk are presented in Fig.\,\ref{fig:spectra:hcn}-\ref{fig:spectra:cn}. \textbf{In MWC\,480, the HCN emission
is weak, but significant at the $7\sigma$ level (see Fig.\ref{fig:map:mwc480hcn}). Channel maps for HCN in MWC\,480 line are only presented in Appendix, which also includes a more complete set of channel maps for all sources, in particular those of the other observed CN hyperfine components.}
We caution that
although the integrated spectra in Fig.\,\ref{fig:spectra:hcn}-\ref{fig:spectra:cn} were obtained from the Clean images,
they underestimate the flux. This is not due to the lack of short spacings, but to the
difficulty of deconvolving channel per channel low signal-to-noise data when the
synthesized beam has relatively high sidelobes. Furthermore, we stress that
this imaging bias has no influence on our derived parameters, since all our analysis
is based on a comparison between model visibilities and actual measurements in the UV plane.
Additional channel maps for CN, and also HCN in MWC\,480 reveal

Calibration errors could directly affect the derived temperature
and molecular column densities.  To minimize the calibration errors, all our flux density
scale is based on a comparison with MWC\,349, a source with precisely known spectral
index of 0.6. In addition, as in \citet{Guilloteau+etal_2010}, MWC\,480 is strong enough
in continuum to provide an additional comparison; the final relative flux scale accuracy
is about 5\%. Our flux density scale here is identical
to that used by \citet{Guilloteau+etal_2010}. However, recent re-calibration of the MWC\,349 flux
density at mm wavelengths indicate that this overall scale may be too low by about 10 \%  (Krips et\,al. in prep.).

The velocity was smoothed to 0.41 $\kms$ in the cases of MWC\,480 and LkCa\,15 to produce
the velocity-channel maps on Fig.\,\ref{fig:map:cn} and to $0.206 \kms$ in the case of
DM\,Tau. We used however full spectral resolution in the model fitting analysis described
below: for CN(2-1), $0.206 \kms$  in LkCa\,15 and MWC\,480, and $0.103 \kms$ in DM\,Tau.
For HCN(1-0), spectral resolutions were $0.52 \kms$ in LkCa\,15, $0.40 \kms$ in MWC\,480 and $0.26 \kms$ in DM\,Tau.
\textbf{The velocity scale is referred to the strongest hyperfine components, i.e.  226.874745 GHz for CN, and
88.631848 GHz for HCN}

Prior to this paper, CN and HCN emission has been already detected in several T-Tauri and
Herbig Ae stars (DM\,Tau, GG\,Tau; LkCa\,15,  HD\,163296, MWC\,480 and V4046 Sgr), but only
with single-dish telescopes, hence at low spatial resolution \citep{Dutrey+etal_1997,
Thi+etal_2004,Kastner+etal_2008},  except for the most nearby star TW\,Hya \citep{Qi+etal_2008}.
A preliminary study of CN(2-1) and HCN(3-2) at $\simeq 2.5''$ resolution has been recently
published by \citet{Oberg+etal_2010}. The observations relevant to our sources are summarized
in Table \ref{tabothers}.

\begin{table}
\caption{Integrated flux of CN and HCN lines}
\label{tabothers}
\begin{center}
\begin{tabular}{lccc}
\hline
\hline
& \multicolumn{3}{c}{S$_\nu$ (Jy\,km\,s$^{-1}$)}\\
& DM\,Tau & LkCa\,15 & MWC\,480\\
\hline
HCN J=1-0 & 0.4$^a$ & & \\
CN J=1-0 & 1.4$^a$ & - & -\\
CN J=2-1 & 8.7$^a$/4.7$^c$ & 6.8$^c$ & 3.3$^c$ \\
HCN J=3-2 & 2.9$^c$ & 5.5$^c$ & 2.3$^c$\\
CN J=3-2 & - & 12$^b$ & 5.3$^b$ \\
HCN J=4-3 & - & 4.6$^b$ & $<1.3^b$ \\
\hline
\end{tabular}\\
\end{center}
\tablefoot{Integrated line intensities from $^a$ IRAM\,30m observation \citep{Dutrey+etal_1997}
and $^b$ JCMT observation \citep{Thi+etal_2004}. The conversion factor
between Jansky to Kelvin (main beam temperature) for the JCMT (beam
$\sim13.5''$) at 345GHz is 18.2\,Jy/K. $^c$ SMA data
from \citet{Oberg+etal_2010}.
}
\end{table}

\section{Method of analysis and results}
\label{sec:model}

Whereas the velocity-channel maps of Fig.\,\ref{fig:map:hcn} and \ref{fig:map:cn}) were produced from the dirty images via Clean-based deconvolution,
 the disk properties were derived directly from the observed
visibilities through a model-fitting. For this, we used the disk model and hypothesis
described in \citet{Pietu+etal_2007}. We assume circular symmetry. The surface density of
molecules follows in radius $r$ a power law of exponent $p$; perpendicularly to the plane
of the disk, the density of molecules follows a Gaussian of half-width at 1/e intensity
(scale height) $h(r)$. We further assume the population of each molecular species can be
represented by a rotation temperature $T$ which depends only on $r$ and follows a power law
of exponent $q$. The disk is thus described by
 \begin{eqnarray}\label{eq:laws}
    T(r) &  = & T_0 (r/R_0)^{-q} \\
    \Sigma(r) &  = & \Sigma_0 (r/R_0)^{-p} \\
    h(r) &  = & H_0 (r/R_0)^{-h},~~~\mathrm{with~}h=-1.25
 \end{eqnarray}
plus a constant turbulent width $\delta V$ and an inner and an outer radius. If the
molecule is at LTE, $T(r)$ is equal to the kinetic temperature.

\subsection{Minimization technique}
The minimization technique used to derive the disk parameters has been described by
\citet{Pietu+etal_2007}. The continuum emission was subtracted in the UV plane
prior to the analysis, rather than being fitted simultaneously.

As explained by \citet{Pietu+etal_2007}, temperature and surface densities can be
derived by the minimization procedure from a single transition, even without hyperfine (hf)
structure, under the assumption of power law distributions, provided the spatial resolution
is high enough to partially resolve the optically thick core. This is because
the shape of the radial dependence of the brightness changes when the line becomes optically thick, from $T_b(r) = T(r)$
in the optically thick region, to a function of  surface density and temperature in
the optically thin domain (e.g. $T_b(r) \propto \Sigma(r)/T(r)$ at high temperatures).
 Lines with hf components yield additional information that may even
allow to test the power law hypothesis.

Handling of the hf components in the modeling requires additional care.
Under our assumption of common excitation temperature,  the apparent hf intensity ratios should
depend only on the line opacities. However, when comparing model and observations, limited spectral resolution and discrete
velocity sampling may introduce a bias. First, the spectral backend delivers
the flux integrated over a channel width, and not just sampled at the channel frequency.
Second, the separation between the hf components is not an integer multiple of the velocity
sampling. In the case of HCN, the intrinsic line widths are comparable to the channel
resolution; then the difference between sampled and integrated values, combined with
different sampling for each hf component, may change the predicted hf intensity ratios, which in turn
leads to incorrect opacities and to erroneous temperatures and surface densities. We
ensured that no such bias is introduced by comparing the results obtained directly from the
disk model by using the same sampling in frequency as in the real data to those derived
with a much higher sampling, followed by a smoothing that simulates the channel widths and
center frequencies. We found almost no differences between the two sets of results, except
for a very small (0.02 km.s$^{-1}$) increase in the line width. The hf line intensity
ratios were almost identical. In the case of CN, the intrinsic spectral resolution, 0.05
km.s$^{-1}$, is small compared to the line widths.

In the minimization procedure, all HCN hf components were fitted together assuming a common
excitation temperature. For CN, the hf components were used in the analysis in two ways:
all together and the intrinsically strong and weak components separately. The 3 sets of
results were then compared.

\subsection{Analysis of CN and HCN lines}

In a first step of the analysis, all geometric parameters of the disk were left free. The
derived parameter values were in complete agreement with those derived by
\citet{Pietu+etal_2007} from CO, but had a lower accuracy as the lines are weaker.
Accordingly, we fixed the inclination and position angle of the disk, as well as the
stellar mass and gas scale height using \citet{Pietu+etal_2007} values (see Table \ref{tabgeom}).
For LkCa\,15, we fix the internal radius $R_\mathrm{int}$ to 38\,AU, a compromise between the CO and dust inner radii from \citet{Pietu+etal_2006,Pietu+etal_2007}.
In a second step only $\Sigma_0$ and $p$ (the value and exponent of the surface density law),
$R_\mathrm{out}$, the outer radius, $T_0$ and $q$ (the value and exponent of the
temperature law) and $\delta V$, the local line width of each molecule, were fitted.

\begin{table}
\caption{Modeling: fixed \textbf{parameters}}
\label{tabgeom}
\begin{center}
\begin{tabular}{lccc}
\hline
\hline
Parameters & DM\,Tau & LkCa\,15$$ & MWC\,480\\
\hline
 P.A.($^o$) & 65  & 150  &  58 \\
$i$($^o$) & -35 & 52  &  37 \\
V$_{lsr}$(km.s$^{-1}$) & 6.00 & 6.28  & 5.02 \\
V$_{\mathrm{100}}$(km.s$^{-1}$) & 2.15 & 2.90 & 4.05 \\
R$_\mathrm{int}$(AU) & 1 & 38  & 1  \\
\hline
\end{tabular}\\
\tablefoot{The parameters are taken from \citet{Pietu+etal_2006,Pietu+etal_2007}.
}
\end{center}
\end{table}

In addition, as many chemical models predict all molecules should be frozen onto
dust grains close to the disk mid-plane, we tested models in which the molecules are
present only above the disk plane. We tested two different parameterizations.
Parametrization (1) is a simple geometric constraint, in which molecules only exist
above $z(r) = \alpha_d h(r)$, with $\alpha_d$ (the ``depletion scale
height'') being a free parameter, i.e. $X_m$ being the molecular abundance,
\begin{equation}\label{eq:alphad}
    X_m(r,z) = 0 \mathrm{~~for~~} z < \alpha_d h(r)
\end{equation}
However, the chemical model of \citet{Aikawa+Nomura_2006} suggests that, in a given disk, molecular abundances depend
essentially on the (molecular hydrogen) column density between the disk surface and the location in the disk.
Thus, at large radii, molecules get closer to the disk plane than at smaller radii. As our measurements
are most sensitive to the 150 -- 400 AU region, assuming a constant $\alpha_d$ as above may be inappropriate.
We thus also considered a second option, where molecules are only present in the upper disk down a
constant depth in (molecular hydrogen) column density, $\Sigma_d$, the ``depletion column density'', i.e.
\begin{equation}\label{eq:sigmad}
X_m(r,z) = 0 \mathrm{~~for~~} \int_z^\infty n(r,\zeta) d\zeta > \Sigma_d
\end{equation}
which in the isothermal approximation yields
\begin{equation}\label{eq:sigmad-h}
X_m(r,z) = 0 \mathrm{~~for~~} \frac{\Sigma(r)}{2} ~\mathrm{erfc}\left(\frac{z}{h(r)}\right) > \Sigma_d
\end{equation}
erfc being the complementary error function. This second option is likely to be more realistic,
as it avoids placing molecules far off the plane at large distances. Note that $\alpha_d = 0$ corresponds to $\Sigma_d$ being large (equal to the H$_2$ column density).

\subsection{Results}
\begin{table*}
\caption{Results of analysis}
\label{tab:results}
\begin{tabular}{c|l|rrrrrrrr}
\hline
\hline
Source & Molecule & $\Sigma$ & $p$ & $R_\mathrm{out}$ & $\delta V$ & $T_k$ & $q$ &  $\alpha_d$ \\ 
 &                & \textbf($10^{12}$ cm$^{-2}$ )& & (AU) & \textbf{(km\,s$^{-1}$)} & (K) &   \\
\hline    
MWC\,480 & HCN 1-0 &  $1.1\pm0.4$ & $2.4\pm0.4$ & [550] & $0.3\pm0.2$ & [30] & [0] & $0.0 \pm 0.7$ \\
 & CN 2-1 (strong) & $11\pm1$ & $2.1\pm0.1$ & $540\pm 40$ &  $0.25\pm0.04$ & $33\pm6$ & [0] & $0.0 \pm0.3$ \\
 & CN 2-1 (weak) & $11\pm1.2$ & $2.1\pm0.15$ & $550\pm 70$ &  $0.26\pm0.07$ & $33\pm6$ & [0] & --  \\
 & CN 2-1 (all) & $10.4\pm0.9$ & $2.1\pm0.1$ & $545\pm 35$ &  $0.25\pm0.04$ & $30\pm4$ & [0]  & --  \\

\hline 
LkCa\,15 & HCN 1-0 & $10.6 \pm 1.5$ & $1.1 \pm 0.2$ & $600\pm40$ & $0.20 \pm 0.03$ & $7.0 \pm 0.6$ & $0.55 \pm 0.25$ & $0.0 \pm 0.8$ \\
  & CN 2-1 (strong) &  $38\pm4$ & $2.3\pm0.2$ & $570\pm20$ & $0.18\pm0.02$ & $10.8 \pm 0.7$ & $0.20 \pm 0.07$ &  $0.15 \pm 0.17$ \\
  & CN 2-1 (weak) &  $115\pm18$ & $0.2\pm0.2$ & $690\pm20$ & $0.26\pm0.03$ & $6.6 \pm 0.3$ & $0.80 \pm 0.07$ & --  \\
  & CN 2-1 (all) &  $58\pm5$ & $0.8\pm0.1$ & $630\pm35$ & $0.18 \pm0.01$ & $8.8 \pm 0.3$ & $0.95 \pm 0.05$ & -- \\
\hline   
DM\,Tau & HCN 1-0 & $6.5 \pm 0.9$ & $1.0 \pm 0.3$ & $660 \pm 20$ & $0.18 \pm 0.01$ & $6.0 \pm 0.4$ & $0.00 \pm 0.12$ &  $0.0\pm0.5$ \\
 & CN 2-1 (strong) &  $26 \pm 4$ & $2.1 \pm 0.15$ & $650 \pm 20$ &  $0.16\pm0.01$ & $8.6 \pm 0.5$ & $0.05 \pm 0.05$ & $0.0 \pm 0.3$\\
 & CN 2-1 (weak) &  $46 \pm 9$ & $0.7 \pm 0.15$ & $550 \pm 25$ &  $0.20\pm0.01$ & $6.5 \pm 0.5$ & $0.50 \pm 0.08$ & $0.0 \pm 0.3$\\
 & CN 2-1 (all) &  $35 \pm 9$ & $0.6 \pm 0.06$ & $620 \pm 15$ &  $0.17\pm0.01$ & $7.5 \pm 0.3$ & $0.60 \pm 0.05$ &  $0.0 \pm 0.3$\\
\hline
\end{tabular}\\
\tablefoot{Column density $\Sigma (r) = \Sigma \times (r/R_0)^{-p}$, external radius
$R_{out}$, local line width $\delta\,V$, temperature $T(r) = T \times (r/r_0)^{-q}$ and
depletion scale height $\alpha_d$ (see text). All values are referred to $r_0 =
300$\,AU. Square brackets indicate fixed parameters.  The error bars are $1\,\sigma$ uncertainties, and do not include uncertainties from calibration}
 \end{table*}

The disks of the two T Tauri stars, DM\,Tau and LkCa\,15, display very similar
properties (see Table \ref{tab:results}). Both show  very low CN and HCN rotation
temperatures, and their strongest hf line components have moderate optical depth
around 100 to 300 AU. The fit to the weakest,
optically thinner, CN hf components yields a nearly flat surface density
distribution and the fit to the stronger hf components a much steeper one. As a consequence
of our single power law hypothesis, the temperature distributions follow the opposite trend, i.e.
the temperature decreases faster for increasing $r$ for the optically thin lines. Such a
behavior may indicate that the single power law is not an appropriate representation of the
CN distribution and that the density falls down faster in the outer parts of the disk
(which are hardly detected in the weaker components) than in the inner parts.
A close examination of the shape of the $\chi^2$ surface, as a function of the exponents
$p$ and $q$, indeed reveals for the strong hf components a broad, curved and relatively
shallow global minimum, with two local minima that correspond to the two (flat and
steep) solutions just described.

We attempted fitting all components with a model including two power laws (crossing
at some radius $R_x$), but did not find any sensibly better results: in most cases, the fit converged towards one
of the two single power laws already identified. The rotation temperatures derived when varying $R_x$ remained within the range of temperatures allowed by the single power-law fits as listed in Table \ref{tab:results}.

Within the hypothesis where molecules are only in a surface layer, the depletion
parameter $\alpha_d$ can be constrained from the observations \citep[see][]{Semenov+etal_2008}, the
distribution of molecules being reasonably well described by two thin disks with slightly
different inclinations, so that their projected kinematic patterns differ. We find
$\alpha_d \simeq 0$ ($\alpha_d = 0 \pm 0.3$ for the more sensitive CN measurements -- see
Table \ref{tab:results}), indicating that the detected molecules are most likely
concentrated close to the disk plane.

For the second approach, we must provide the underlying H$_2$ density profile.
We assumed two different disk models for the H$_2$ density based
on the analysis of the millimeter continuum emission performed by \citet{Guilloteau+etal_2010}.
The first is a power law distribution; however, although the millimeter continuum emission
indicates outer radii of 200-300 AU, we extrapolate the H$_2$ surface density up to 800 AU,
which leads to a model with unlikely high densities at large radii.
The second one uses the exponentially tapered solution of self-similar viscous evolution.
Both solutions have similar surface densities near 100 AU, but differ in the outer parts, i.e.
beyond 200-250 AU. The H$_2$ surface density is about $4\times10^{23}$ cm$^{-2}$ near
100 AU. It decreases as $r^{-0.8}$ for the power law model, while the exponential taper leads
to much steeper decline beyond 300 AU.
For CN, the derived ``depletion column density'', $\Sigma_d$
is somewhat dependent on the disk model. For DM\,Tau, we find $\Sigma_d = 3 ^{+1}_{-0.7} \times 10^{22}$ cm$^{-2}$
for the power law, and $\Sigma_d = 8 \pm 1 \times 10^{22}$ cm$^{-2}$ for the viscous model.
In the former case, the error is highly asymmetric: the 3 and 4 $\sigma$ lower limits are : $\Sigma_d = 2$ and $1 \times 10^{21}$ cm$^{-2}$ respectively.

For the Herbig star (MWC\,480), the sensitivity of our observations is not good enough to derive both $T$ and $q$.
Nevertheless, assuming a slope $q=0$, a value of $T=30$\,K is fitted for the CN emission. That is
in accordance with the expected kinetic temperature of the molecular layer in the theoretical models. Using $q=0.4$ instead results in a temperature $T (300 \mathrm{AU}) = 14$~K, but affects neither the derived surface density
(by less than 5\%), nor the exponent $p$ (by less than 0.1). This is because in the 10 - 30 K temperature range,
the CN J=2-1 line brightness is weakly dependent on the temperature \citep[see][their Fig.\,4, given for CO
but also applicable to CN]{Dartois+etal_2003}. For HCN, which has low signal-to-noise ratio, we assumed the same temperature
($q=0$, $T=30$ K) and outer radius than derived from CN.
Although arbitrary, this choice is justified by the results found for DM\,Tau and LkCa\,15, where CN
and HCN display similar rotation temperatures and outer radii.

\begin{figure} 
 \centering
\includegraphics[width=6cm]{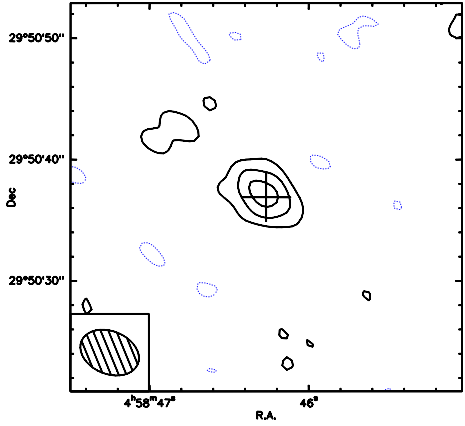}
\caption{Signal to noise image for the HCN J=1-0 integrated emission in MWC\,480 (after removal of the continuum emission), with a velocity weighting profile equal to
the best fit integrated model profile shown in Fig.\,\ref{fig:spectra:hcn}. Contours are in steps of $2 \sigma$.}
\label{fig:map:mwc480hcn}
\end{figure}

The detection of HCN in MWC\,480 is not easily illustrated, because the line intensity
is spread over the three hyperfine components of the J=1-0 line. The best fit
profile is shown in Fig.\,\ref{fig:spectra:hcn}. Fig.\,\ref{fig:map:mwc480hcn} shows a signal-to-noise
image obtained using the optimal filtering as described by \citet{Dutrey+etal_2007}, i.e. weighting
each channel by its modeled integrated intensity prior to averaging in velocity. There, the peak signal-to-noise
is around 6, and clearly points toward the MWC\,480 star, thereby showing unequivocally the reality of the
line emission from HCN.

All temperatures may be increased by 10\% to account for the calibration issues mentioned in Sec.\ref{sec:observations}. An effect of similar magnitude, although not as linear as for the temperature, is expected
on the molecular surface densities. However, the calibration uncertainties do not affect $q$, $p$, $R_\mathrm{out}$, $\alpha_d$ or $\Sigma_d$, which depend only the shape of the brightness distribution.

\section{Chemistry modeling}

\begin{table}
\caption{Elemental abundances with respect to total hydrogen}
\centering
\begin{tabular}{lc}
\hline
\hline
Element & abundance\\
C  & $1.38 \times 10^{-4}$\\
O  & $3.02 \times 10^{-4}$\\
N  &$7.954 \times 10^{-5}$\\
Mg &$1.0 \times 10^{-8}$\\
S  &$2.0 \times 10^{-6}$\\
Si &$1.73 \times 10^{-8}$\\
Fe &$1.7 \times 10^{-9}$\\
\hline
\end{tabular}
\label{tab:abon-chimi}
\end{table}

Generic chemical models of circumstellar disks have been published by
\citet{Aikawa+Herbst_1999,Willacy+Langer_2000,Aikawa+etal_2002,vanZadelhoff+etal_2003,Aikawa+Nomura_2006,Willacy+etal_2006,Fogel+etal_2011}.
Each of these works introduce a different degree of complexity. 
\citet{Aikawa+Herbst_1999} consider gas phase chemistry, plus sticking and desorption, and
assumed a vertically isothermal disk model following the ''Minimum Mass Solar Nebula'' extended
to 800 AU. The study from \citet{Willacy+Langer_2000} uses the thermal structure derived
from the \citet{Chiang+Goldreich_1997} two-layer approximation, as well as enhanced photodesorption yields
following \citet{Westley+etal_1995}. \citet{Aikawa+etal_2002} used a D'Alessio disk model including vertical
temperature gradients
and self-consistently variable flaring, although dust and gas are assumed to be fully thermally coupled.
All three models used a 1+1D approximation for radiative transfer, in which the stellar UV is attenuated
along the line of sight to the star, while the ISRF impacts isotropically on the disk surface.
\citet{vanZadelhoff+etal_2003} improved the UV treatment, by using a 2-D radiative transfer code
to solve for the UV field inside the disk, as \citet{Fogel+etal_2011}. Shielding by H$_2$ is treated in an approximate way, however.
\citet{Aikawa+Nomura_2006} further expanded the models by considering the effect of grain growth and \citet{Fogel+etal_2011}
investigate the effects of dust settling following the prescription by \citet{DAlessio+etal_2006}.
With the exception of \citet{Fogel+etal_2011} and \citet{Willacy+etal_2006}, none of these models include any grain surface chemistry, which may be important in such environments. However, all of them predict that CN is constrained to a warm molecular layer,
sometimes well above the disk plane, and that molecules are depleted in the disk plane.  The column densities
of CN and HCN do not vary much with radius. For a more direct comparison with with our results, we fitted
the column densities predicted by the chemical models by power laws as function of radius, i.e. we derived
$\Sigma_0$ and $p$. %

The model from \citet{Willacy+etal_2006} is somewhat different, as it also includes turbulent
diffusion, and will be discussed later. A more elaborate set of grain surface chemistry has
been included by \citet{Walsh+etal_2010}, who also find similar layered structure.

Existing chemical models are not necessarily tailored to the sources we have observed.
To better understand the impact of some unknown properties of the observed disks (e.g. UV flux, mass,
dust size), we performed new chemical network calculations.  The objective of these calculations is not to find an appropriate
model for each of the observed disks, but to illustrate common properties of disk chemical models, and pinpoint the
dependency between the disk parameters and the predicted molecular column densities. We focus here
on CN, which is predicted to be formed mainly in the ``warm'' photo-dissociation layer, where chemical reactions are fast.
This allows us to use a time-independent chemical model, rather than a more computer-intensive time dependent one, as
would be needed for molecules formed deeper in the disk.

\subsection{PDR code}
We use the PDR code from the Meudon group \citep{LeBourlot+etal_1993,LePetit+etal_2006}
modified to account for a non uniform grain size
distribution \citep[see][]{Chapillon+etal_2008}. The chemical model assumes a one-dimensional stationary plane-parallel slab
of gas and dust illuminated by an ultraviolet (UV) radiation field. The radiative transfer
in the UV takes into account self and mutual shielding of H, H$_2$ and CO lines
and absorption and diffusion of the continuum radiation by
dust grains in a 1-D geometry.
Molecular abundances and, optionally, thermal balance of the dust grains and
the gas are calculated iteratively at each point in the cloud. The chemical network is
similar to that of \citet{Goicoechea+etal_2006}. The chemical model does not consider the
freeze-out of atoms and molecules onto grains and surface chemistry reactions, to the
notable exception of those leading to the formation of H$_2$.

The assumed grain size distribution is a standard power-law $n(a) \propto a^{-\gamma}$,
where \app~and \amm~ are the maximum and minimum cutoff radii, respectively. The
introduction of a range of grain sizes affects the UV extinction curve, the rate of H$_2$
formation and the thermal balance of the gas.
The resulting extinction curve is calculated using the Mie
theory for homogeneous isotropic spherical particles. The composition of the dust is 50\%
silicate and 50\% graphite. We assume $\gamma=-3.5$ (the value observed in
nearby interstellar clouds) and $a_-=3$nm in all calculations. This minimum grain radius
$a_-=3$nm is chosen small enough to properly account for the photoelectric heating
process, the UV extinction, and the formation of H$_2$.

Keeping the dust mass constant, we have simulated grain growth by varying the maximum
radius \app (0.3$\mu$m -- 0.1mm) while keeping the exponent $\gamma$ constant
for the sake of simplicity. The amount of small grains therefore is reduced to the benefit
of large grains, modifying accordingly the extinction curve. With $\gamma=-3.5$ kept
constant, the UV opacity scales as $1/\sqrt{a_+}$ for $a_+ > 10 \mu$m  \citep[see][]{Chapillon+etal_2008}.

The chemistry code is one-dimensional and calculates the temperature, radiation
field and molecular abundances in the direction perpendicular to the surface, that we
assume here perpendicular to the disk plane. The justification for such an approach is that
small dust grains in the upper disk atmosphere scatter (half of) the incident UV flux
towards the disk. We mimic the second dimension by resuming the calculation at different
disk radii (12 to 15 different radii from 50\,AU to the external radius R$_{\mathrm{out}}$),
the output being a 1+1D model \citep[see][for more details]{Chapillon+etal_2008}.

\subsection{Set of PDR models}

We explore a range of parameters to understand the effects of UV field intensity, grain size
and cosmic ray ionization rate. Three basic disk models (with different density and
temperature structure) are used to represent the 3 sources DM\,Tau, LkCa\,15 and MCW\,480.
The input disk structure is similar to those of \citet{DAlessio+etal_1999}.
The density and (dust) temperature distributions were computed self-consistently from the
disk masses and stellar properties of each source by D.Wiebe \& D.Semenov \textit{(private
communication)} assuming uniform grain size of 0.1$\,\mu$m; the disk models display a vertical temperature gradient, the temperature
increasing with height  \citep[e.g.][]{Chiang+Goldreich_1997}. Except
when otherwise noted, we assumed the gas temperature to be equal to the dust temperature.
The elemental abundances used are given in Table \ref{tab:abon-chimi}.

\begin{figure*} 
 \centering
\begin{tabular}{ccc}
\includegraphics[angle=270,width=6cm]{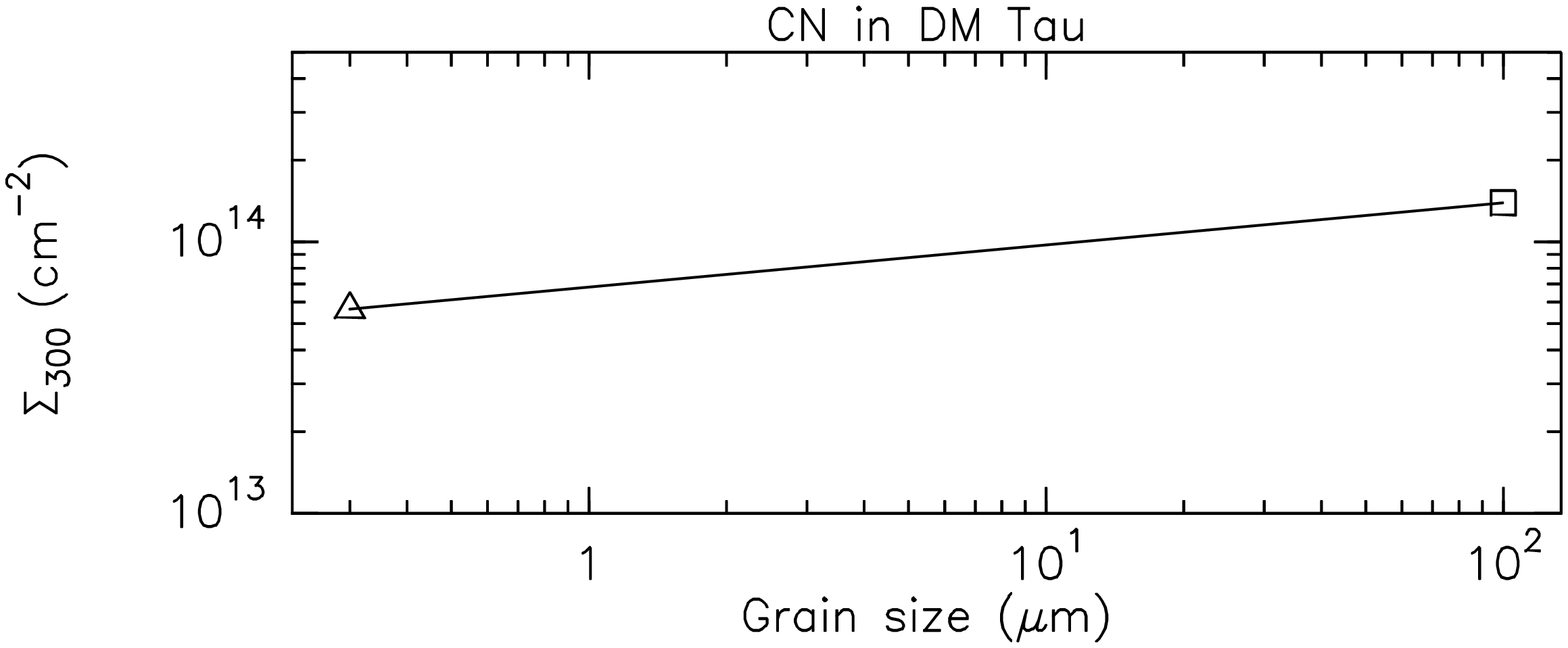} & \phantom{1234} & \includegraphics[angle=270,width=6cm]{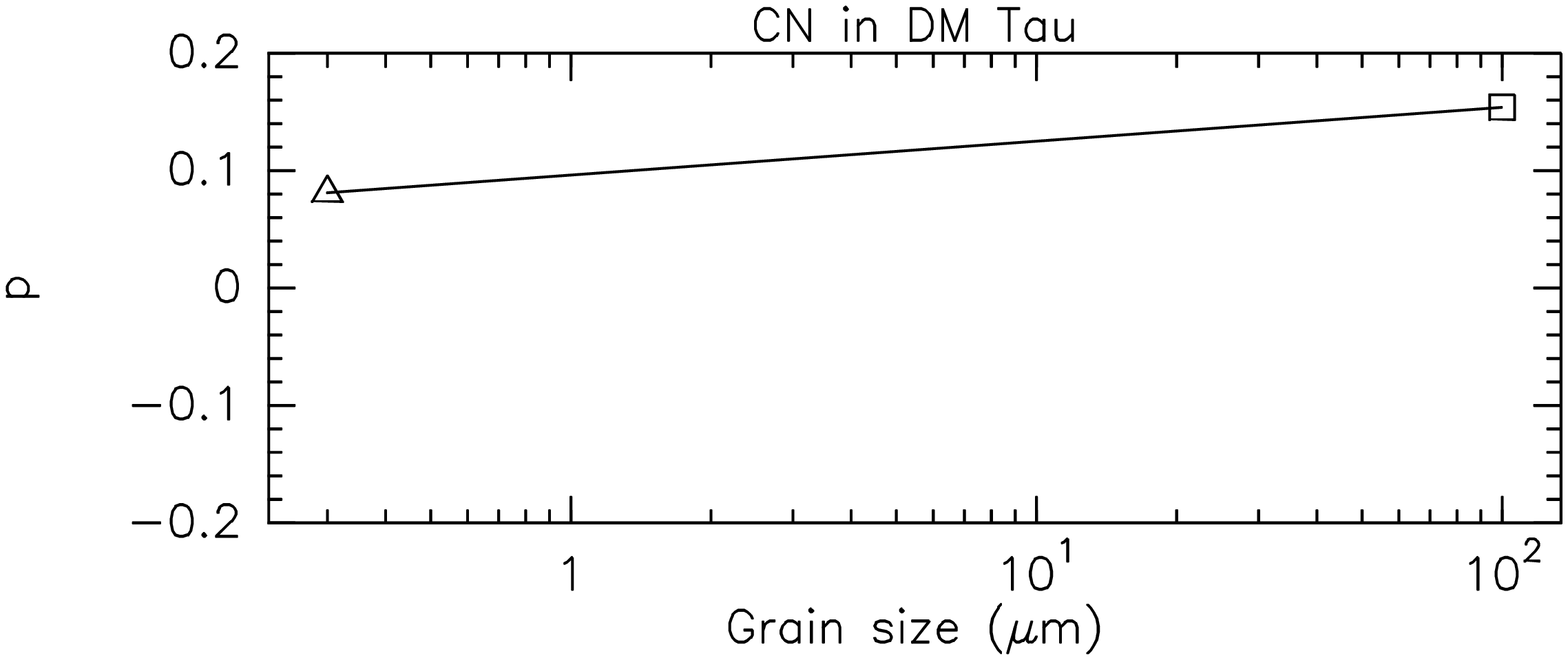}\\
\includegraphics[angle=270,width=6cm]{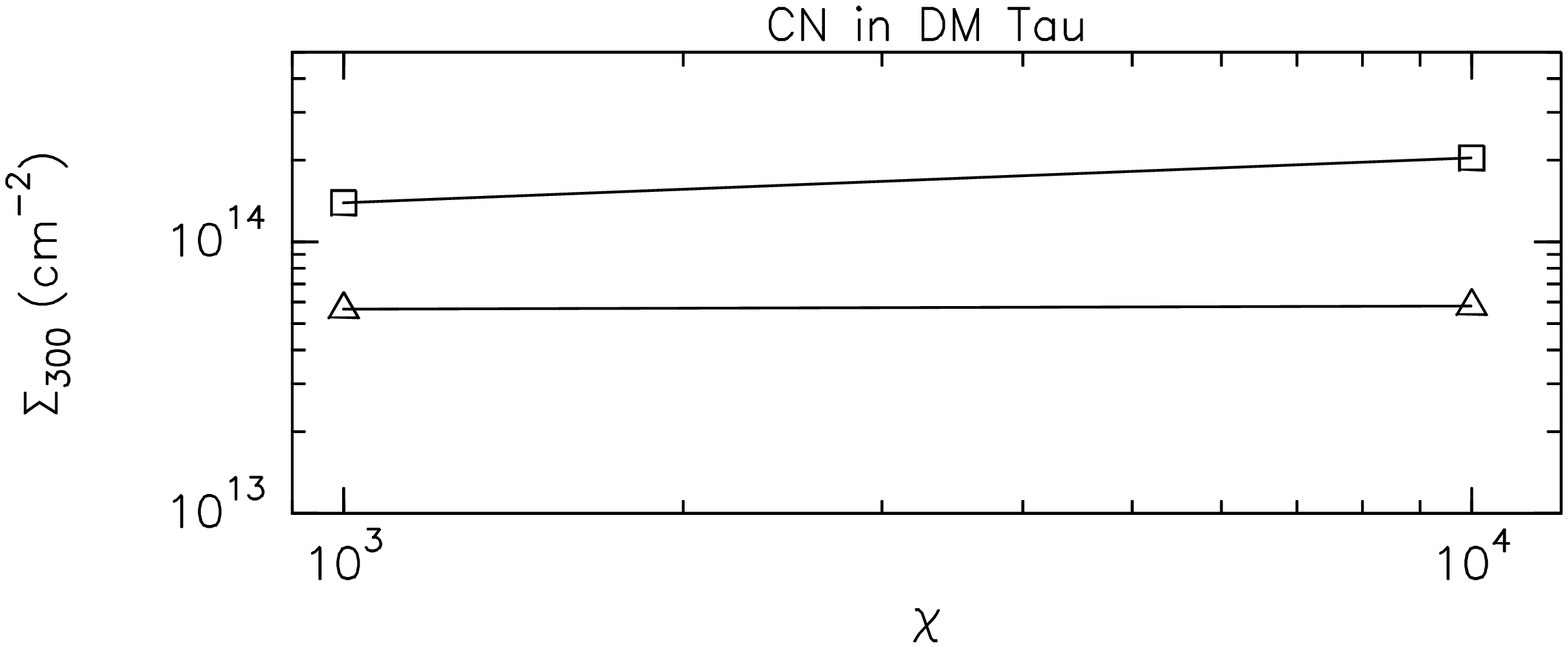} & & \includegraphics[angle=270,width=6cm]{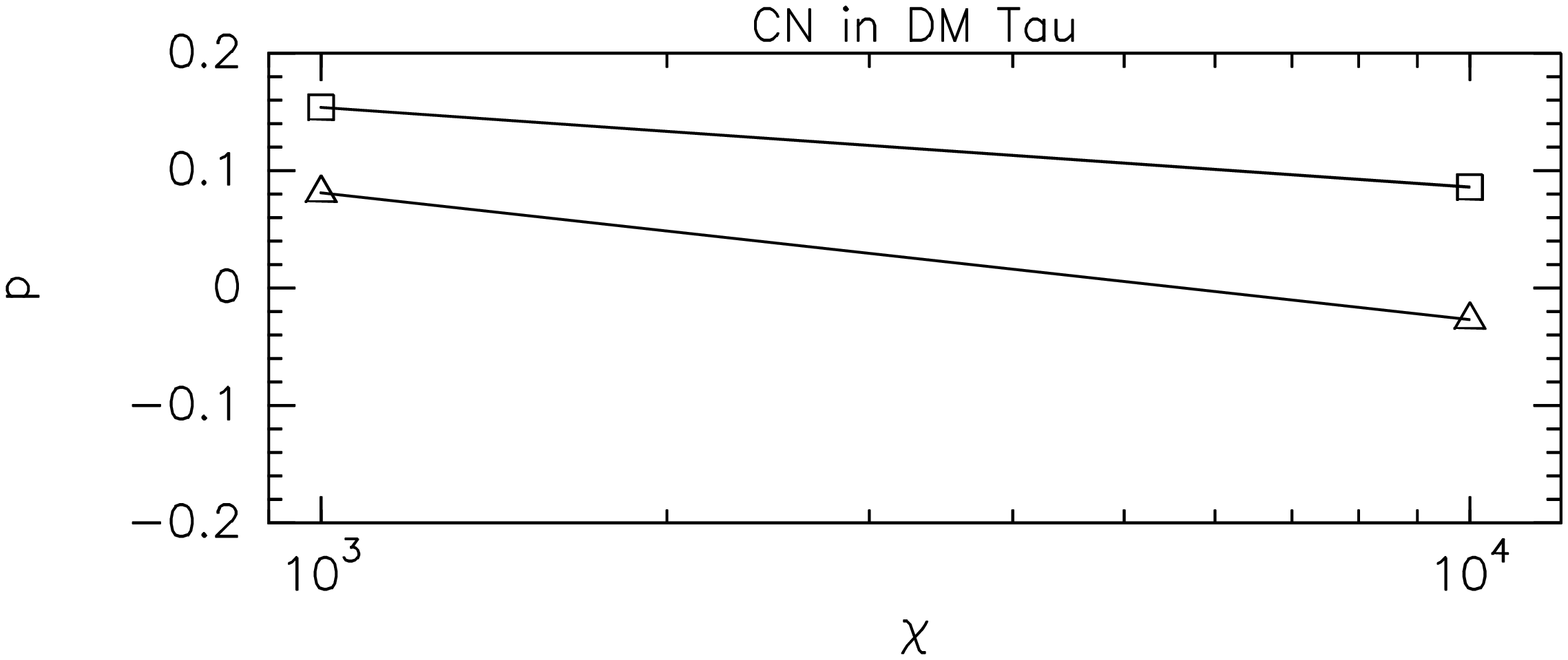}\\
\includegraphics[angle=270,width=6cm]{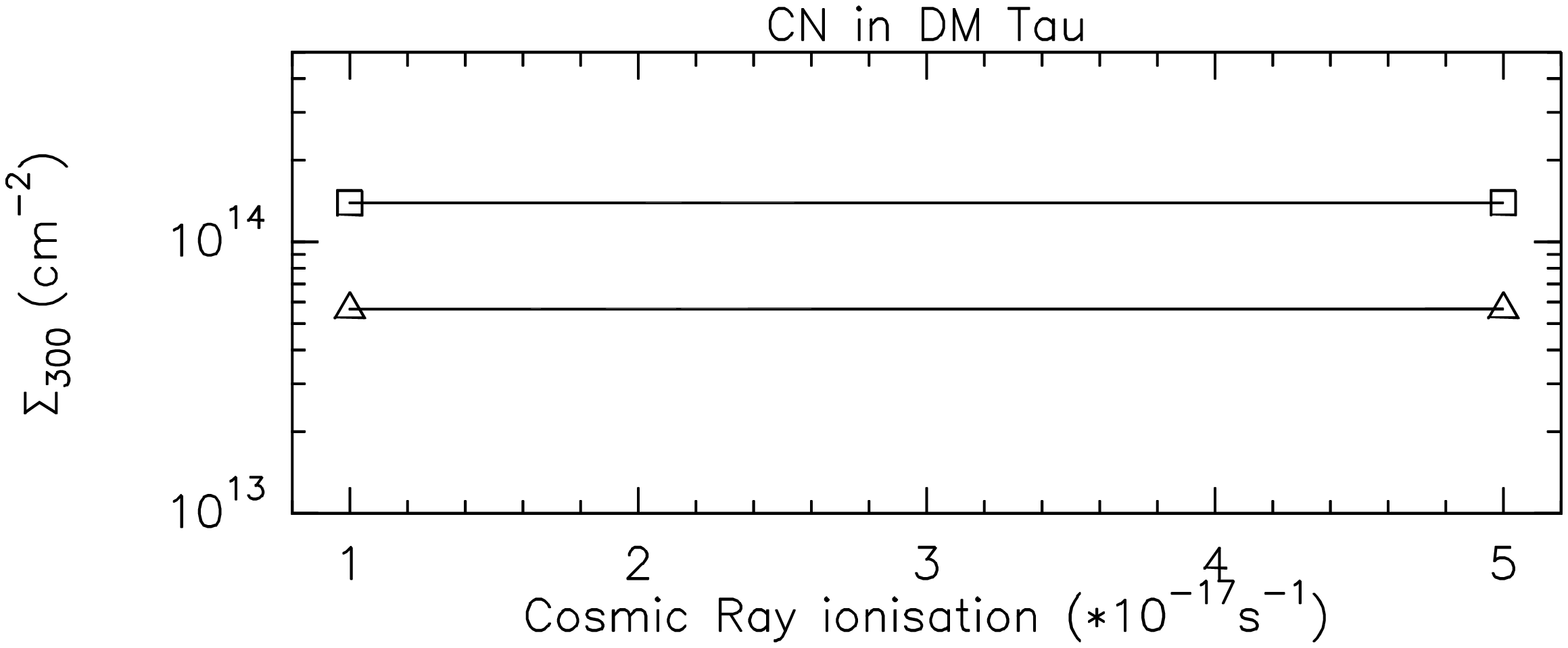} & & \includegraphics[angle=270,width=6cm]{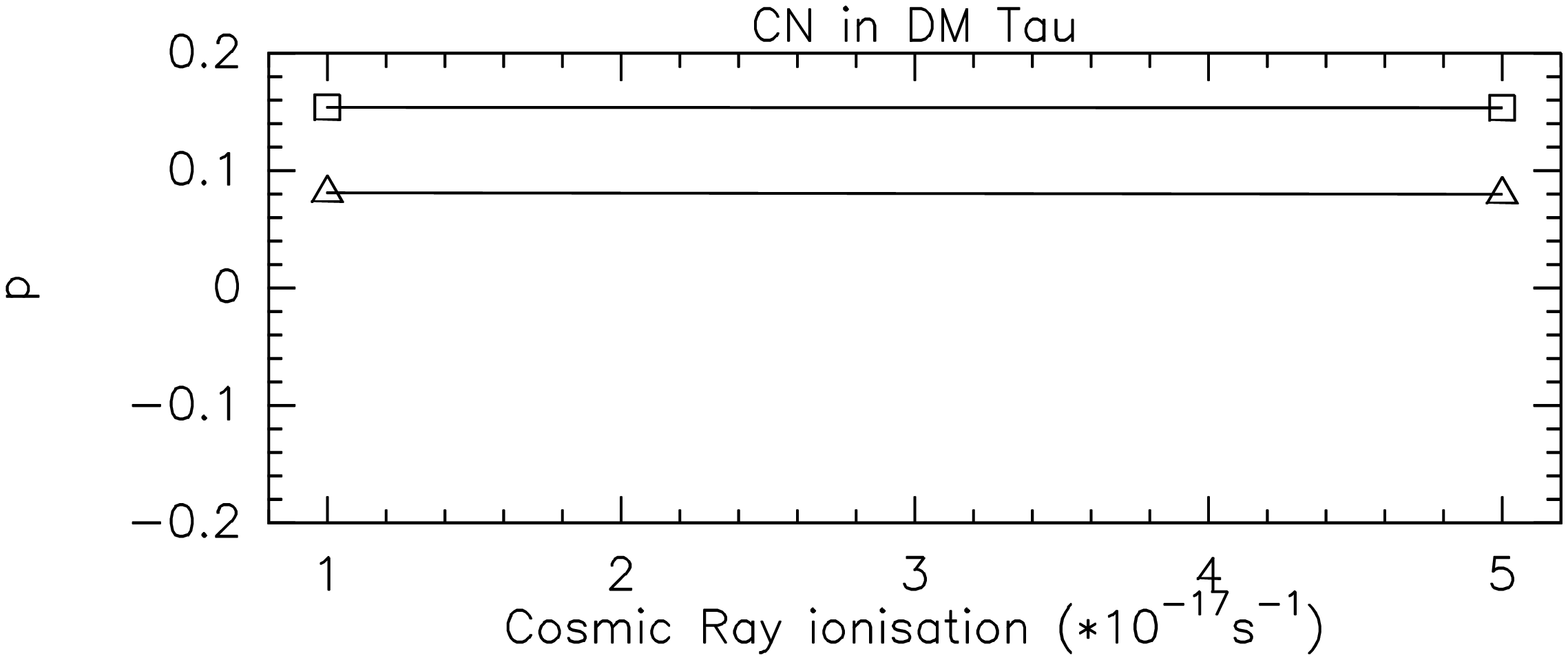}\\
\includegraphics[angle=270,width=6cm]{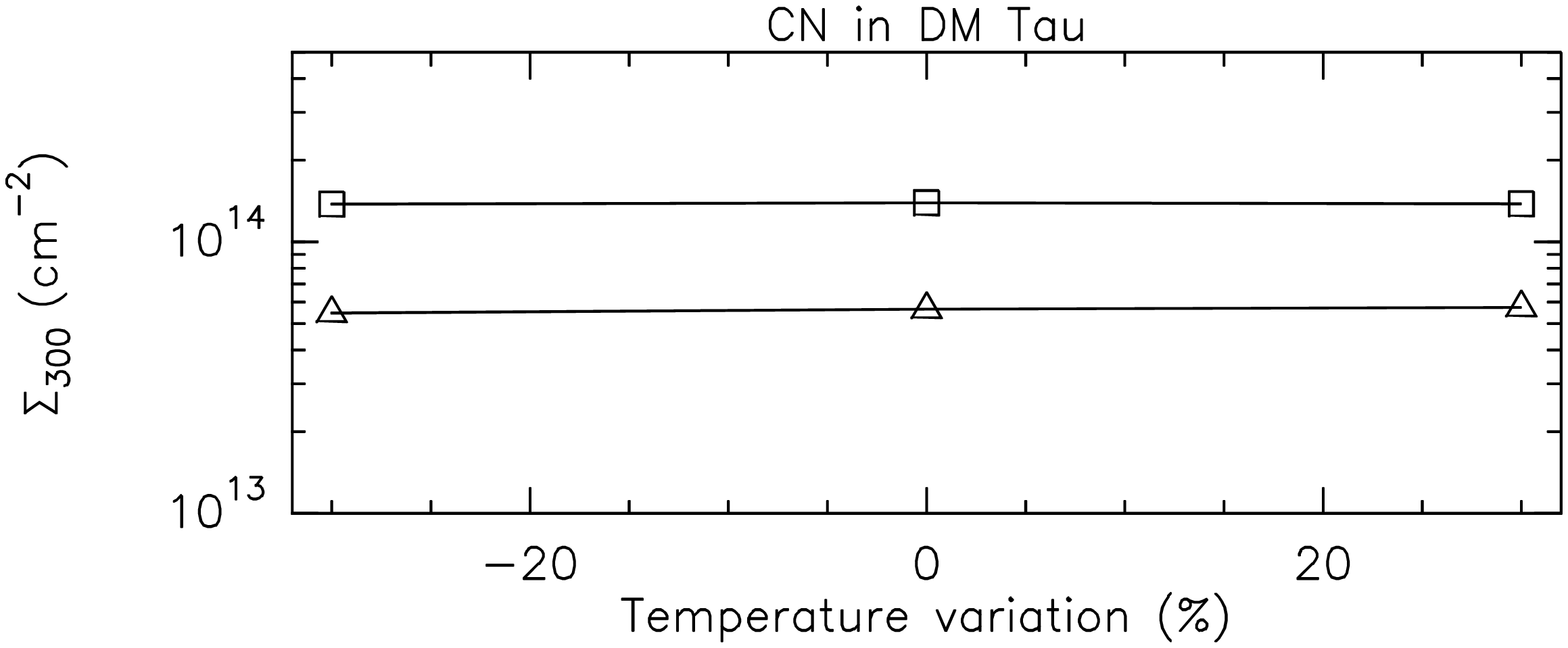} & & \includegraphics[angle=270,width=6cm]{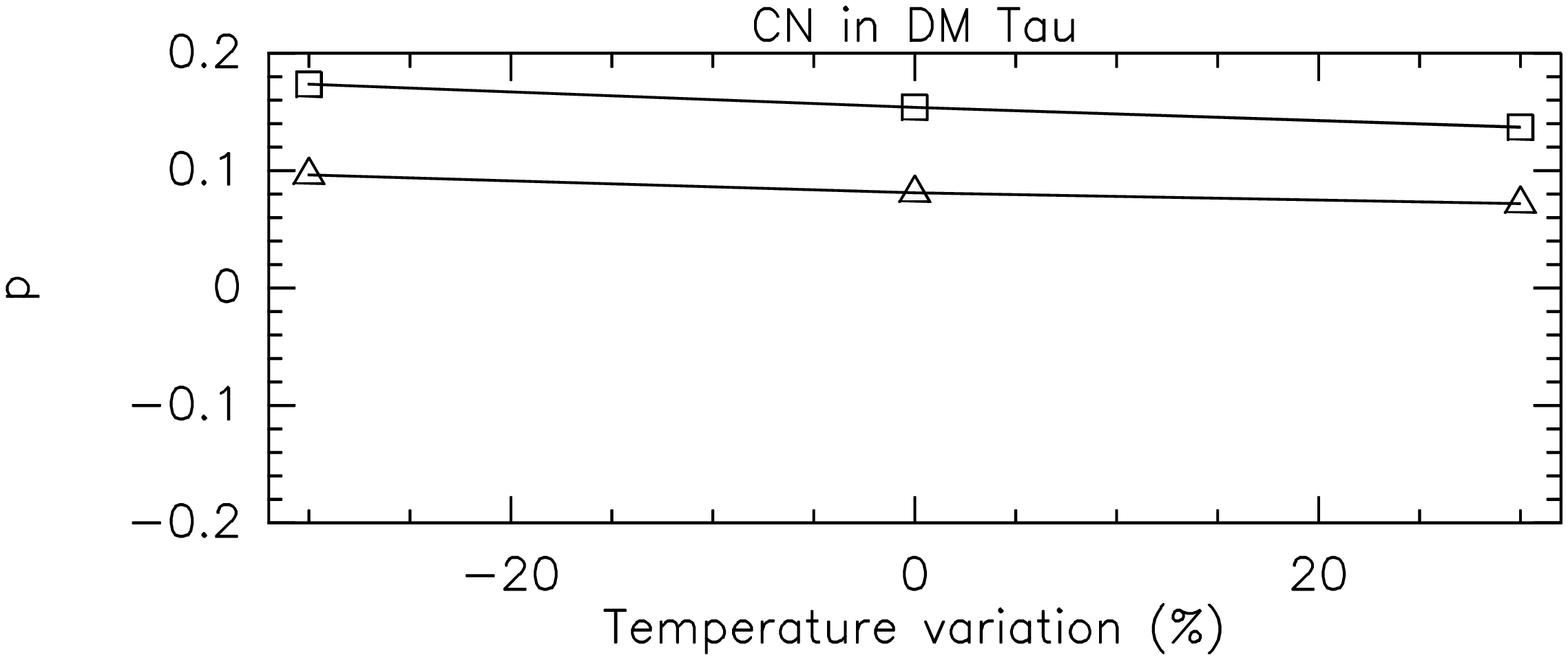}\\
\includegraphics[angle=270,width=6cm]{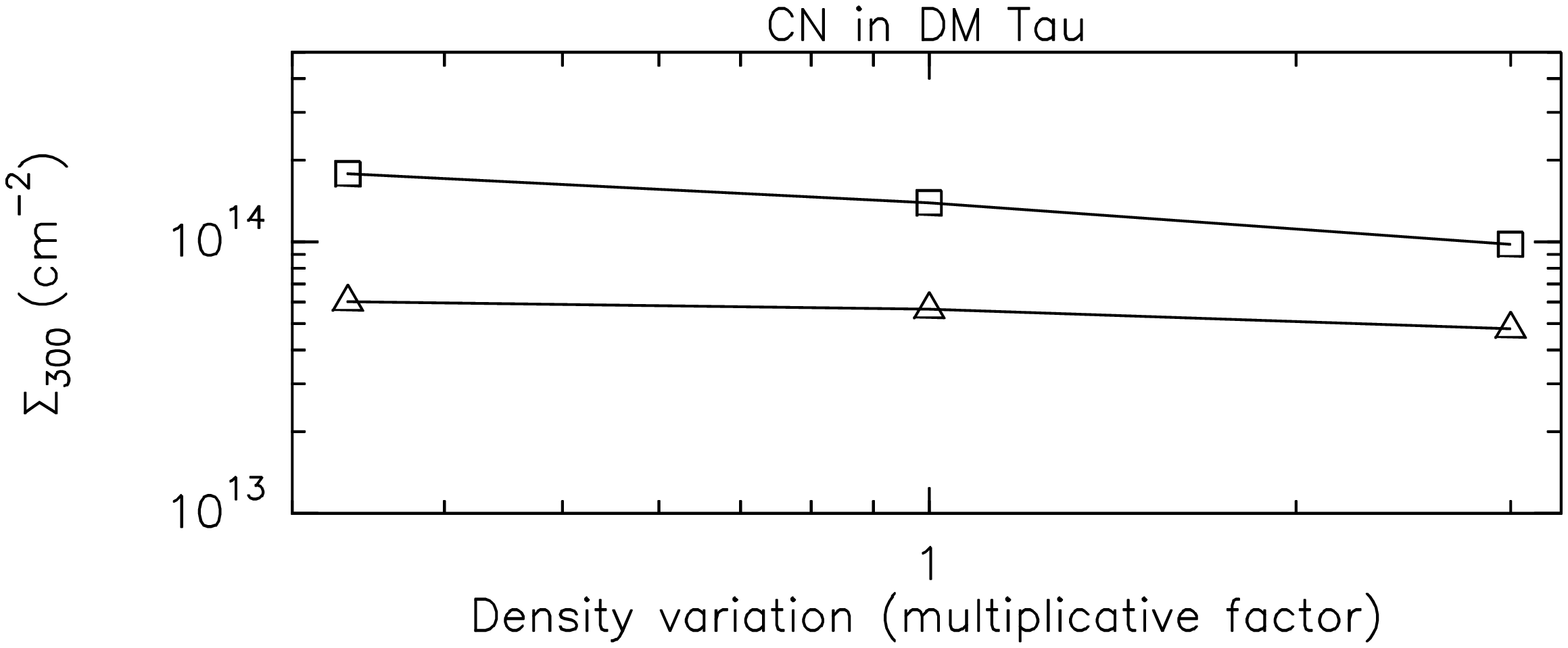}  & & \includegraphics[angle=270,width=6cm]{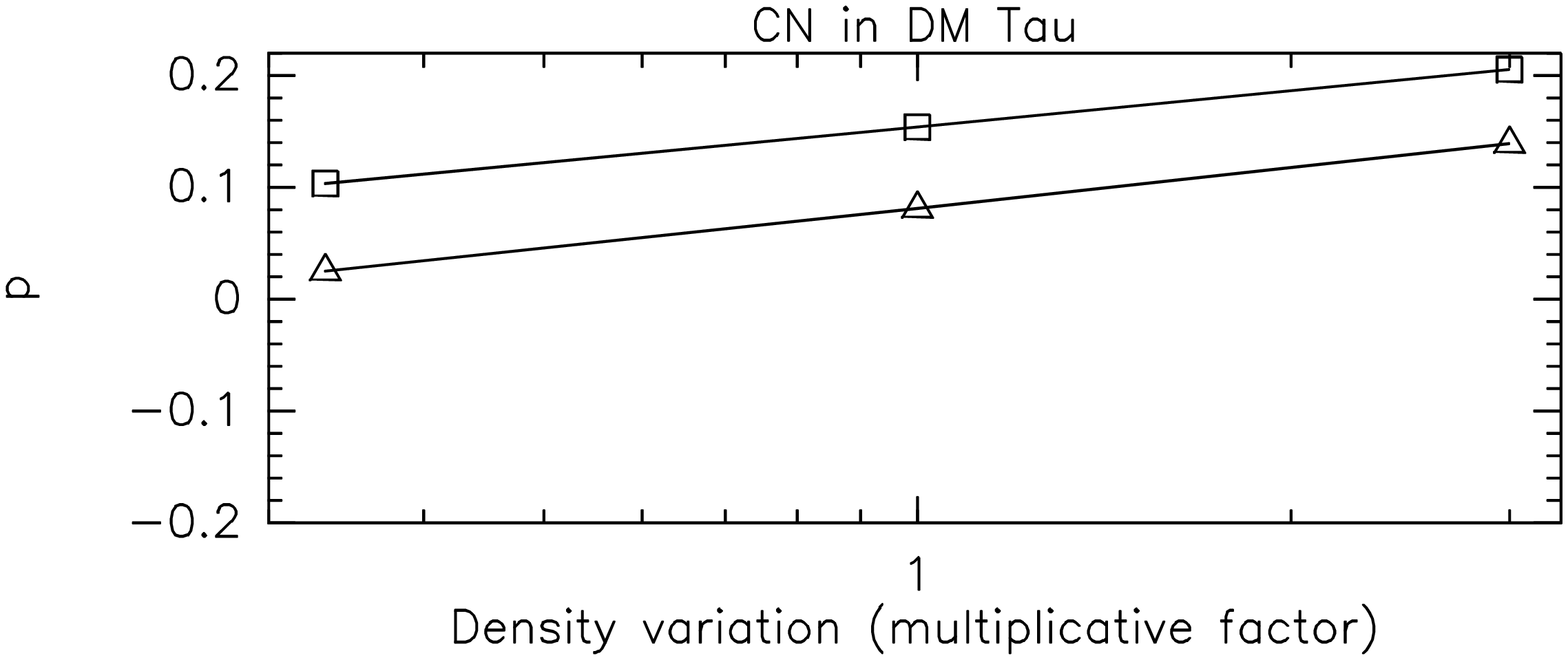}\\
\end{tabular}
\caption{Variations of the radial distribution of CN ( $\Sigma_{300}$ and $p$ parameters) according to the \textbf{maximum} grain size \app, the UV field intensity ($\chi$ factor), the rate of cosmic ray ionization and modification of the physical structure. The ``standard'' values are $\chi=10^3$, CR=$10^{-17}$s$^{-1}$ and nominal temperature and density structure. Squares: big grains (\app=0.1\,mm), triangles: small grains (\app=0.3\,$\mu$m).}
\label{fig:var:cn}
\end{figure*}

\begin{figure*} 
 \centering
\begin{tabular}{ccc}
\includegraphics[angle=270,width=6cm]{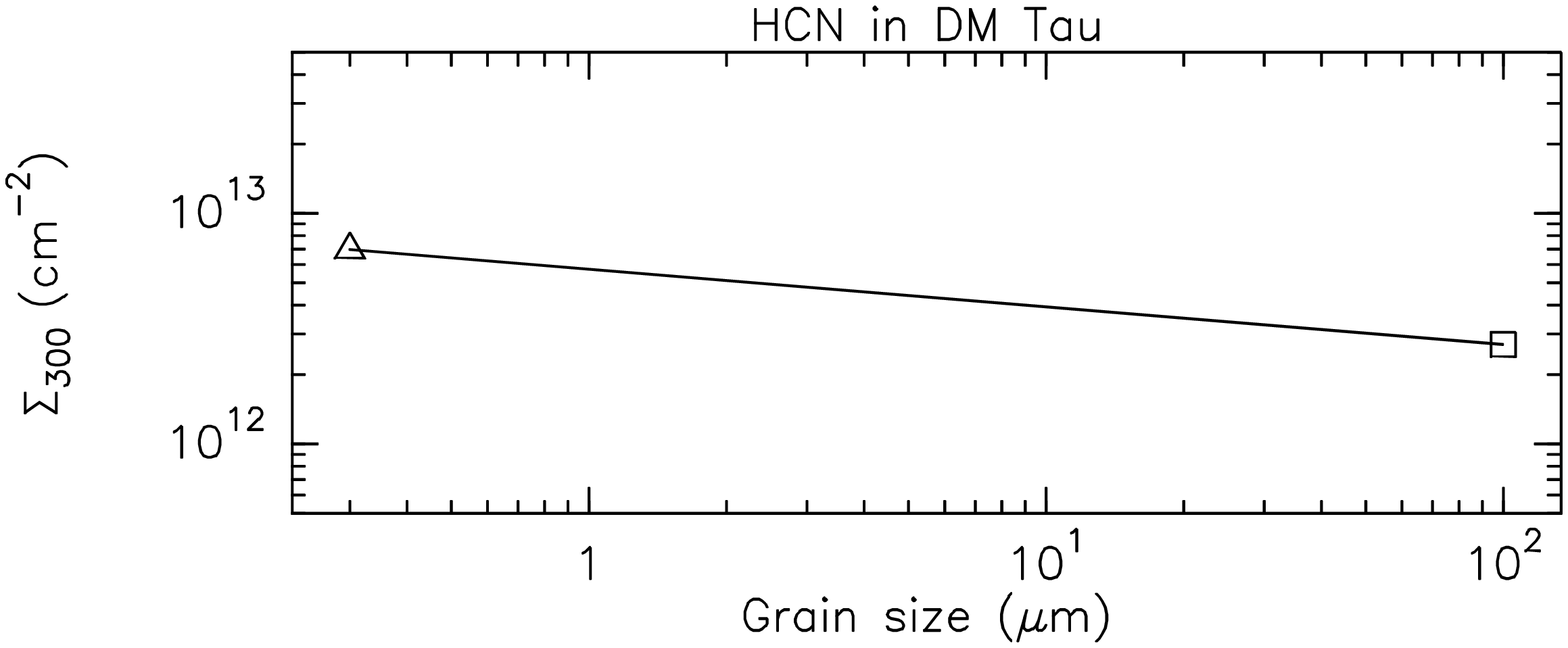} & \phantom{1234} & \includegraphics[angle=270,width=6cm]{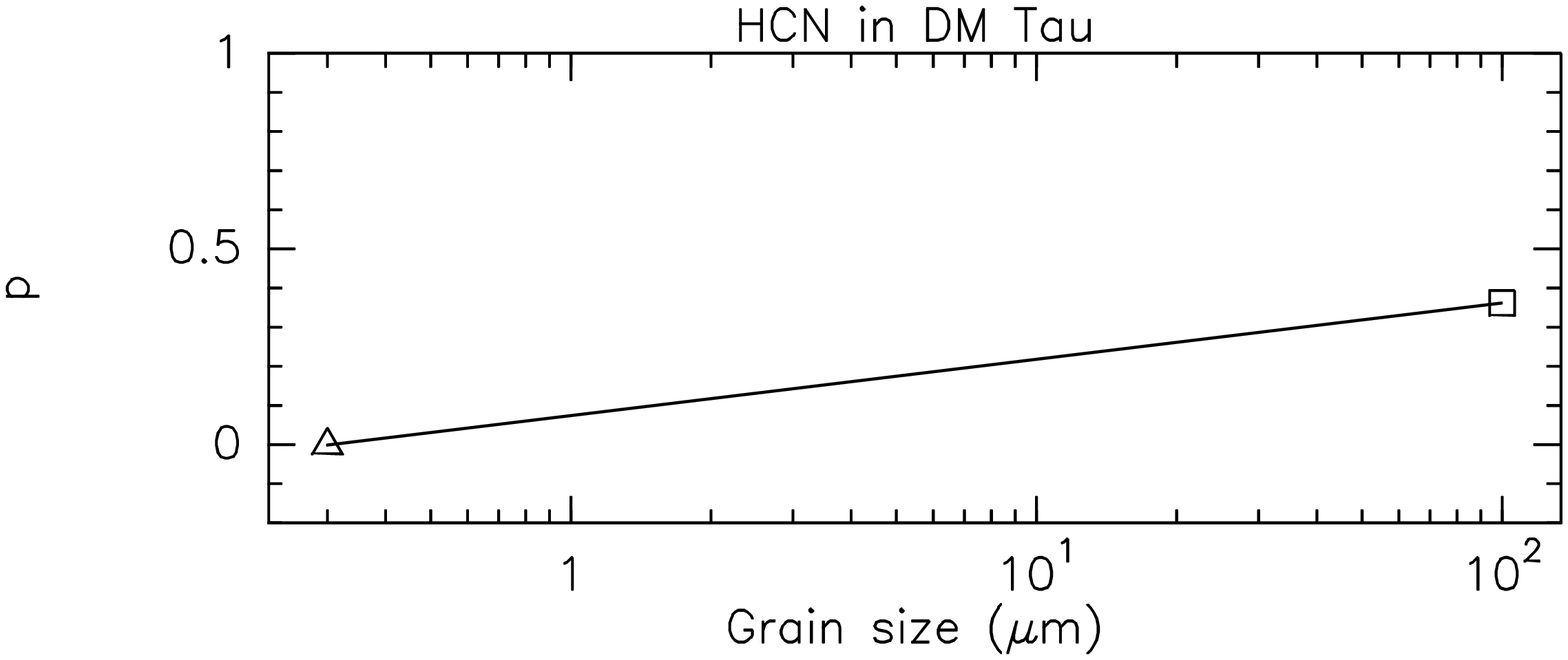}\\
\includegraphics[angle=270,width=6cm]{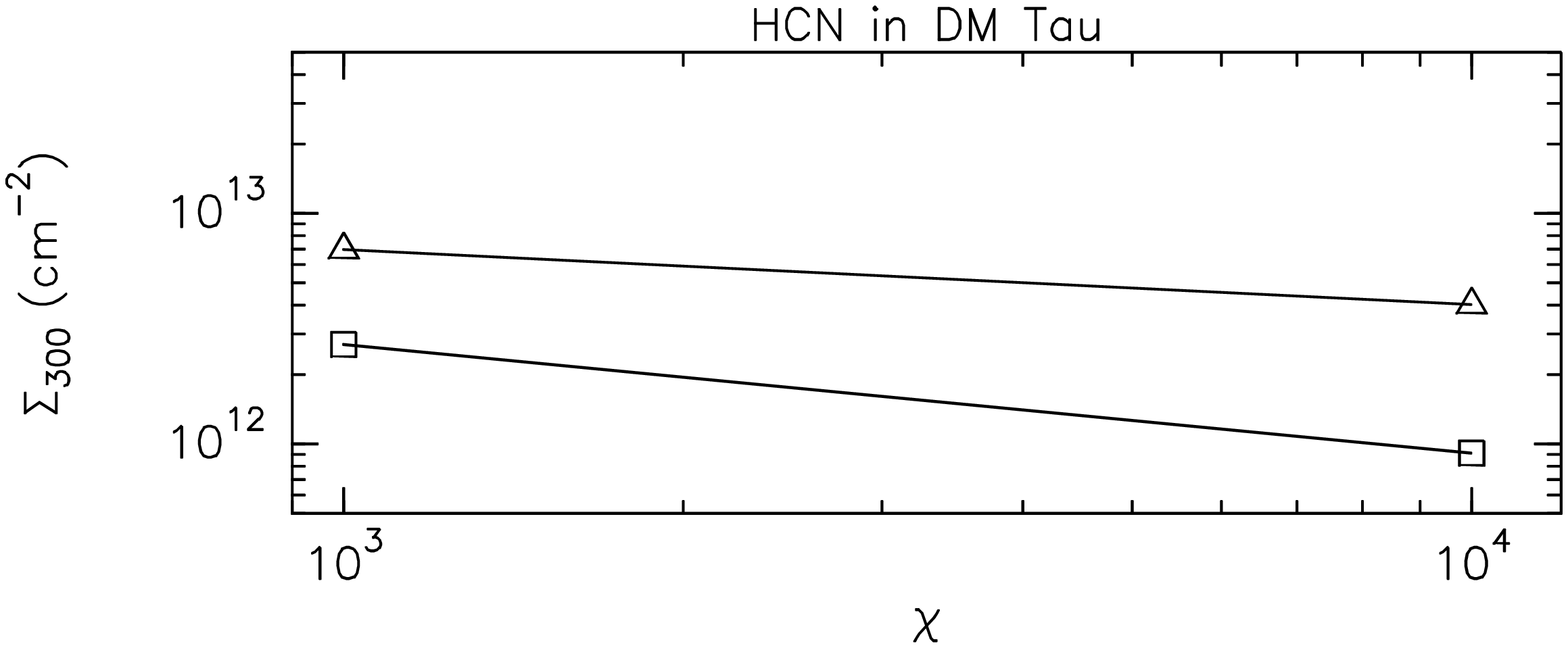} & & \includegraphics[angle=270,width=6cm]{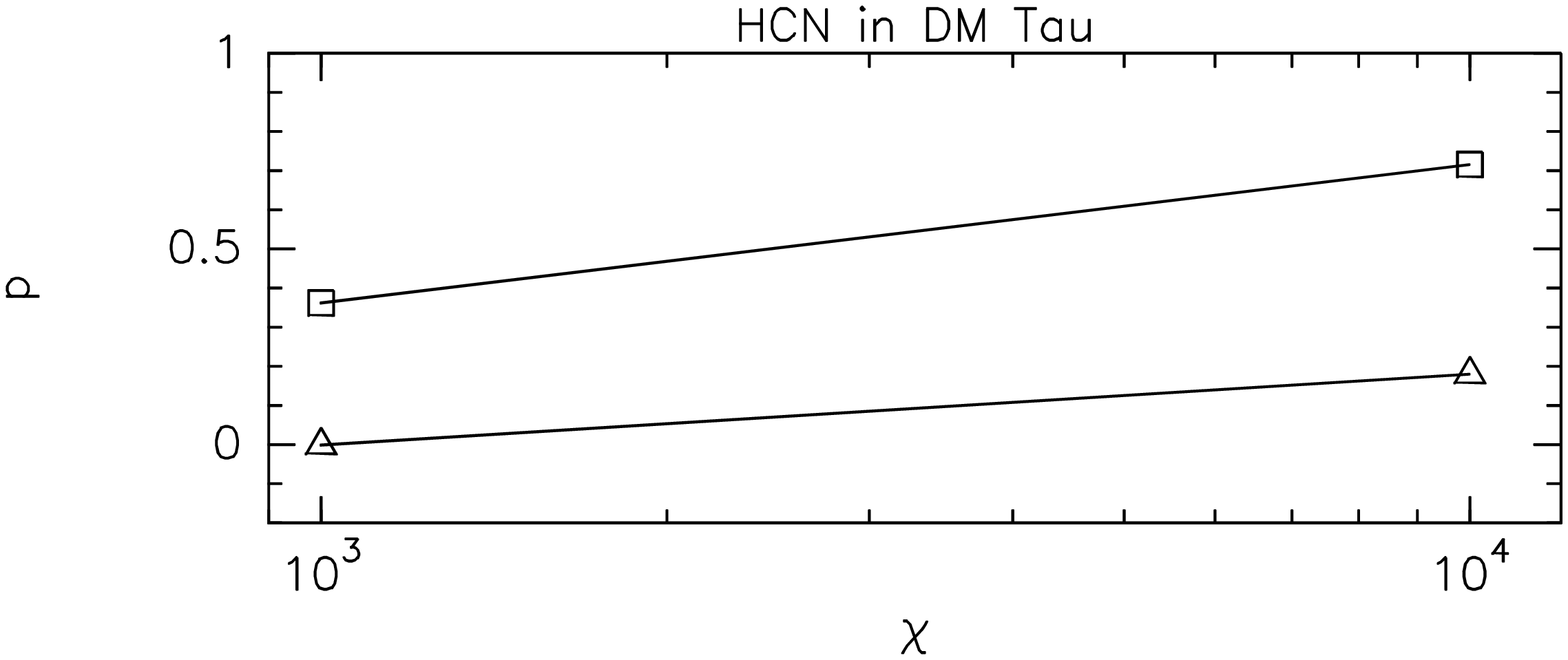}\\
\includegraphics[angle=270,width=6cm]{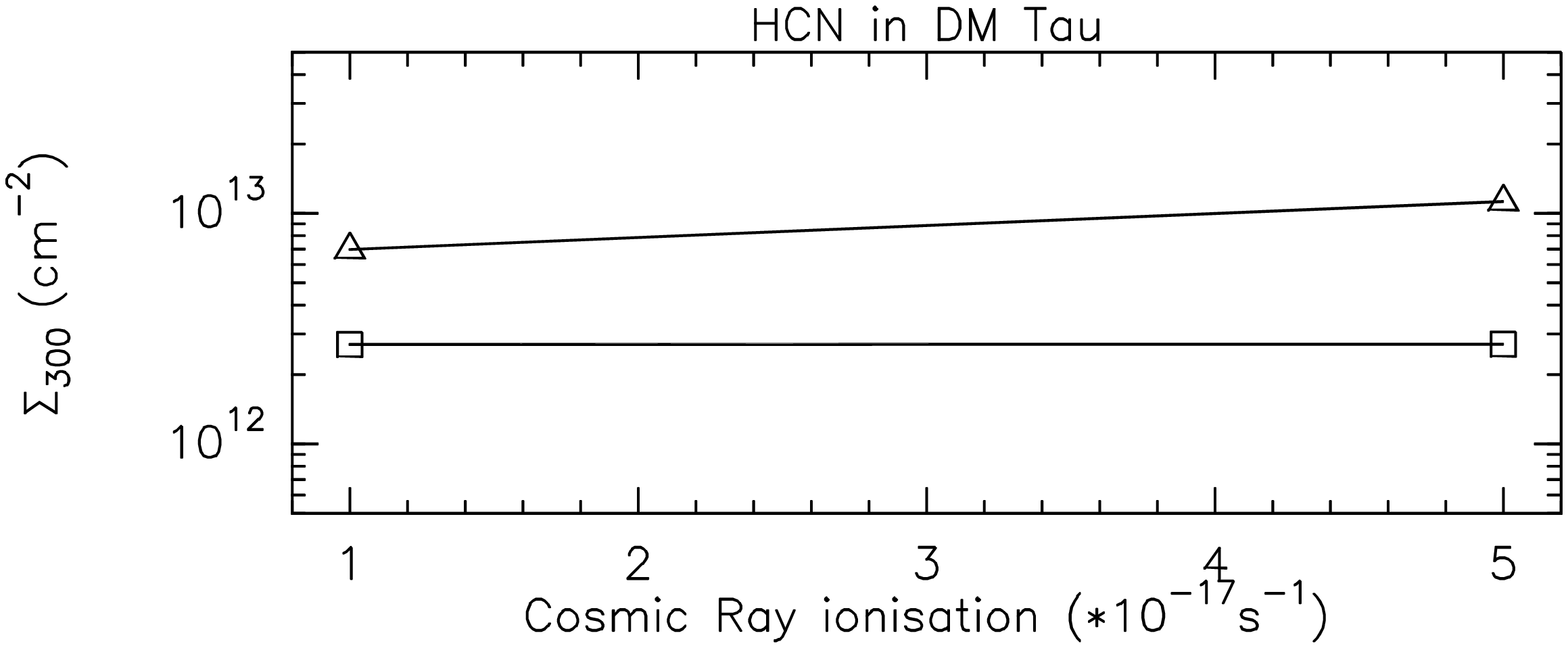} & & \includegraphics[angle=270,width=6cm]{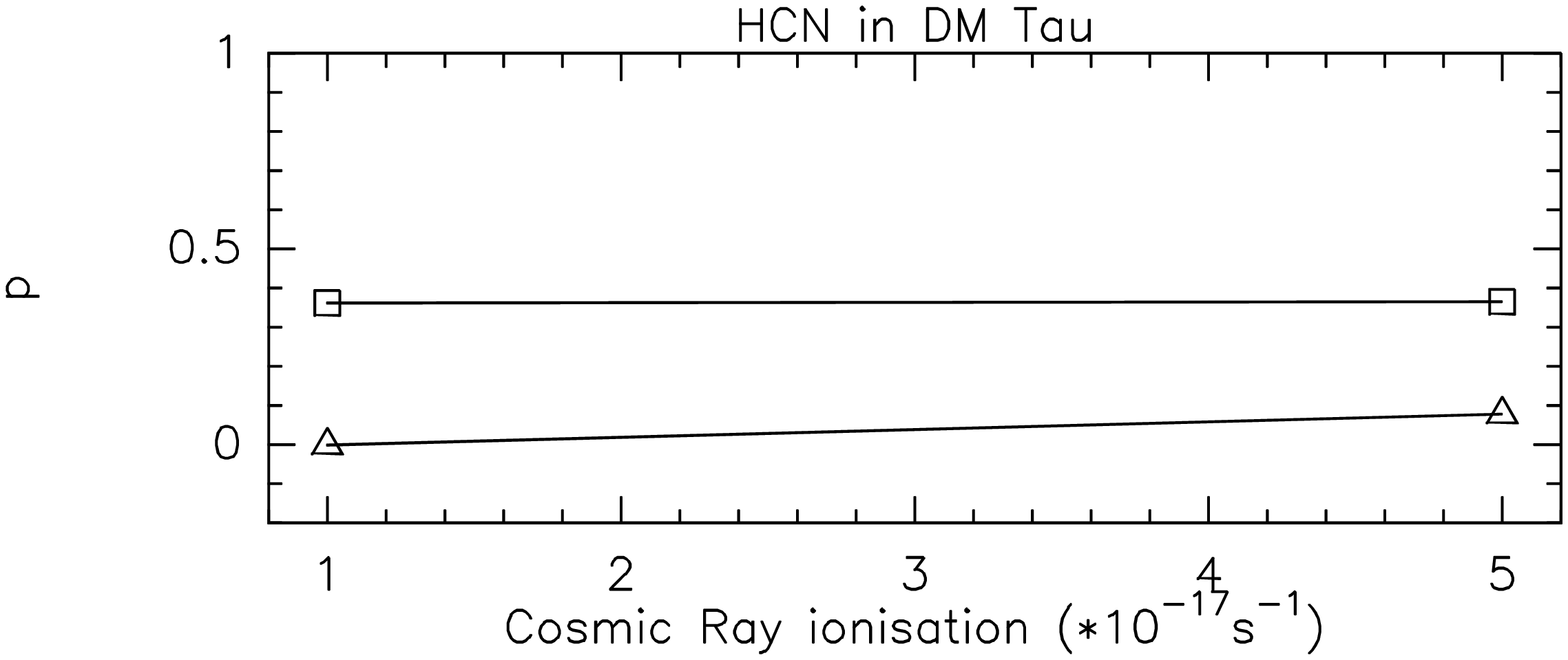}\\
\includegraphics[angle=270,width=6cm]{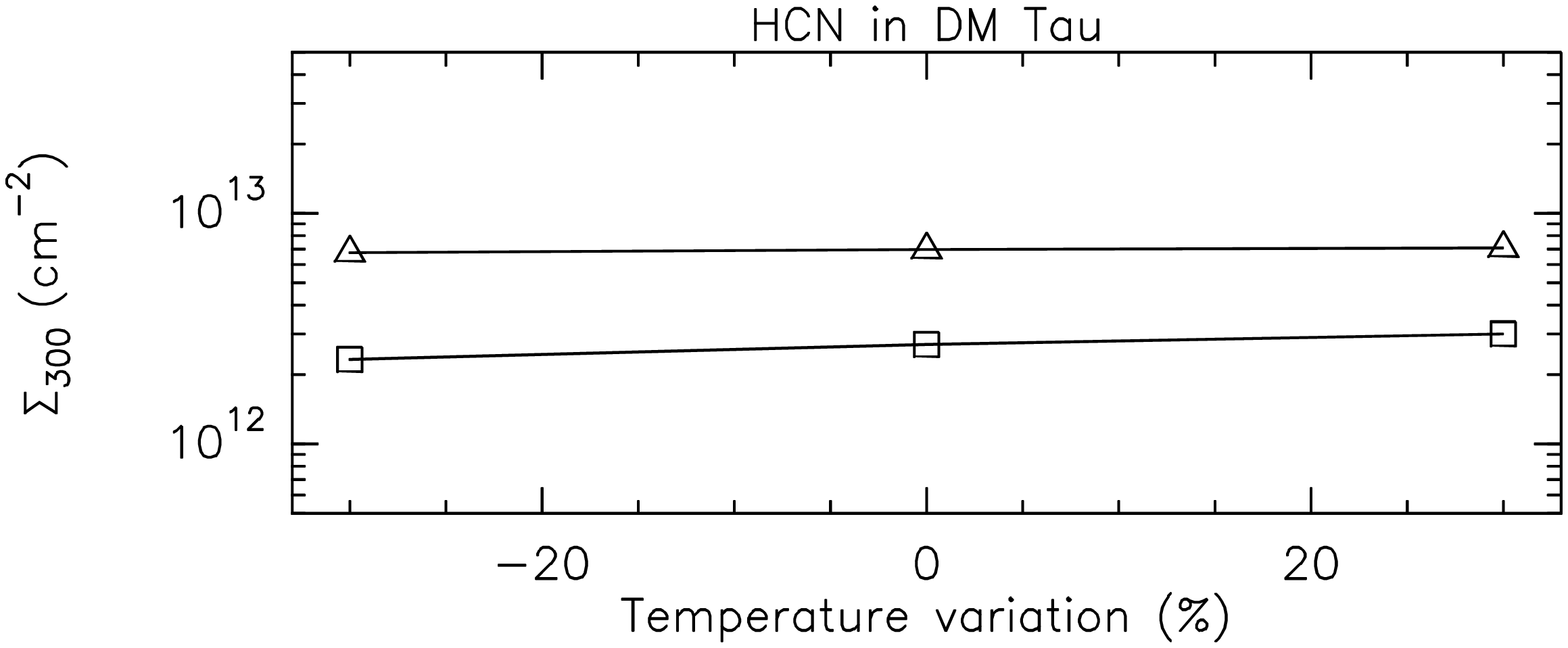} & & \includegraphics[angle=270,width=6cm]{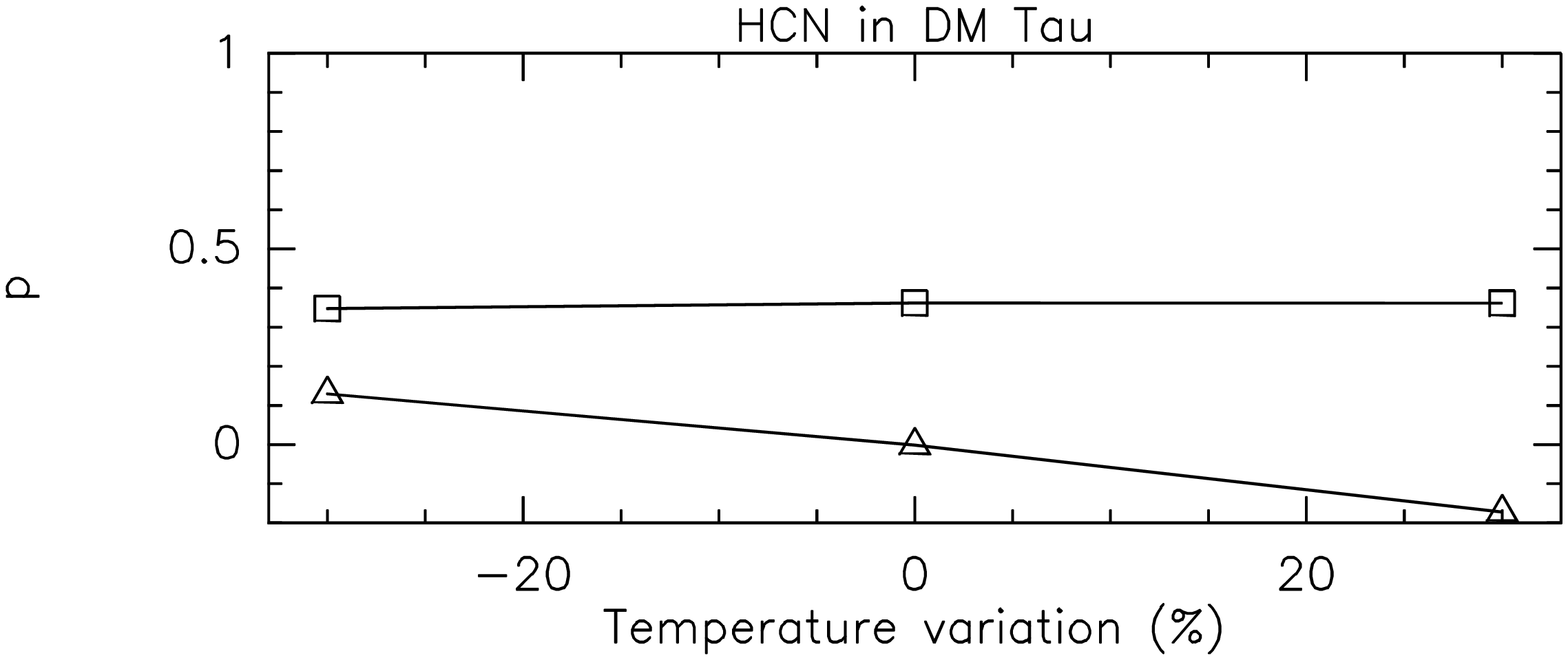}\\
\includegraphics[angle=270,width=6cm]{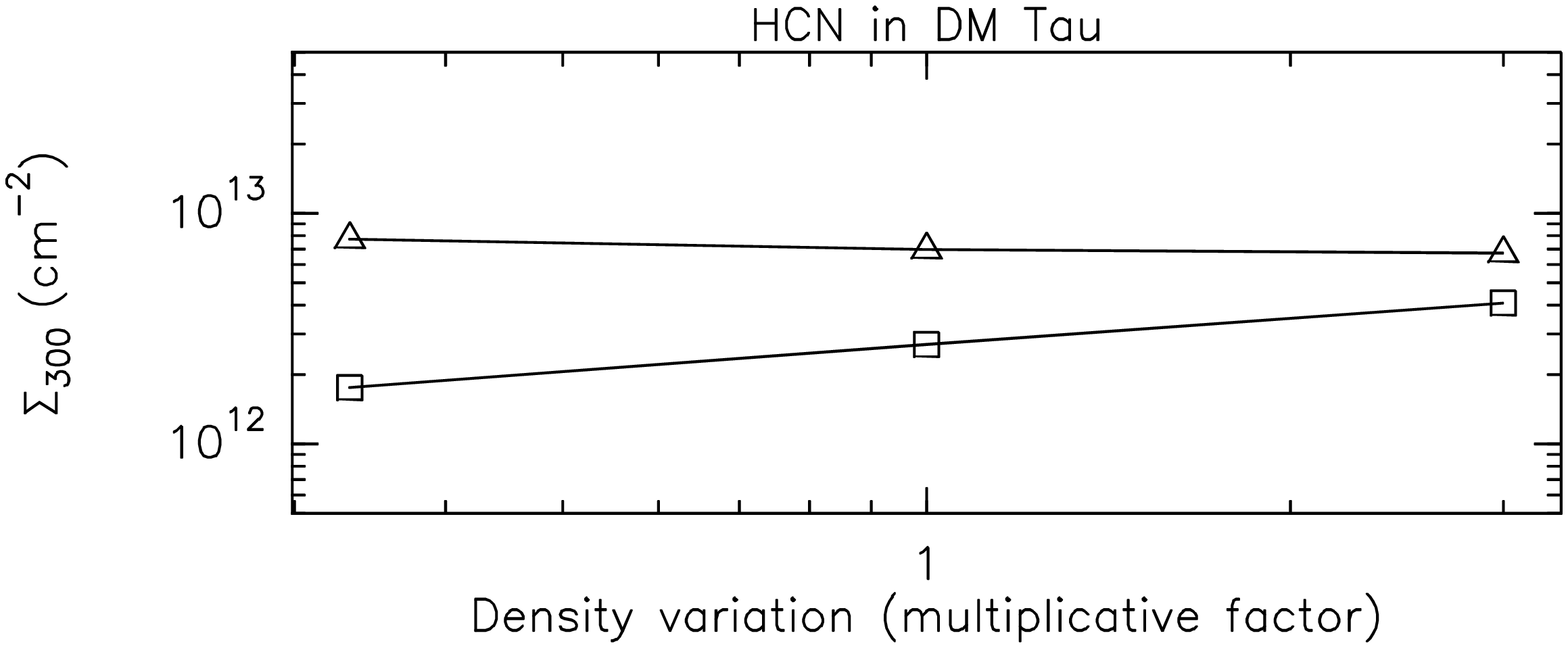} & & \includegraphics[angle=270,width=6cm]{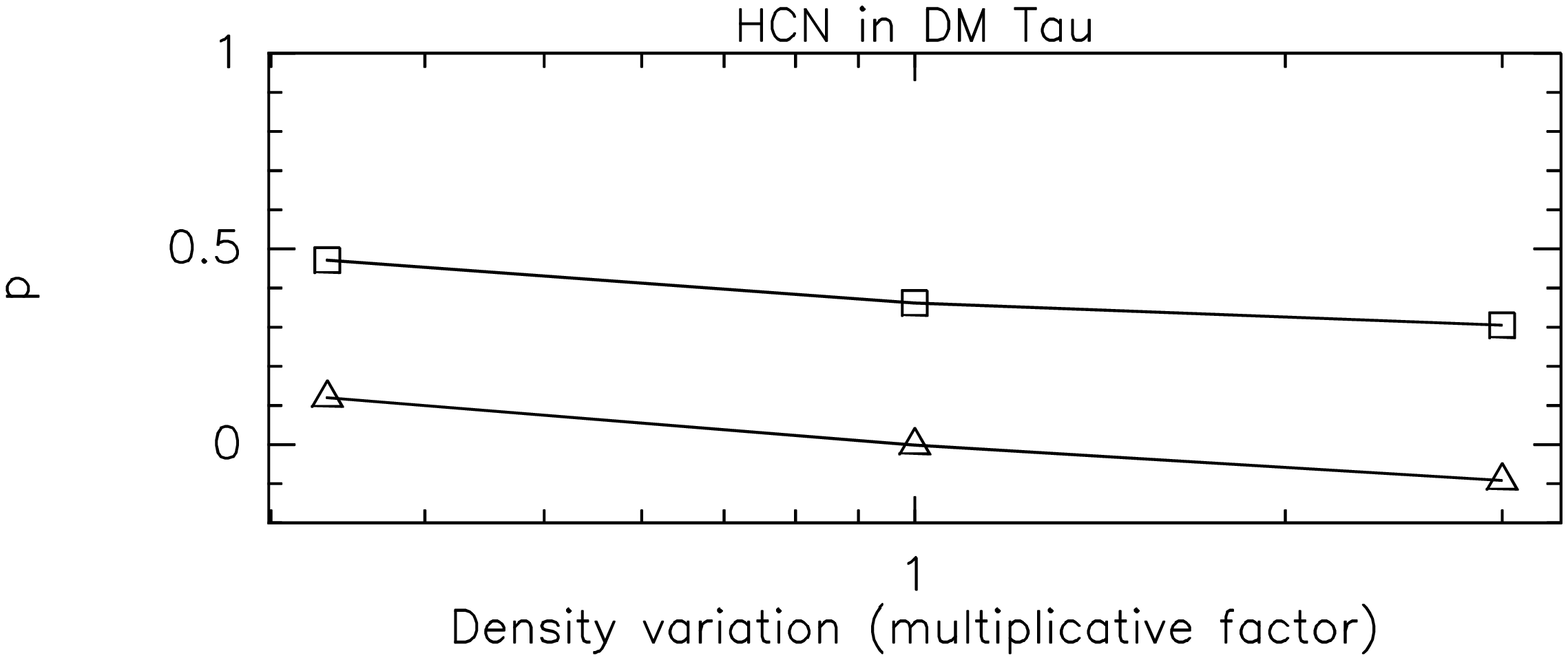}\\
\end{tabular}
\caption{Variations of the radial distribution of HCN ($\Sigma_{300}$ and $p$ parameters) according to the \textbf{maximum} grain size \app, the UV field intensity ($\chi$ factor), the rate of cosmic ray ionization and modification of the physical structure. The ``standard'' values are $\chi=10^3$, CR=$10^{-17}$s$^{-1}$ and nominal temperature and density structure. Squares: big grains (\app=0.1\,mm), triangles: small grains (\app=0.3\,$\mu$m).}
\label{fig:var:hcn}
\end{figure*}

For each disk, we vary several parameters.

{\bf Grain distribution:}  We have simulated grain growth by varying the maximum radius \app : in the ``big grains'' case \app=0.1\,mm, in the ``small grains'' case \app=0.3\,$\mu$m.

{\bf The incident UV field} \textbf{Illuminating} the disk surface is the sum of the interstellar
UV field (ISRF, due mainly to the surrounding massive stars) and the stellar UV field coming from
the central star. T-Tauri spectra present strong UV excess due to shock accretion of material onto
the star. This UV field is assumed to follow the shape of an ISRF Draine field
with an intensity scaling factor $\chi$ at $r=$ 100 AU. 
The UV field  decreases with increasing
radius as $1/r^2$. According to \citet{Herbig+Goodrich_1986} $\chi=10^4$ is appropriate, but
\citet{Bergin+etal_2003} find a much lower value, $\chi=500$.
Using the measurements from \citet{Bergin+etal_2003}, \citet{Henning+etal_2010} derived UV fluxes for DM\,Tau ($\chi=300$), LkCa\,15 ($\chi=1850$) and MWC\,480 ($\chi=5000$) (see their Table\,1). Accordingly we explore the range $\chi=10^3-10^4$ and run also some additional models with $\chi=10^2$
to see the impact of the stellar UV field intensity on the CN/HCN
chemistry.

In addition, a couple of models were run for DM\,Tau with the photodissociation rates calculated without UV excess, but for a black body at 4000\,K using rates from \citet{vanDishoeck+etal_2006}. Results are presented only in Fig.\,\ref{fig:tot}.

{\bf The Cosmic Ray ionisation rate (CR):} We performed calculations with the standard
value of $10^{-17}$s$^{-1}$ and with an increased rate of $5\,10^{-17}$s.$^{-1}$.

{\bf Physical structure:} In addition to the values intrinsic to the basic disk models, we
vary the density (by a factor of 3) and the temperature (by $\pm 30$\%) around those
values. In the later case, the scale height is not recomputed, but kept equal to that of the basic model.

{\bf The thermal balance} was computed
only at R=300\,AU in a few models (those presented on Fig.\,\ref{fig:cn-t2}),
the incident UV field being the main energy source.  The dust temperature
was derived following \citet{Burton+etal_1990} and used to compute the energy exchange
between the gas and the grains, as described in \citet{Burke+Hollenbach_1983}. The gas
temperature is then calculated, taking into account heating and cooling processes \citep[see][for more details]{LePetit+etal_2006}. Again, the scale height is
that of the basic model and is not modified consistently.

{\bf \lya\ line:} \citet{Bergin+etal_2003} pointed out the possible importance of
the \lya\ line on the chemistry. Essentially, HCN can be photo-dissociated by \lya\ radiation
while CN cannot. As the \lya\ line can be very strong in the spectra of T-Tauri (carrying up to
a few 10\% of the UV energy), this affects the photodissociation rate, and thus the CN/HCN chemistry.
To test the influence of \lya\ radiation, we implemented its impact on the chemistry
following \citet{vanDishoeck+etal_2006}, and ran a few models in which the \lya\ line represents
1/3rd of the total UV luminosity.

Unless otherwise specified, the model parameter values are as follows: \app=0.3$\mu$m (i.e. small grains), a ISRF UV field with $\chi=10^3$ and no \lya\ included, CR=$10^{-17}$s$^{-1}$, nominal density and temperature and no thermal balance calculated.

\subsection{PDR modeling results}

\begin{figure*}[t]  
  \centering
      \includegraphics[angle=270,width=8.0cm]{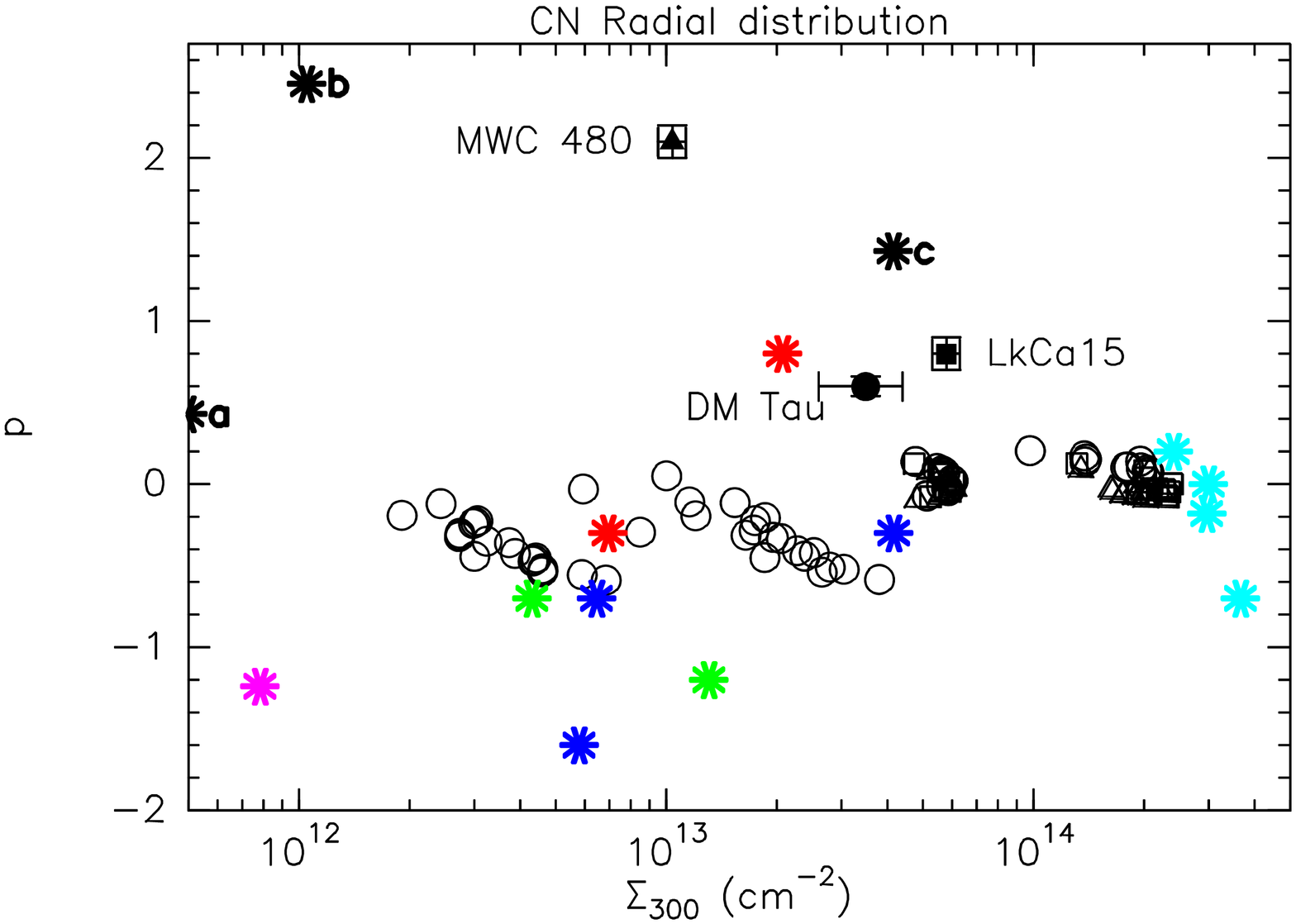}
       \hspace{1.0cm} \includegraphics[angle=270,width=8.0cm]{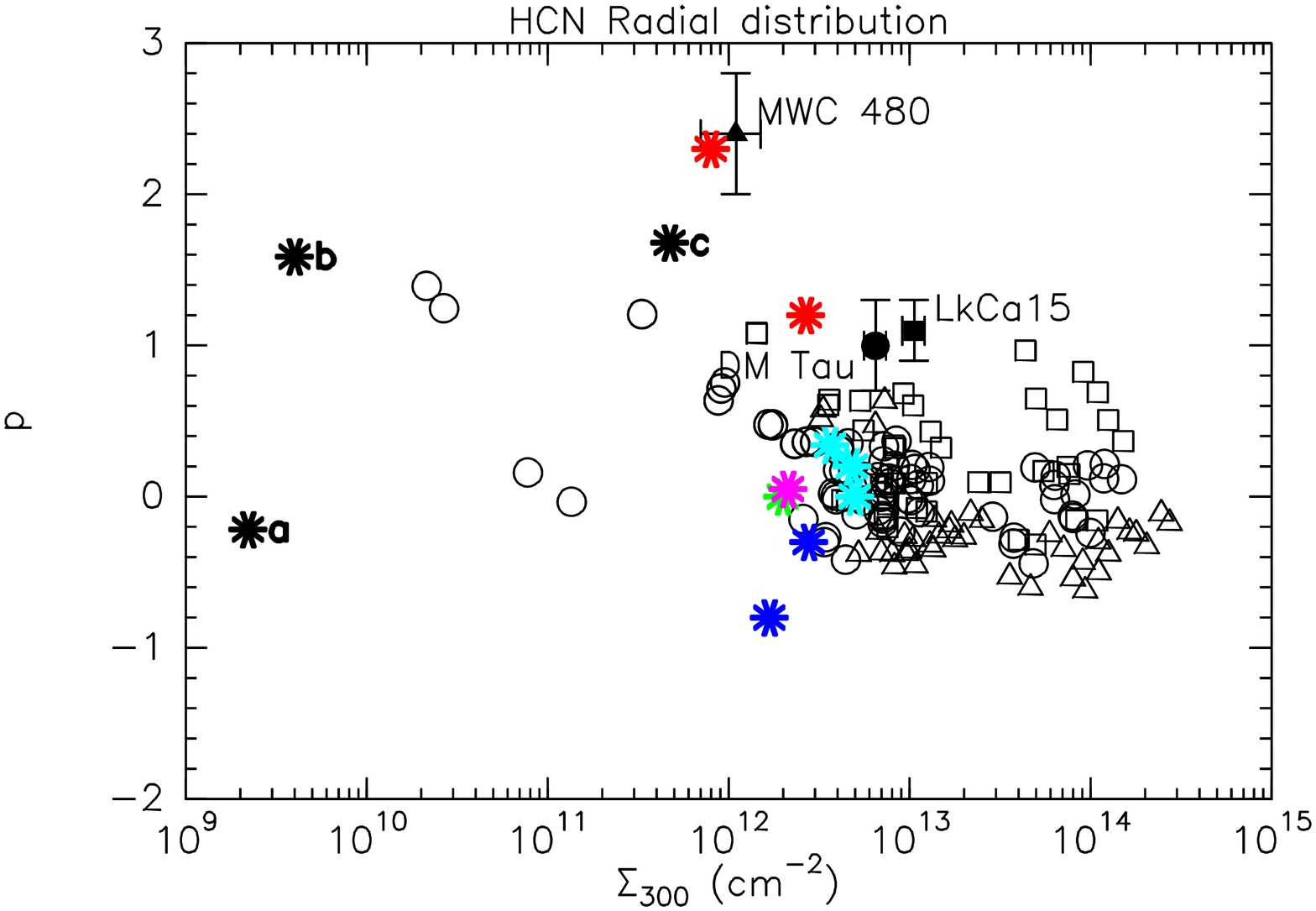}
\caption{Value of $\Sigma_{300}$ and $p$ for the three sources DM\,Tau (circle),
LkCa\,15 (square) and MWC\,480 (triangle). Our observations are plotted
in \textbf{filled} black markers, the results from the PDR chemical modeling are in open
markers. In the DM\,Tau case, we also plot the results with the photodissociation rates calculated for a black body at 4000\,K from \citet{vanDishoeck+etal_2006}.
Results from the literature are marked with stars, red:
\citet{Aikawa+Herbst_1999}, green: \citet{Aikawa+etal_2002}, blue:
\citet{vanZadelhoff+etal_2003}, cyan: \citet{Aikawa+Nomura_2006},
magenta: \citet{Willacy+Langer_2000} and black: \citet{Willacy+etal_2006}
(a, b and c correspond to their $K=0, K=10^{16}$ and $K=10^{18}$ cases).
The error bars show the power-law fitting errors ($1\,\sigma$).}
\label{fig:tot}
\end{figure*}

The column densities predicted by the chemical model are fitted by power laws as functions of radius,
and are thus characterized by the values of the $\Sigma_{300}$ and $p$, like for the observations.
Power laws offer fairly good approximations to the resulting distribution of column densities.

The calculated $\Sigma_{300}$ and $p$  as a function of the various input parameters are illustrated for DM\,Tau in Fig.\,\ref{fig:var:cn} for CN and Fig.\,\ref{fig:var:hcn} for HCN.
\textbf{The impact of the model parameters is shown in a more compact form for all sources in Fig.\,\ref{fig:cn-hcn}}. A comparison with other chemical models, fitted in the same
way, is displayed in Fig.\,\ref{fig:tot} together with the observed values.

The main results are:
\\
- The calculated column densities (at 300 AU) $\Sigma_{300}$ match reasonably well the
observed values (within a factor of a few for the best set of parameters), but the radial
dependence (see parameter $p$) is weaker in the models ($p(\mathrm{mod}) < p(\mathrm{obs})$).\\
- The most influential parameters are the UV flux and the grain sizes. These two parameters control the
photodissociation process. \\
- As a consequence a high CN/HCN ratio is observed for large grains. \\
- A modification of the CR  influences the HCN/CN chemistry only in the case of
small grains, i.e. when UV penetration is efficiently blocked.\\
- Despite the lack of molecule freeze out and grain chemistry, our PDR code results for
CN and HCN are in broad agreement (flat radial distribution and similar value of column densities within a factor of a few)
with published results of models that take these processes into account (see
\citet[][Fig.\,7--9]{Aikawa+Herbst_1999}, \citet[][Fig.\,3]{Aikawa+etal_2002}, \citet[][Fig.\,5
and 8]{vanZadelhoff+etal_2003}, \citet[][Fig.\,9]{Aikawa+Nomura_2006}). This suggests that the freeze-out of molecules onto grains, as implemented in these works, does not affect significantly HCN and especially CN, which remain
dominated by photo-chemistry in these models. Note, however, that \citet{Walsh+etal_2010} find out the
grain surface chemistry decreases the HCN column density, but only mildly affect CN.

 As in the other chemical studies, the CN molecule is located within a narrow layer close to the surface (A$_\mathrm{V}\sim$0.1 corresponding to z$\sim\,200$\,AU (n$\sim\,5 \times 10^5$cm$^{-3}$) in the small grain case and z$\sim\,50$\,AU (n$\sim\,7 \times 10^6$cm$^{-3}$) in the big grain case) whereas HCN is buried more deeply inside the disk. The vertical distribution of molecules at $r\,=\,300$ AU are presented in Fig.\,\ref{fig:abundance} for small and big grains and two different values of the incident UV field. As we do not include sticking onto grains, HCN remains fairly abundant in the mid-plane contrary to models including sticking onto grains where HCN is not present under \textbf{$z/r\sim$}0.1 \citep[e.g.,][]{Aikawa+Nomura_2006}.

\begin{figure}[th] 
\includegraphics[width=\columnwidth]{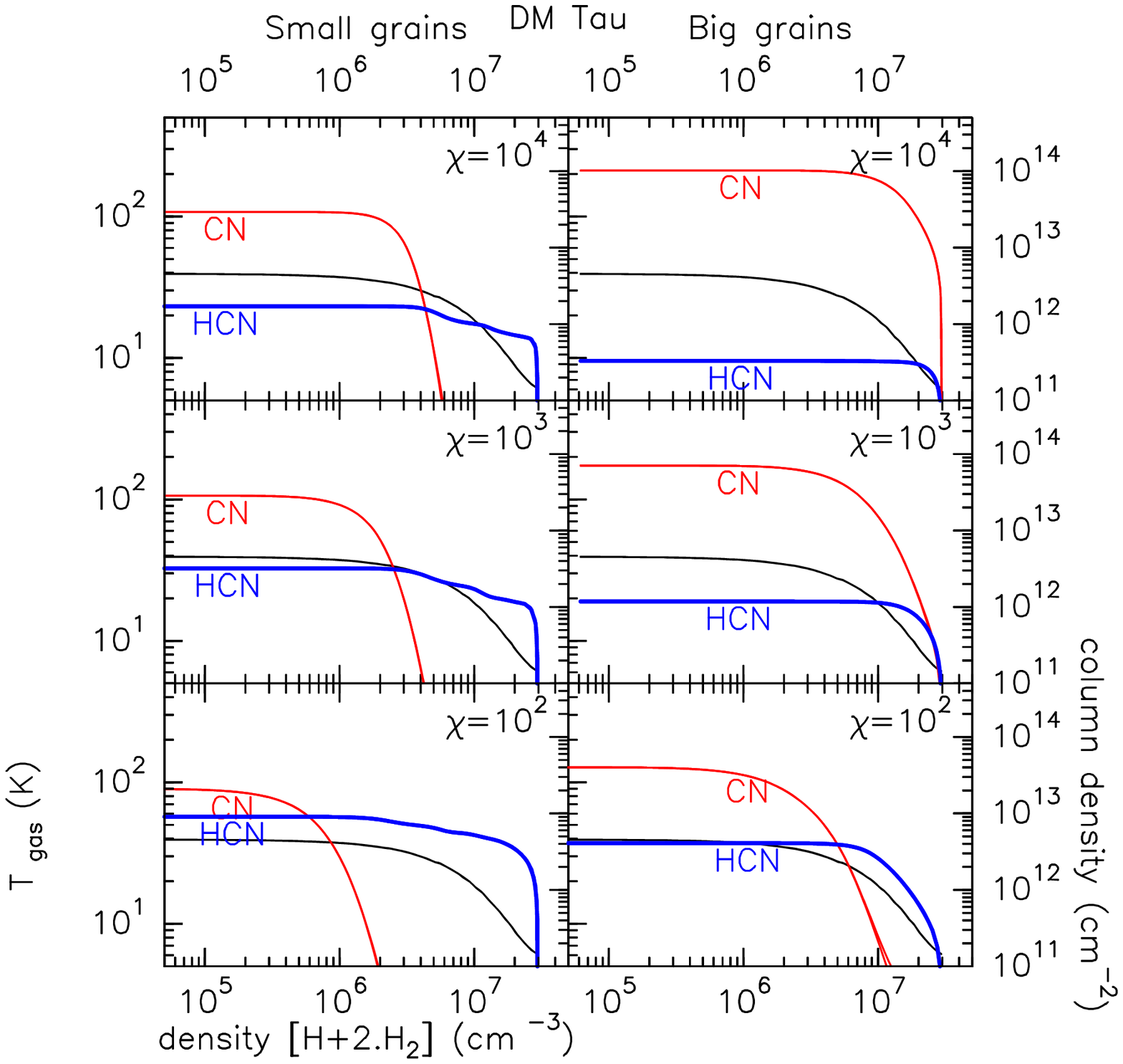}
\caption{Cumulative CN (red) and HCN (blue) column densities and temperature (black) from the mid-plane to the atmosphere as a function of the density at a radius of 304\,AU in the DM\,Tau disk. Models with temperature fixed from the structure file. This samples only half of the disk, so the column densities have to be multiplied by 2 to retrieve the $\Sigma$ values.}
\label{fig:cn-t1}
\end{figure}

Relative abundances can be a misleading indicator of the location of the bulk of molecules,
because of the strong vertical density gradient. To better indicate where molecules appear,
we trace in Fig.\,\ref{fig:cn-t1} the cumulative column density from the disk mid-plane to
the surface as a function of the gas density. The latter can be converted into height above
the disk mid-plane using the vertical distribution. 
The  temperature (which, in this model, applies equally to gas and dust) is also displayed.
Fig.\,\ref{fig:cn-t1} shows the temperatures and densities at which the bulk of the column density is built up,
allowing to predict what is the average line excitation temperature.
For essentially all models the CN column density is mostly build up in a region where the kinetic
temperature is $\geq$ 30\,K and at a density $n \geq 10^6\mathrm{cm}^{-3}$, the only exception
being the small grains, low UV ($\chi = 10^2$) case, where only half of the CN column density is
in these conditions.

\begin{figure}[th] 
\includegraphics[width=\columnwidth]{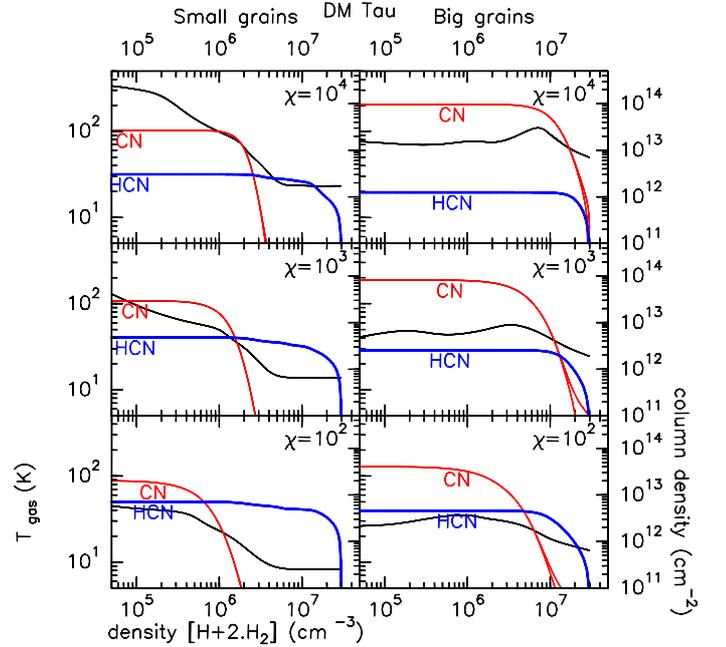}
\caption{
Same as Fig.\,\ref{fig:cn-t1} for models with thermal balance computed.}
\label{fig:cn-t2}
\end{figure}

Similar results are obtained in models for which the gas and dust thermal
balance is computed: the vertical distributions of CN and HCN abundances are not significantly affected,
although the gas temperature is in general somewhat higher in this case (Fig.\,\ref{fig:cn-t2}).
The chemistry is not significantly affected because
most reactions depend weakly on the temperature in this temperature range.

\begin{figure} 
\includegraphics[angle=270,width=\columnwidth]{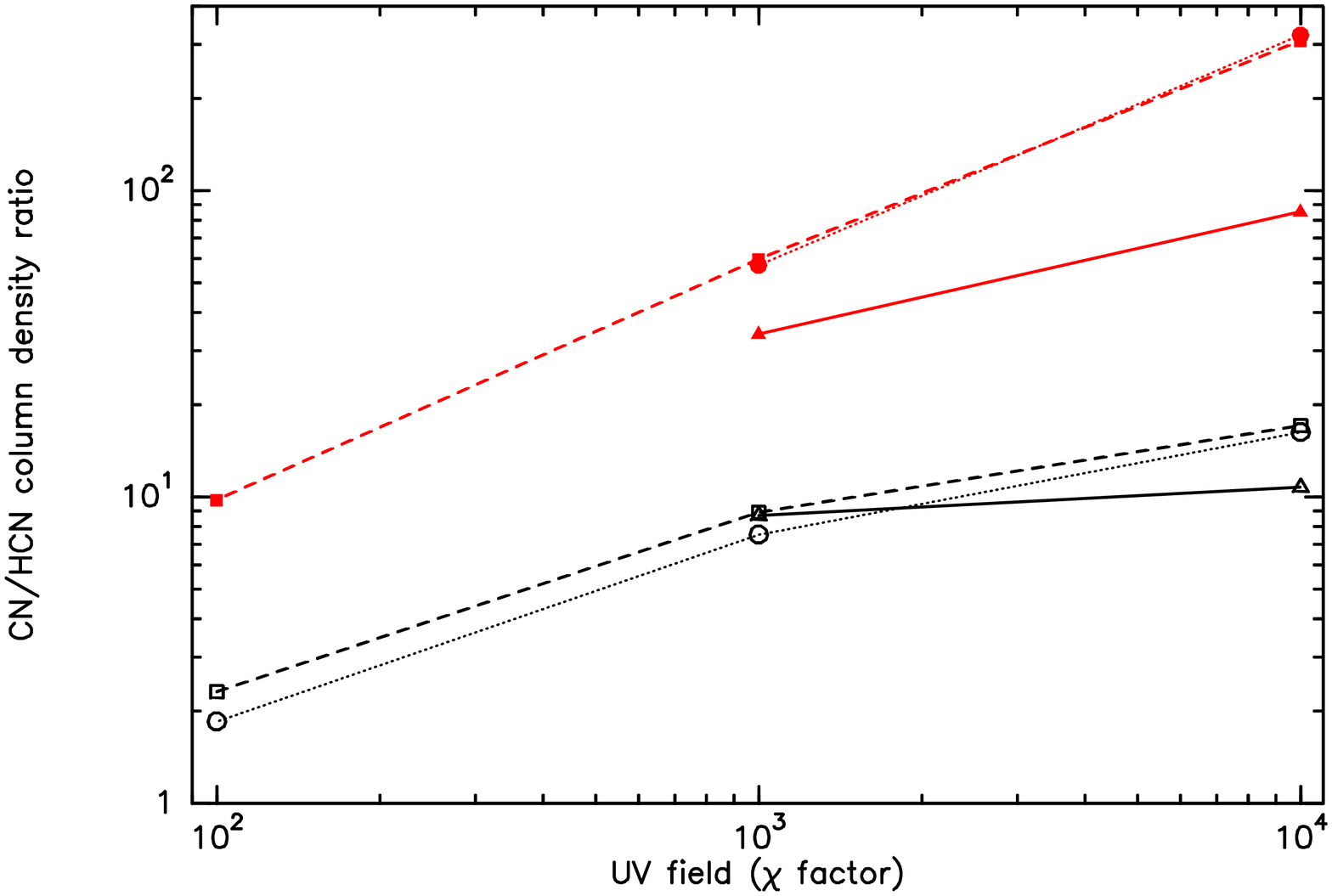}
\caption{CN/HCN column density ratios as a function of the UV field at R\,=\,97\,AU (triangles), 304\,AU (squares), and 602\,AU (circles) for the models with   small grains (thick black lines) and big grains (thin red lines)}
 \label{fig:cn-t3}
\end{figure}

\section{Discussion}

The results from our power law model fitting of the observations of the T-Tauri stars
DM\,Tau and LkCa\,15 appear to contradict the chemical model predictions on several aspects.
\begin{enumerate}
\item While $^{12}$CO indicates temperatures in the range 20 to 30 K at 100 AU for the upper layers
\citep{Pietu+etal_2007}, the derived CN and HCN rotation temperatures,  which apply mainly to radii 100-300 AU,
where our observations are the most sensitive, are very low (8 to 12 K, including calibration
errors).
\item In addition, the constraint on $\alpha_d \simeq 0$ (equivalent to large $\Sigma_d$)  suggests that CN molecules appear
closer to the disk than expected. This is coherent with a low temperature but in apparent contradiction
with the location of the predicted molecular layers.
\item Most models predict slopes $p$ which are much lower than observed; the main exception
being the results of \citet{Aikawa+Herbst_1999} for CN.
\end{enumerate}
For the Herbig star (MWC\,480), the temperature may be adequate, but the slope problem remains.
This leads to several open questions :
\begin{enumerate}
\item Do we understand the basic physical structure (density, temperature) of
protoplanetary disks correctly ?
\item How a misleading structure can affect the line formation ?
\item Do we understand the chemistry, in particular that of CN well enough ?
\end{enumerate}

\subsection{Excitation conditions}

CN is not the only photo-dissociation sensitive molecule displaying surprising
low excitation temperatures. From a spatially resolved, two-transition study, \citet{Henning+etal_2010}
found that C$_2$H display the very same behavior, which cannot be represented with current chemical models.
While for CN the debate remains possible, the chemistry
of C$_2$H is relatively well known, so the coincidence points towards a general problem to retain
molecules at low temperatures in proto-planetary disks.

In all chemical models, HCN is formed closer to the mid-plane than CN, at densities largely
sufficient to thermalize the HCN J=1-0 transition ($5 \times 10^5$ cm$^{-3}$,
\cite{Guilloteau+Baudry_1981,Monteiro+Stutzki_1986})
The HCN excitation temperature derived from our observations is then expected
to reflect the kinetic temperature of the disk mid-plane in the region where the J=1-0 line is
optically thick, that is up to about 150 AU in DM\,Tau and LkCa 15.
The values derived from the observations are $\simeq 6$ K for DM\,Tau and $\simeq 12$\,K for
LkCa 15, the excitation temperature for MWC\,480 remaining unconstrained. CN, on the other hand,
is expected to be mainly formed (at least largely formed) in the upper layer, close to
the disk surface. Although the overall finding that $T($CN$) > T($HCN$)$ in both
sources is in qualitative agreement with this expectation, the observed values are surprisingly low.

From Fig.\,\ref{fig:cn-t1}, for all models with small grains, 80 \% of the CN column density builds up in
a region where the kinetic temperature is $\geq$ 30\,K. Furthermore, 50 \% of this column density originates
from regions where the density is at least $5-8 \times 10^5 \mathrm{cm}^{-3}$, and CN is not present at all for
densities exceeding about $3 \times 10^6\mathrm{cm}^{-3}$, i.e. below 2 scale heights. This is in contradiction
with the observations which indicate that the bulk of CN is near the disk plane ($\alpha_d < 1$ at the $3\sigma$ level).

While all published disk chemical models clearly have high CN abundances only at large heights, deriving
the densities and temperatures at which the bulk of the CN column density is built is often not possible, as most
studies focussed more on the general behavior of the chemistry than on detailed line emission prediction. Nevertheless,
we note that \citet{Walsh+etal_2010} only have substantial amount of CN for $T > 20$ K and  $n > 10^6$\,cm$^{-2}$,
at rather large z/r, as in our model.

The comparison with other transitions, CN(1-0) and CN(3-2) show that for MWC\,480, our
model predict a stronger CN(3-2) line (9.0 Jy.km/s) than observed. This can be ascribed to partial
sub-thermal excitation for this transition, as its critical density is $3 - 6\times 10^6$ cm$^{-3}$.
For DM\, Tau, our predictions for the CN J=1-0 agree with the observations of \citet{Dutrey+etal_1997},
but there is no published data on the CN(3-2) line. For the disk of LkCa15, our best fit model underestimates
by 30 \% the line flux of the CN(3-2) line. Raising the rotation temperature to about 25 K brings the
predicted line flux in better agreement for this transition. This suggests that a fraction of the CN
molecules lie in a warmer (and dense enough) temperature region. However, although the exact collision rates for CN with H$_2$
are unknown, the densities are high enough to be close to thermalization for the observed lines, and it
is likely that $T_\mathrm{rot} > T_k/2$, so explaining temperatures below 15 K by significant sub-thermal
excitation appears difficult.

\subsection{Surface densities of CN and HCN}
\label{sec:reaction}

A general prediction of all chemical models is that the surface densities of CN and HCN are nearly constant
as a function of radius $p \simeq 0$.  The observed slopes are larger,
$\simeq 1$, and the separate hyperfine CN analysis even suggests a steepening  of the surface density
distribution at larger radii. Furthermore, changing the model disk mass to 1/3$^\mathrm{rd}$ or 3 times the nominal value
only changes the surface density of CN and HCN by about 10-20 \% (see Fig.\,\ref{fig:var:cn}, \ref{fig:var:hcn} and \ref{fig:cn-hcn}).

\begin{figure}  
\includegraphics[angle=270, width=\columnwidth]{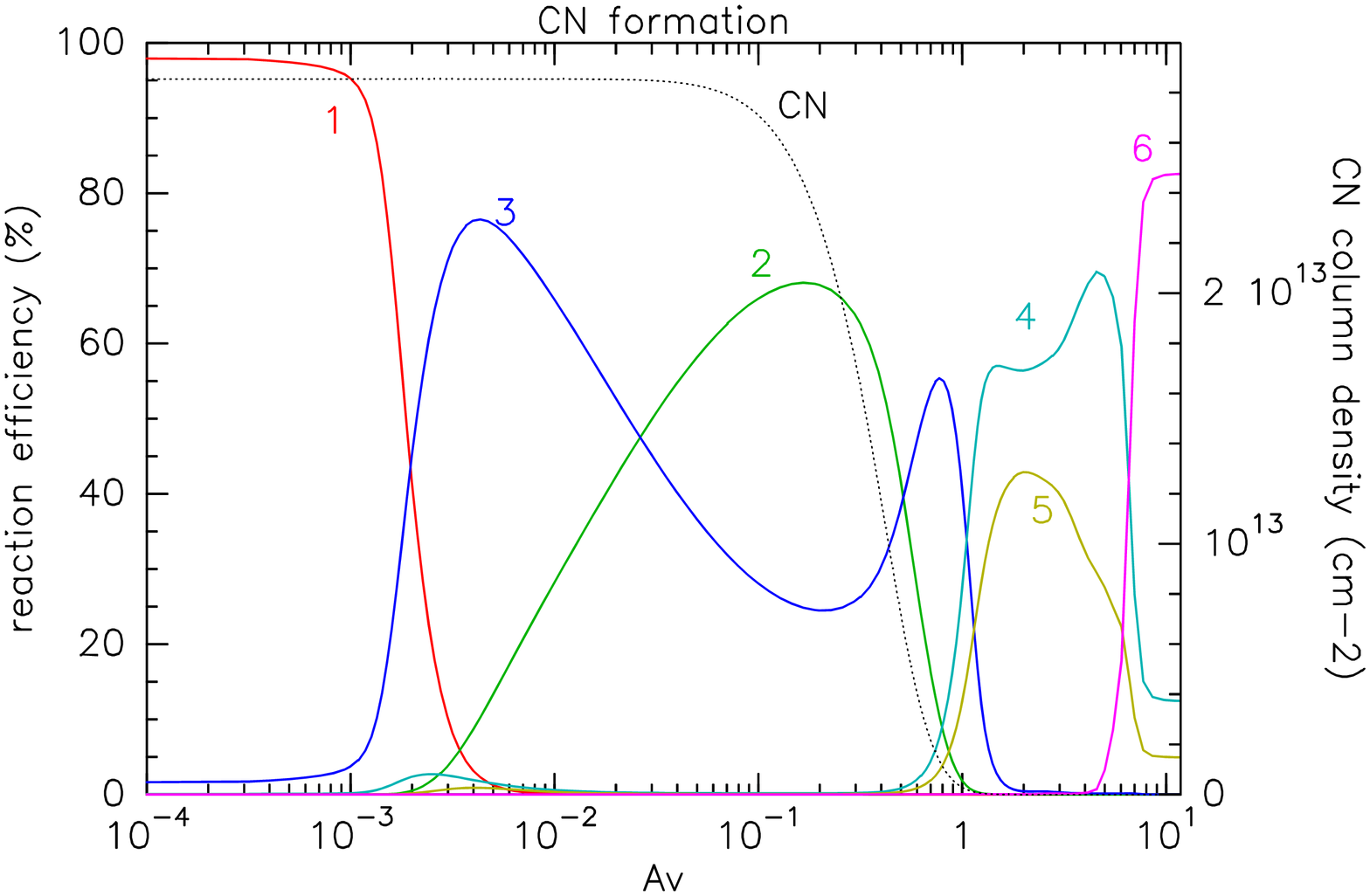}
\caption{Efficiency of CN formation reactions at R=304\,AU for DM Tau with small grains, $\chi=10^3$, CR=$10^{-17}$s$^{-1}$. Only the main reactions are plotted (i.e. more than 10\%). 1 (red): H~+~CN$^+~\rightarrow$~H$^+$~+~CN, 2 (green): N~+~C$_2$~$\rightarrow$~CN~+~C, 3 (blue):  N~+~CH~$\rightarrow$~CN~+~H, 4 (cyan): HCN~+~photon~$\rightarrow$~CN~+~H, 5 (yellow):  HNC~+~photon~$\rightarrow$~CN~+~H, 6 (magenta): OCN~+~photon~$\rightarrow$~O~+CN. The cumulative CN column density is plotted in black.}
\label{fig:reac}
\end{figure}

\citet{Bergin+etal_2003} showed that \lya\ radiation significantly affects the relative photodissociation rates of HCN and CN.
In our model, the impact of the \lya\ line on CN is quite small. This is because the direct production of CN from the photodissociation of HCN ($\mathrm{HCN} + h\nu \rightarrow \mathrm{CN} + \mathrm{H}$) is not in general the main production path of CN. Figure \ref{fig:reac} presents the main chemical reaction leading to the formation of CN at a radius of 304\,AU. Photodissociation of HCN is the most efficient CN production mechanism only in a small region with high HCN abundance. Upwards in the disk, the main CN formation reactions are $\mathrm{N} + \mathrm{CH}
\rightarrow \mathrm{CN} + \mathrm{H}$ and $\mathrm{N} + \mathrm{C}_2 \rightarrow \mathrm{CN} + \mathrm{C}$, so that
the overall abundance of CN is controlled by the abundance of atomic Nitrogen. However, HCN decreases with Ly$\alpha$ intensity, so that the CN/HCN ratio depends on the Ly$\alpha$ flux \citep[as also found by][]{Fogel+etal_2011}. We find however that the effect is very small when dust is composed of small dust grains. With large grains,
the HCN column density is divided by a factor $\simeq 2.5$ for a \lya\ line contributing to  1/3rd of the UV luminosity.
This is consistent with the \citet{Fogel+etal_2011} results where the effect is of \lya\ is visible only for settled grain models (i.e. models with a reduced UV extinction in the upper layers, as for large grains).

Finally, PDR models also predict that the CN/HCN ratio increases with grain size because large grains allow UV flux to penetrate deeper in the disk. This effect is illustrated in Fig.\,\ref{fig:cn-t3}. The high CN/HCN ratio (4 to 10 at 300 AU)  in the three sources is in agreement with the presence of large grains in the three disks. On the other hand, \citet{Aikawa+Nomura_2006} results suggest a rather weak dependency of the CN/HCN ratio on the grain sizes. However, this study explored much bigger grain sizes than we do here, but the differences between their 10 $\mu$m and 1 \,mm case is similar to our findings.
\citet{Walsh+etal_2010} also found that the CN/HCN ratio increases when surface chemistry is included, as
Nitrogen can be incorporated into larger molecules in the disk mid-plane.

\subsection{Toward a cold molecular Layer ?}

We investigate here mechanisms which would partially provide cold molecular material close to the mid-plane.

\subsubsection{Large grains} Grain growth can bring molecules closer to the disk mid-plane, by allowing
UV radiation to penetrate deeper in the disk structure. Fig.\,\ref{fig:cn-t1} also indicates that the only
case in which CN is produced in sufficient quantities at low temperatures is the high UV ($\chi =10^4$),
big grains $a_+ = 0.1$~mm case, where 50\% of the CN column density builds up at temperatures below 10 K,
and the rest between 10 and 20~K. However, neglecting freeze-out and surface chemistry in this case is unlikely to be valid.
Moreover, in this case, the disk temperature and density structure was imposed \textit{a priori}, and is
not consistent with the impinging UV field and its attenuation.  If, using the same density structure, we
compute the thermal balance assuming the incident UV field is the only heating source, the picture changes
completely (Fig.\,\ref{fig:cn-t2}). The whole gaseous disk becomes warm, the minimum temperature being 60 K.
Furthermore, 90 \% of the CN  column density is build up at densities above 10$^6$ cm$^{-3}$, thus
sub-thermal excitation becomes also unlikely.

Enhanced photo-desorption would bring larger amounts of (cold) CN, HCN and C$_2$H near the
disk mid-plane. \citet{Willacy+Langer_2000} checked this possibility using moderate grain growth but
it does not provide enough cold gas. Doing the same calculation within larger grains appears however
as an interesting clue to partially reconcile observations and models.

\subsubsection{Turbulent mixing} Turbulent transport could bring CN and C$_2$H formed in the warm upper
layers down to the disk mid-plane, where they would cool very quickly.  The main difficulty is to
avoid sticking onto cold dust grains. \citet{Hersant+etal_2009} explored this for CO molecules, and
found that turbulent diffusion does not affect much the total amount of CO, but could bring a small
fraction of it to low temperatures. \citet{Willacy+etal_2006}
managed to get a substantial amount of CN towards the disk plane, but only using a large diffusion coefficient,
$10^{-18}$ cm$^2$.s$^{-1}$, which is 10 times larger than expected from our current estimates of the
viscosity ($\alpha$ viscosity parameter below 0.01) in disks. We note however that the model predictions are (at least for CN) not too far from
the observed column densities. The low mixing efficiency, which is a result of dilution \citep[the gas density in
the upper layer being small in comparison with the disk mid-plane, see also][]{Semenov+etal_2006})
may thus not be a major problem, as we only want to bring the existing CN molecules to low temperature
regions, but not to increase significantly the total amount of CN.

\subsubsection{Lowering the gas-to-dust ratio}
In our resolved observations, $\Sigma_d$ is actually geometrically constrained by measuring the height
above the disk mid-plane where the molecules are located; it is expressed as a column density by
simply scaling to the total disk surface density (see Eq.\ref{eq:sigmad-h}). When assuming
$\kappa_\nu(230 \mathrm{GHz}) = 2$ cm$^2$.g$^{-1}$ (of dust) and a canonical gas-to-dust ratio of 100, the
measured ``depletion'' column density, $\Sigma_d$, appears $\sim 4$ -- 10 times larger than predicted from
chemical models \citep[see in particular][]{Aikawa+Nomura_2006}.  \citet{Aikawa+Nomura_2006} also show
that $\Sigma_d$ does not strongly depend on the grain size: so what controls the location
of the CN layer (and those of other molecules) is essentially the column density of hydrogen towards the
disk surface.  Lowering the gas-to-dust ratio to 20 (or reducing the hydrogen surface density by a factor 5),
we bring to first order the observed $\Sigma_d$ in rough agreement with the model prediction.
In such a case, CN would be present in regions where the density is low enough (a few $10^5$  cm$^{-3}$)
to lead to substantial sub-thermal excitation.

We checked this possibility by using our PDR and radiative
transfer codes in a disk 6 times less massive than our nominal assumption. The excitation is computed using
the modified LVG approximation as described in \citet{Pavlyuchenkov+etal_2007} who also show that this is a
reasonably accurate solution for the radiative transfer in disks. The collision rates for CN were assumed
identical to those of HCN with He \citep{Green+Thaddeus_1974}. For DM\,Tau, using the tapered edge surface
density profile, we indeed find out \textbf{$\Sigma_d = 1.7 \pm 0.1 \times 10^{21}$} cm$^{-2}$, and a kinetic temperature of
15 - 20 K. Given the uncertainties on the collision rates, such a solution appears compatible with the
expected temperature range in the molecular rich region. Such a low gas-to-dust ratio would also provide
an appropriate explanation for the low temperatures also derived from C$_2$H by \citet{Henning+etal_2010}.

Unfortunately, with our chemical model, such low disk masses result in predicted HCN and CN column
densities which are substantially lower than observed, so that this mechanism is not fully satisfactory
to explain the observations.  Imaging higher lying transitions, which have higher
critical densities than the J=2-1 line observed here, could give more accurate constraints.

Note that, rather than changing the gas-to-dust ratio, raising the mm wave dust absorption coefficient
by a similar factor is also a strictly equivalent alternative, and leads to the same low disk masses ($0.003 \msun$
for DM\,Tau). However, our assumed value of $\kappa(1.3\mathrm{mm}) = 2$\,cm$^2$.g$^{-1}$ (of dust) already appears
large \citep[see e.g.][and references therein]{Draine_2006}.

\subsubsection{Low temperature chemistry}

Finally, \citet{HilyBlant+etal_2008} have observed very cold $^{13}$CN in the nuclei of a few
dense cores of known temperature and density (T $\sim 6-10$\,K, $n \sim 10^4-10^6 \mathrm{cm}^{-3}$), which
present temperature and densities relatively similar to those encountered in the mid-planes of TTauri disks.
They find that CN is not strongly depleted in the nuclei of the cores, contrary to CO, and can be a fair column
density tracer in these extreme conditions. However, \citet{HilyBlant+etal_2010} were  unable to account
for the near constancy of the CN/HCN ratio in their chemical models. These
observations lead also open the possibility that cold chemistry is not fully understood at least for CN.
\citet{HilyBlant+etal_2010} conclude that our knowledge of
the low temperature chemistry of CN is limited by the extrapolation of the rate coefficients
measured at high temperatures to the 10 K regime for many of the important reactions, including
those mentioned in Sec.\ref{sec:reaction} \citep{Pineau+etal_1990,Boger+Sternberg_2005}.
However, at such low temperatures and high densities, chemistry also depends on reactions on the grain surfaces,
as shown by \citet{Walsh+etal_2010} in particular for HCN.
Hence, another viable alternative would be inappropriate knowledge of the grain surface chemistry.

\section{Summary}

We present sensitive, high spatial and spectral resolution observations of HCN J=1-0 and CN J=2-1 in 2 T Tauri disks
(DM Tau and LkCa 15) and one Herbig Ae (MWC\,480). Column densities and excitation conditions
are recovered through a minimization scheme which utilizes all the information provided by the hyperfine structure
of the observed transitions. Although we find out $T(\mathrm{CN}) > T(\mathrm{HCN})$ as expected from a layered disk structure, the analysis indicates that the location of CN and its apparent excitation temperature appears incompatible
with the expectations from current chemical models. Lowering the gas-to-dust ratio might partially solve the
conflict on temperature and localization of the molecules, but may not produce large enough quantities of CN and HCN.
The observations thus suggest that a substantial fraction of the molecules are created in the disk plane
at low temperatures, perhaps by surface chemistry. Sensitive, spatially resolved images of higher excitation lines of CN and HCN would be required to provide a complete diagnostic to determine the excitation conditions in disks allowing
us to decide between the possible options.

\begin{acknowledgements}
We acknowledge all the Plateau de Bure IRAM staff for their help during the observations. We thank
Pierre Hily-Blant for his modifications of a previous version of the PDR code. We also acknowledge
Franck Le Petit for many fruitful discussions about the PDR code. This research was supported by the
program PCMI from INSU/CNRS.
\end{acknowledgements}

\bibliography{bib-cn}
\bibliographystyle{aa}

\newpage
\begin{appendix}
\section{Channel maps}
\begin{figure*}[!h]  
       \includegraphics[width=0.8\textwidth]{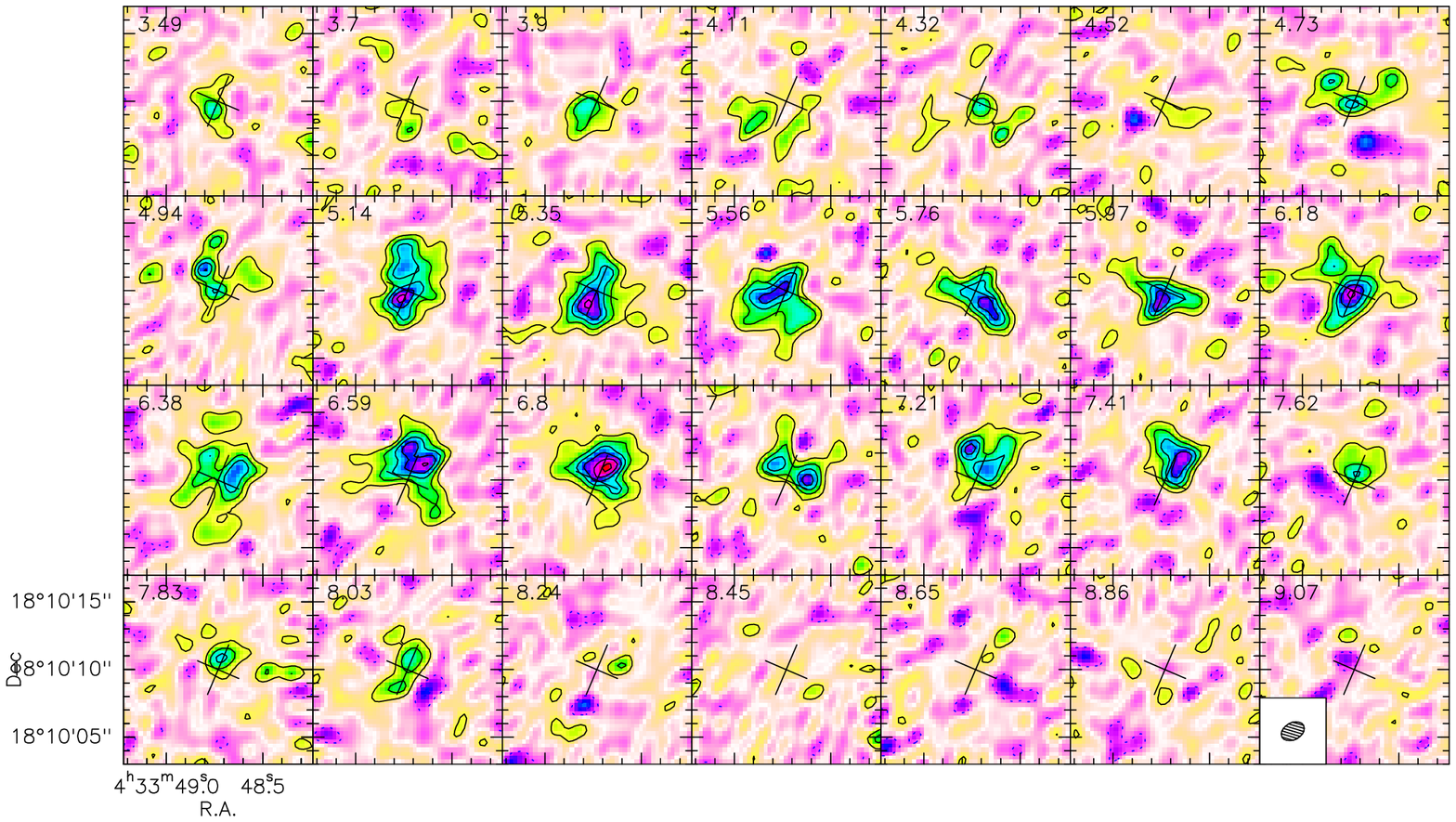}
       \includegraphics[width=0.8\textwidth]{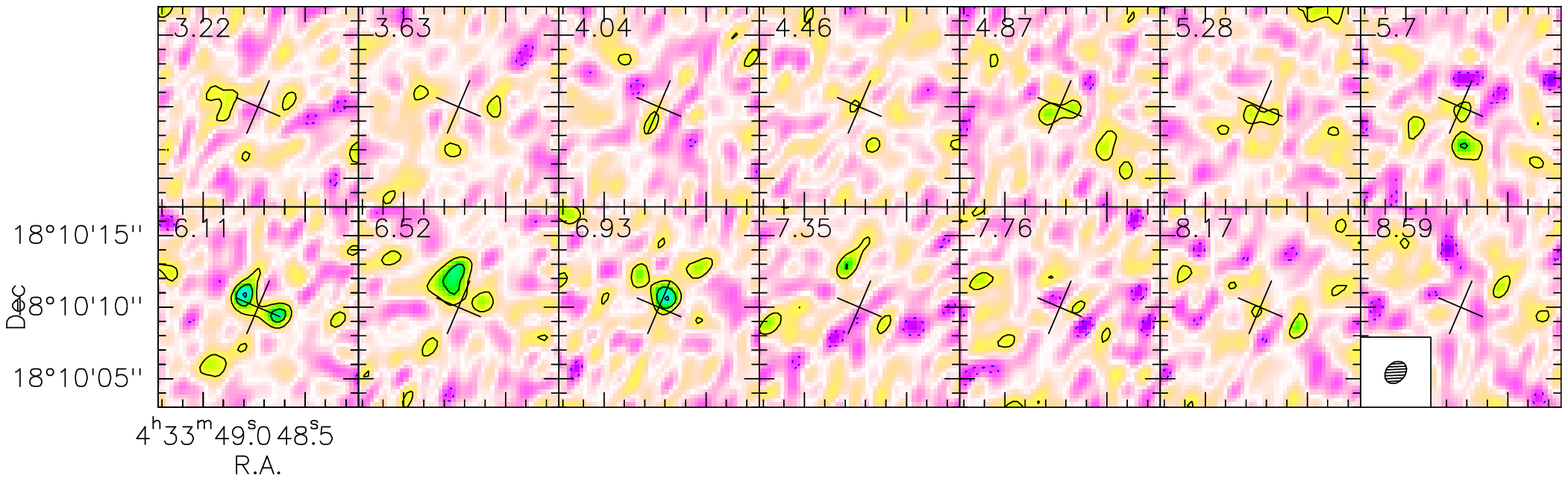}
       \includegraphics[width=0.8\textwidth]{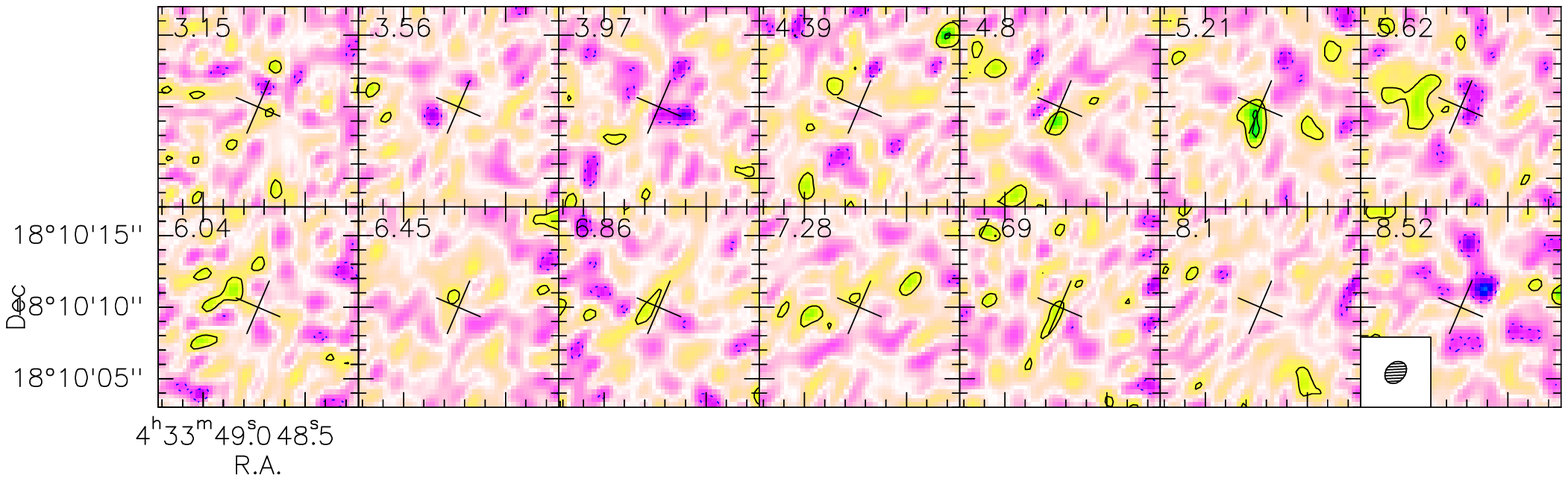}
       \includegraphics[width=0.8\textwidth]{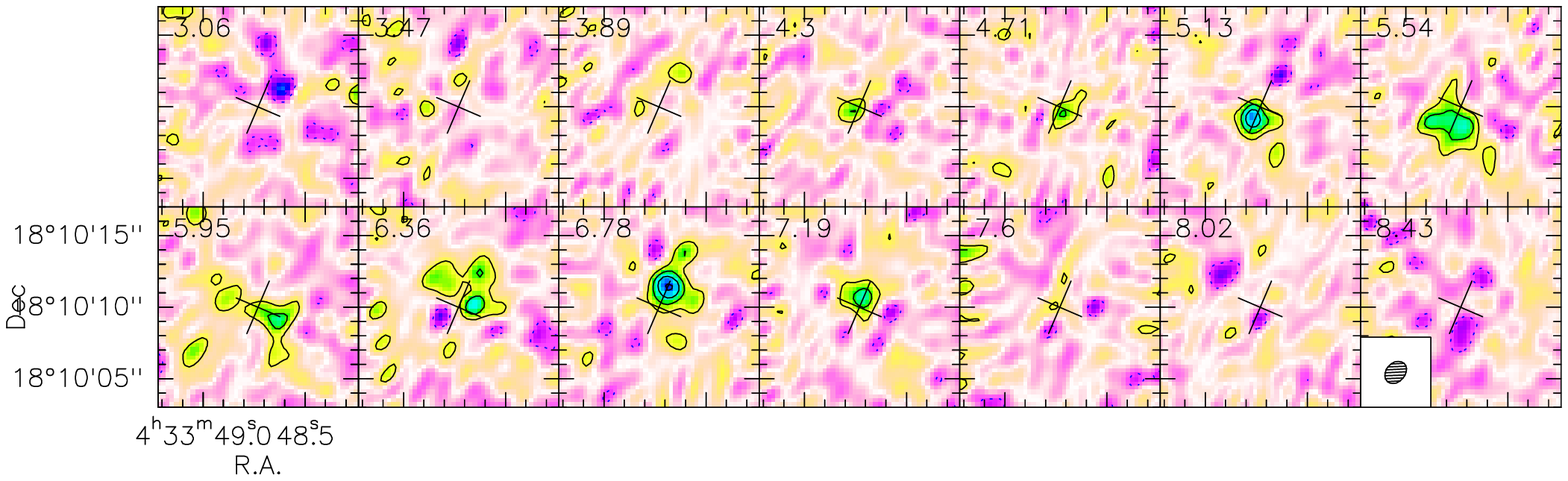}
\caption{Channel maps of CN hyperfine components towards DM Tau. From
top to bottom, the velocity scales are referred to  226.874745 GHz (3 blended components),
 226.679382 GHz, 226.663703 GHz and 226.659575 GHz. Note the better
 spectral resolution in the top panel. the spatial resolution is
   $1.7\times 1.2''$ at PA $120^\circ$, contour spacing is 53 mJy/beam, or 0.61 K, corresponding to
   $2.0 \sigma$ in the upper panel, and $2.3 \sigma$ in the lower ones.}
\label{fig:cn-dmtau}
\end{figure*}

\begin{figure*}[!h]  
       \includegraphics[width=0.8\textwidth]{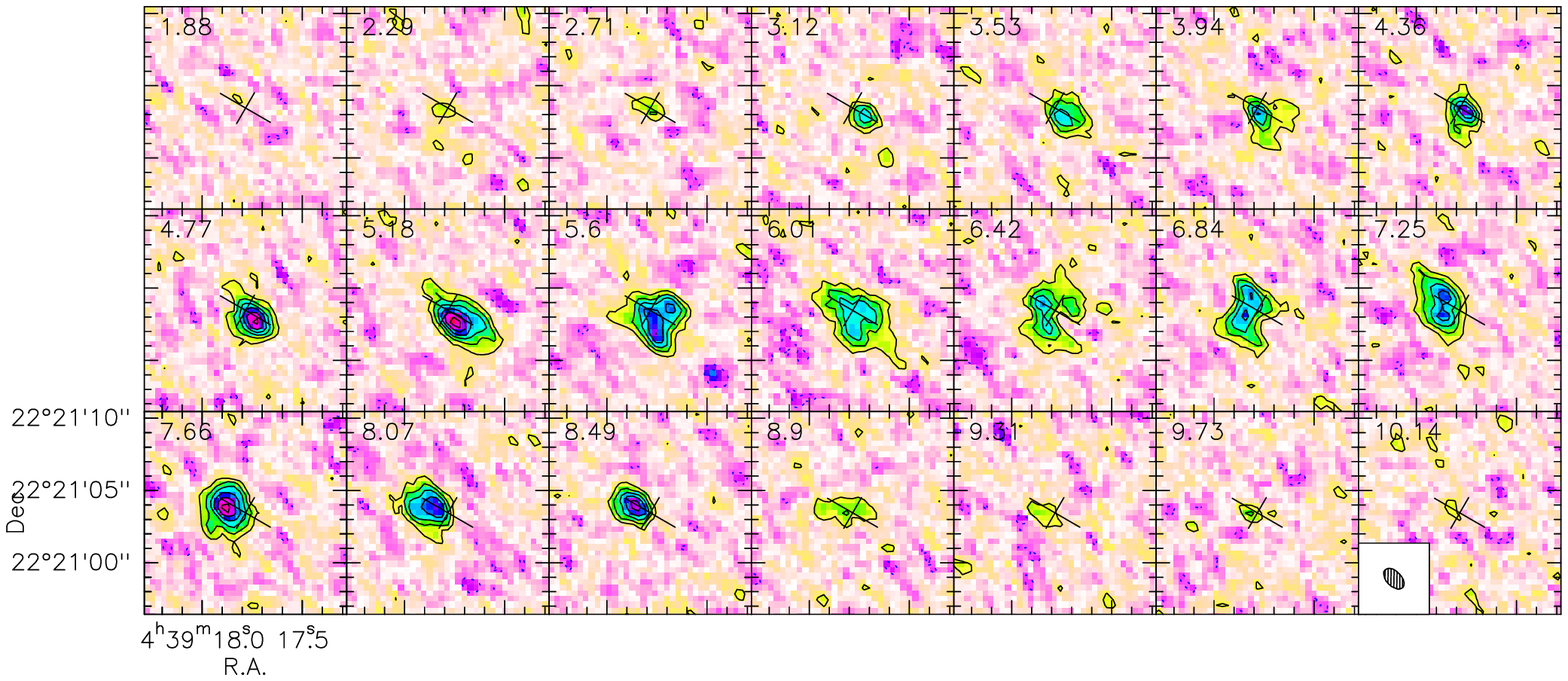}
       \includegraphics[width=0.8\textwidth]{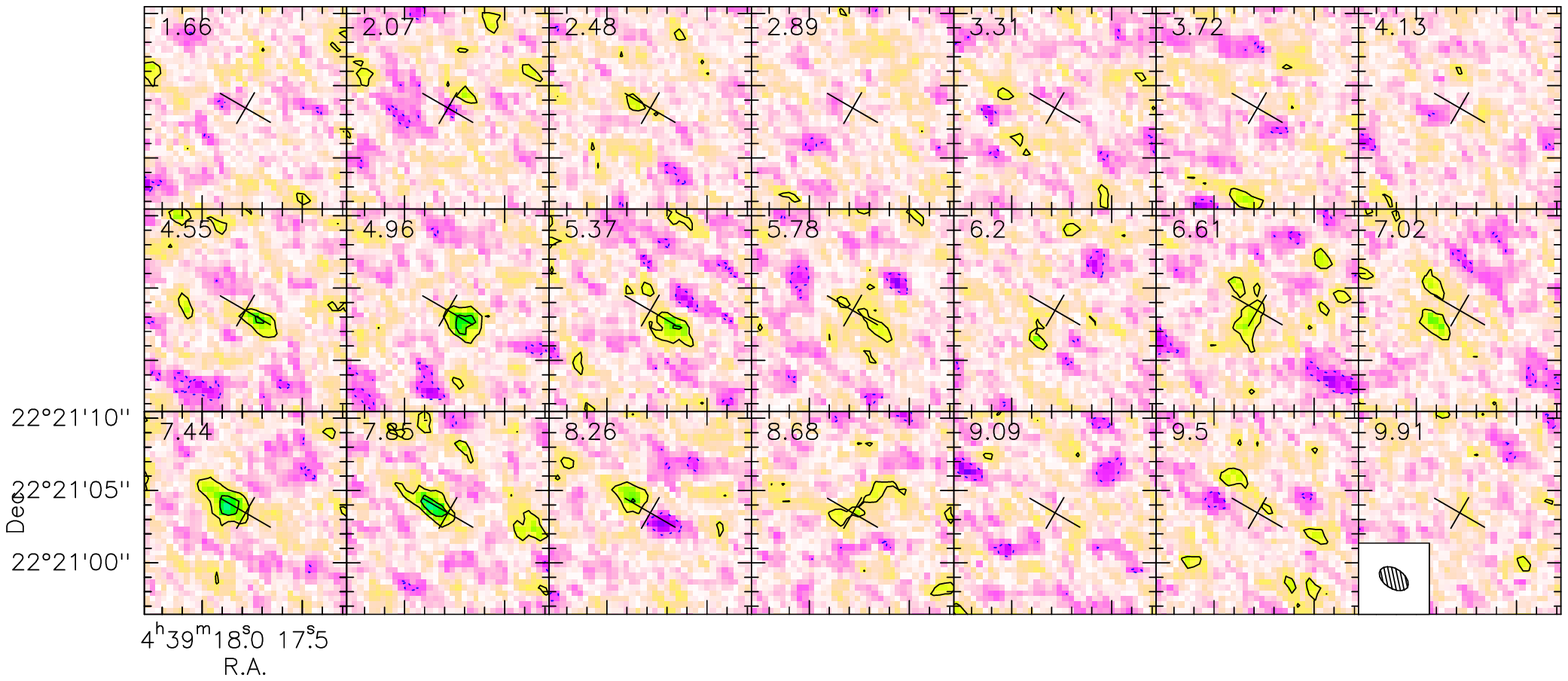}
       \includegraphics[width=0.8\textwidth]{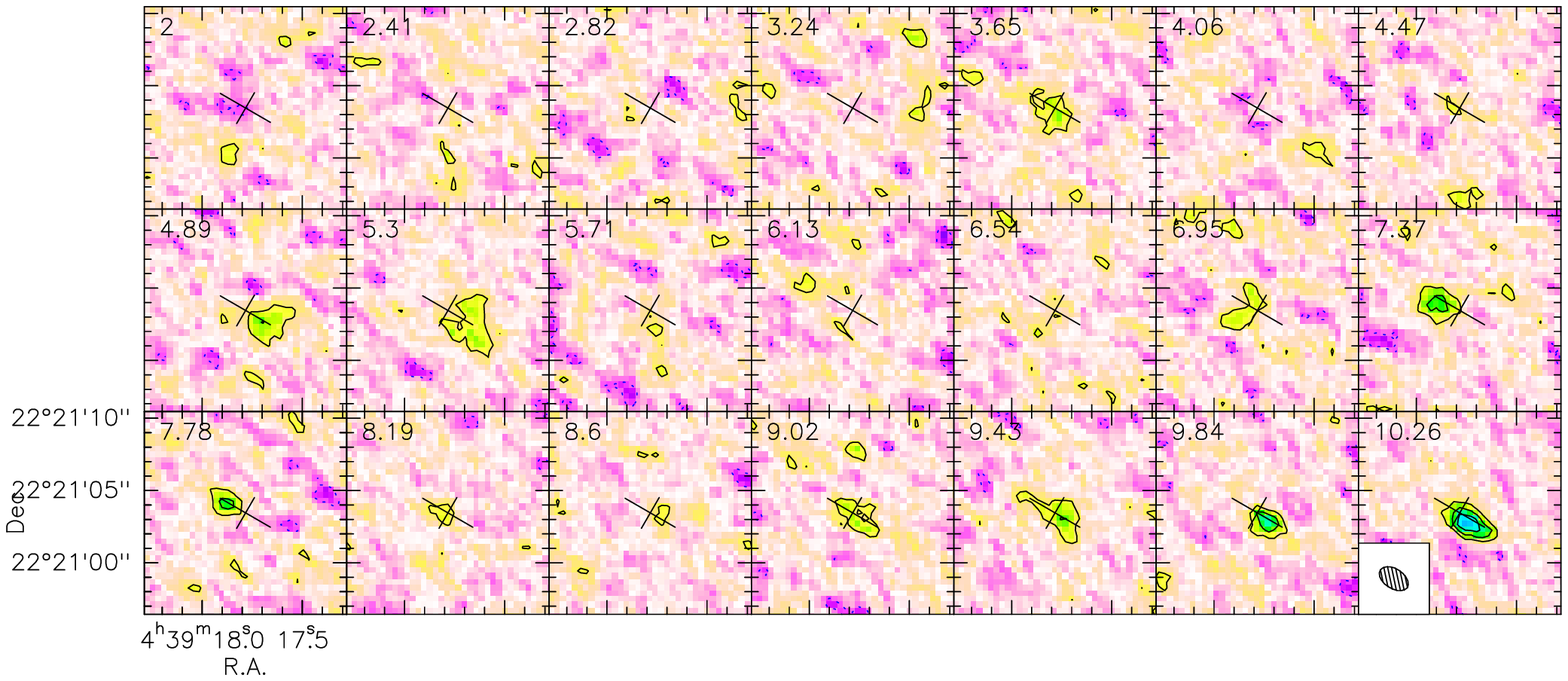}
       \includegraphics[width=0.8\textwidth]{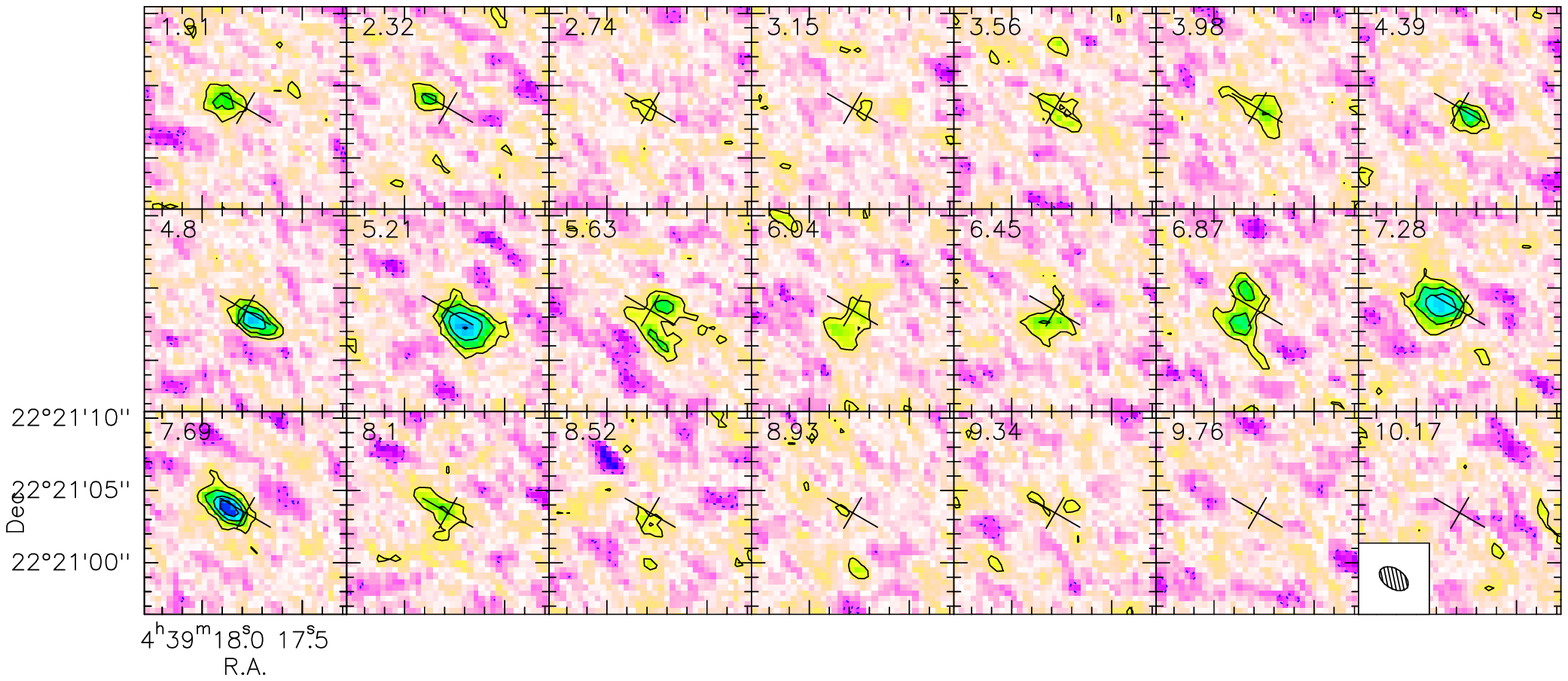}
\caption{As Fig.\ref{fig:cn-dmtau}, but towards LkCa15. All panels
use the same spectral resolution. Note the repeated channels
in the two bottom panels, as the velocity spacing between these components
is 5 km/s, i.e. 12 channels. The spatial resolution is $1.7\times 1.0''$ at PA $44^\circ$, contour spacing is
   60 mJy/beam, or 0.81 K and $2.1\sigma$. }
\label{fig:cn-lkca}
\end{figure*}

\begin{figure*}[!h]  
       \includegraphics[width=0.8\textwidth]{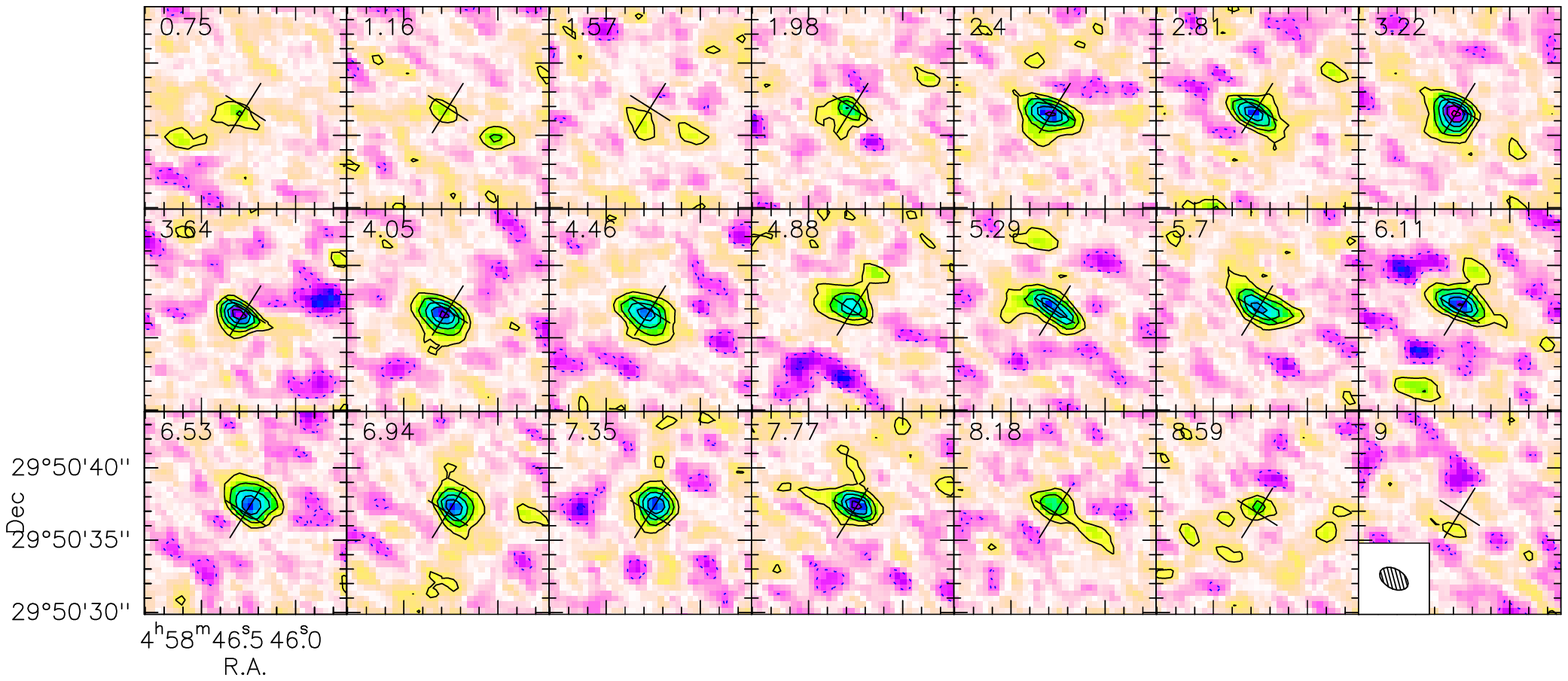}
       \includegraphics[width=0.8\textwidth]{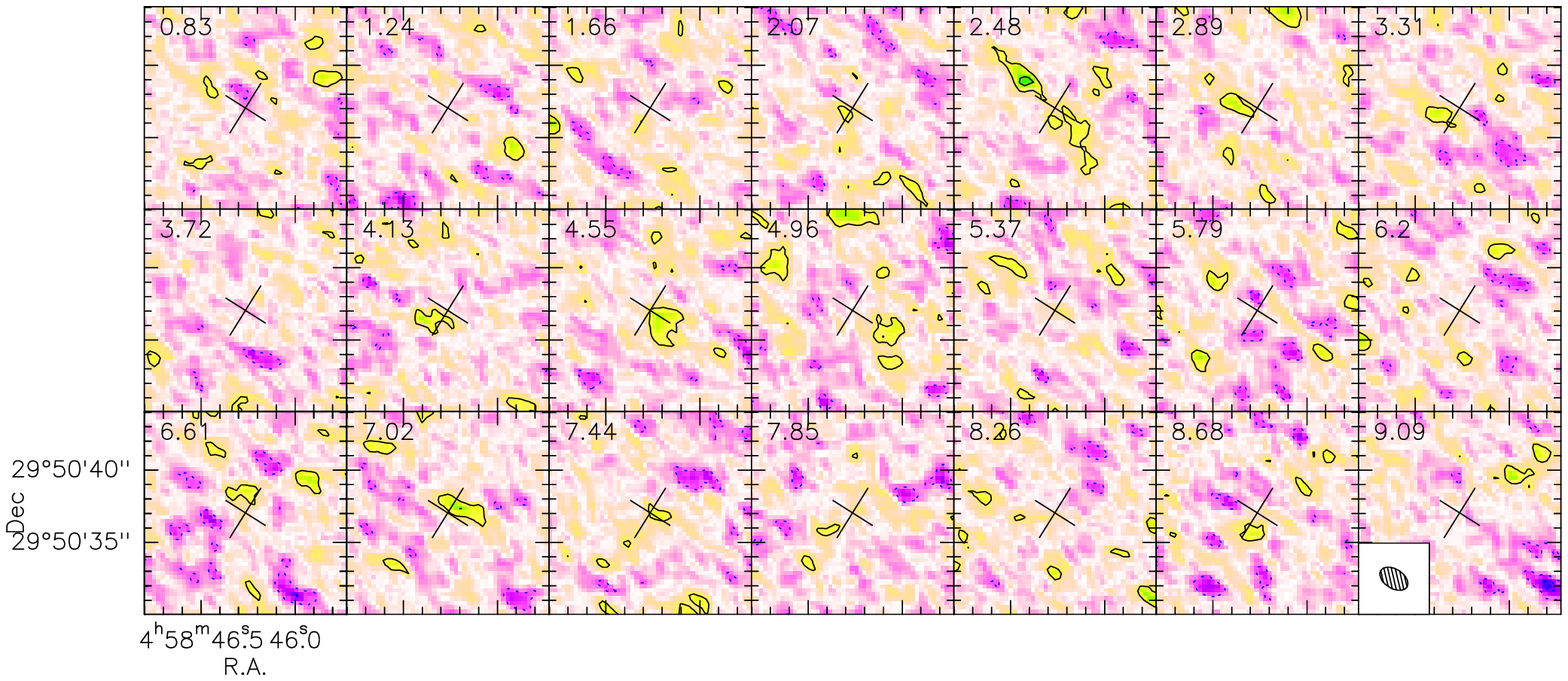}
       \includegraphics[width=0.8\textwidth]{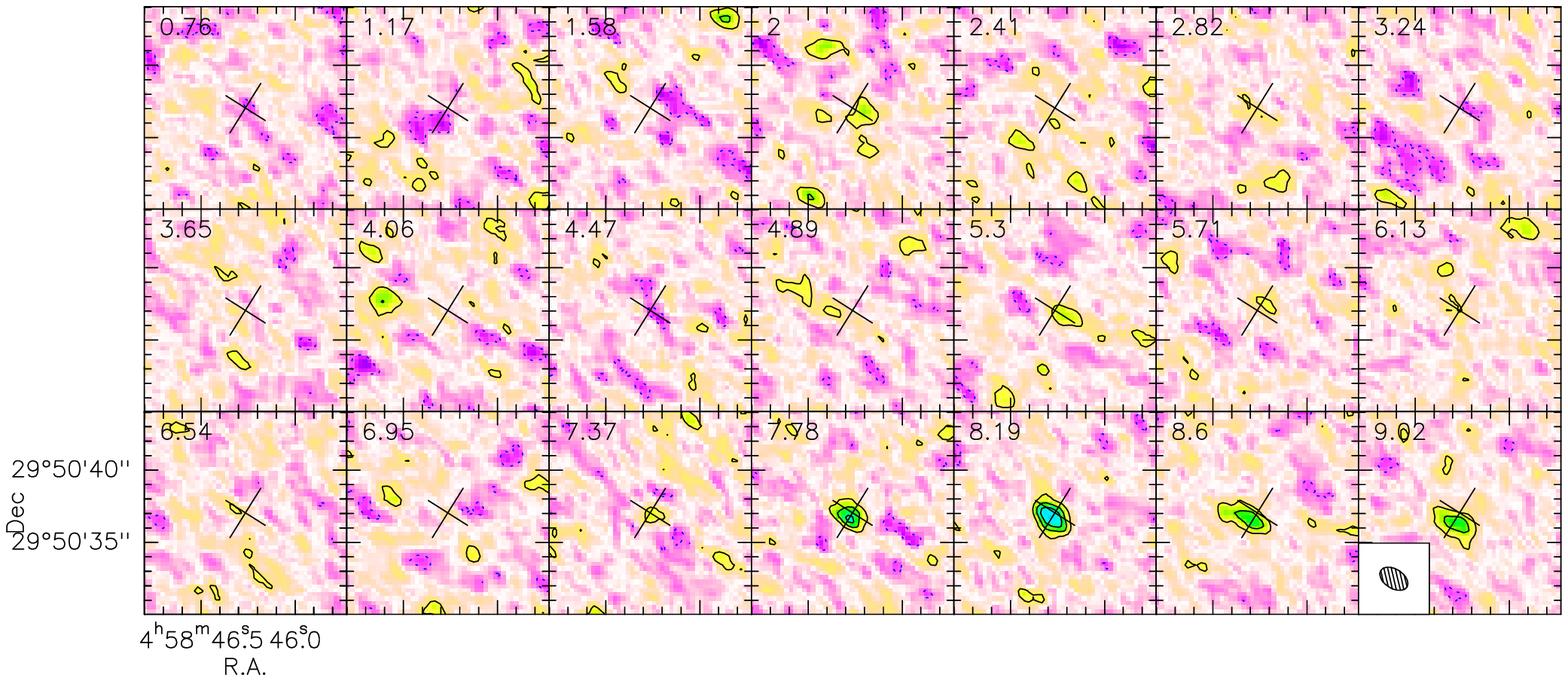}
       \includegraphics[width=0.8\textwidth]{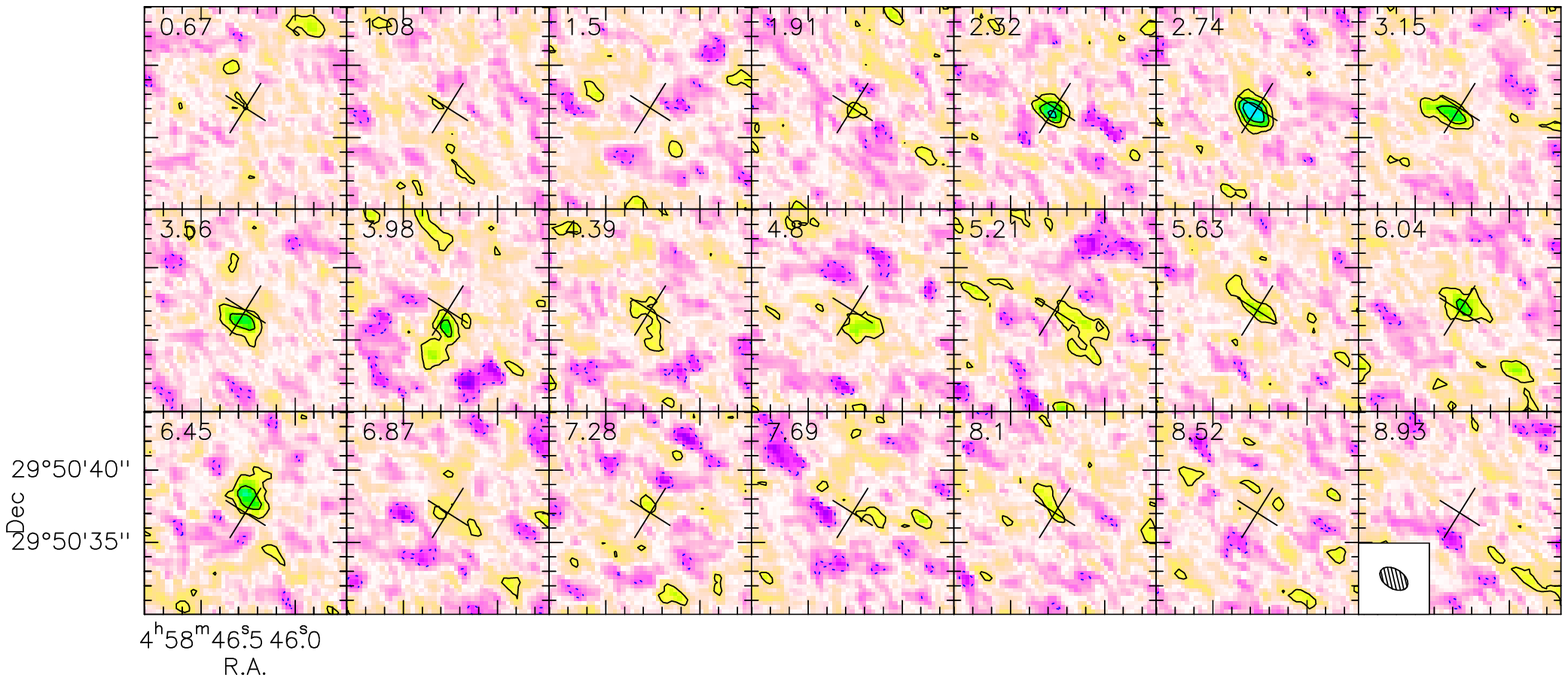}
\caption{As Fig.\ref{fig:cn-lkca}, but towards MWC 480.
The spatial resolution is $2.2\times1.35''$ at PA $60^\circ$, contour spacing is 55 mJy/beam, or 0.45 K and $2 \sigma$.
The large rotation velocity result in an overlap of emission between the two hyperfine components in the bottom panels.}
\label{fig:cn-mwc480}
\end{figure*}

\begin{figure*}[!h]  
       \includegraphics[width=\textwidth]{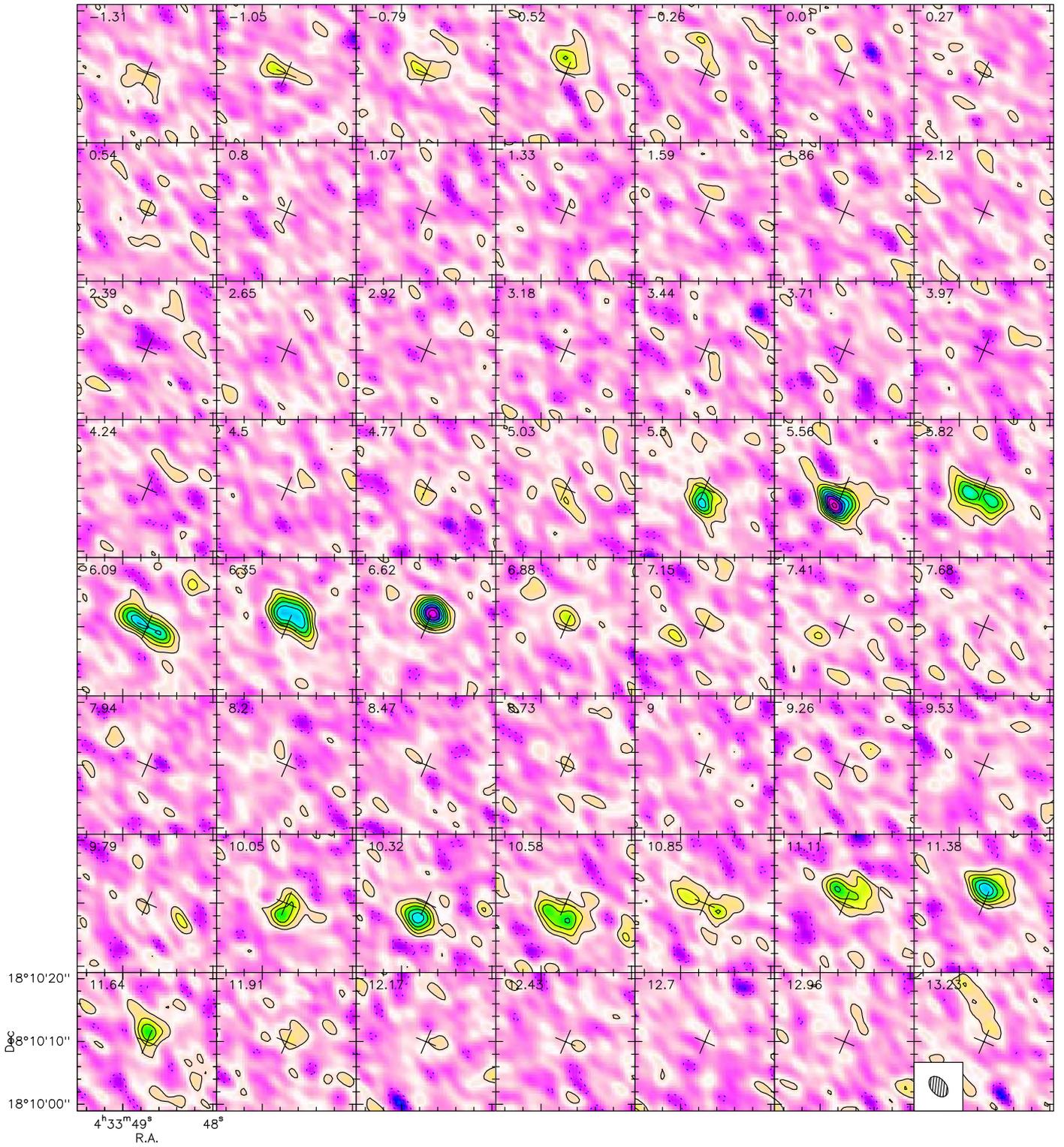}
\caption{Channel maps of HCN hyperfine components towards DM\,Tau. The spatial resolution is
   $3.7\times 2.4''$ at PA $39^\circ$, contour spacing is 12 mJy/beam, or 0.21 K, $2.0 \sigma$.}
\label{fig:hcn-dmtau}
\end{figure*}

\begin{figure*}[!h]  
       \includegraphics[width=0.8\textwidth]{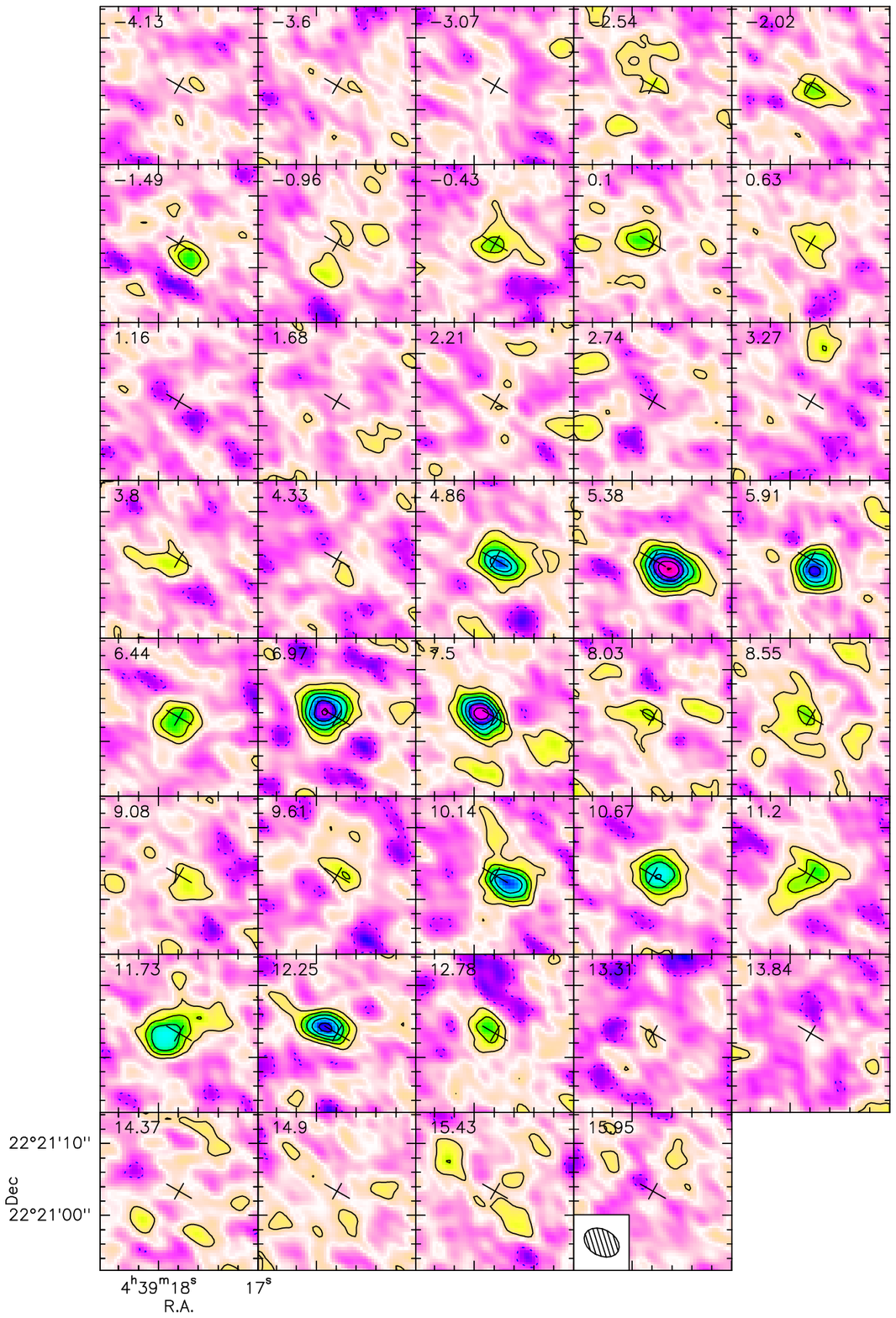}
\caption{Channel maps of HCN hyperfine components towards LkCa15. The spatial resolution is $4.1\times 2.9''$ at PA $51^\circ$, contour spacing is 16 mJy/beam, or 0.21 K and $2.0 \sigma$.}
\label{fig:hcn-lkca}
\end{figure*}

\begin{figure*}[!h]  
       \includegraphics[width=0.8\textwidth]{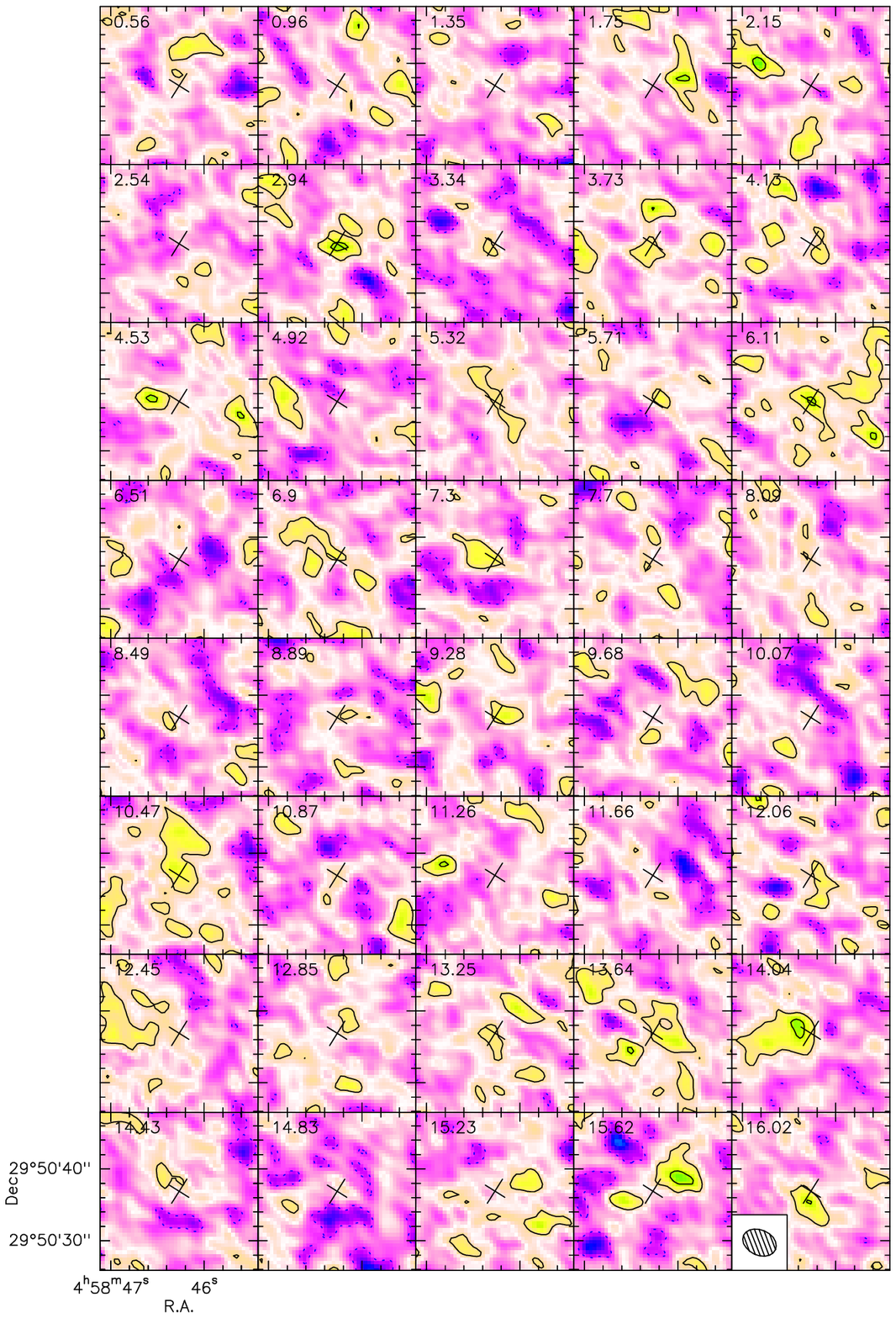}
\caption{Channel maps of HCN hyperfine components towards MWC\,480. The spatial resolution is
$4.5\times3.6''$ at PA $66^\circ$, contour spacing is 16 mJy/beam, or 0.15 K and $1.8 \sigma$.}
\label{fig:hcn-mwc480}
\end{figure*}

\section{Chemical model results}

\begin{figure*}[!h]  
  \centering
\begin{tabular}{c}
  DM\,Tau \\
       \includegraphics[angle=270,width=9cm]{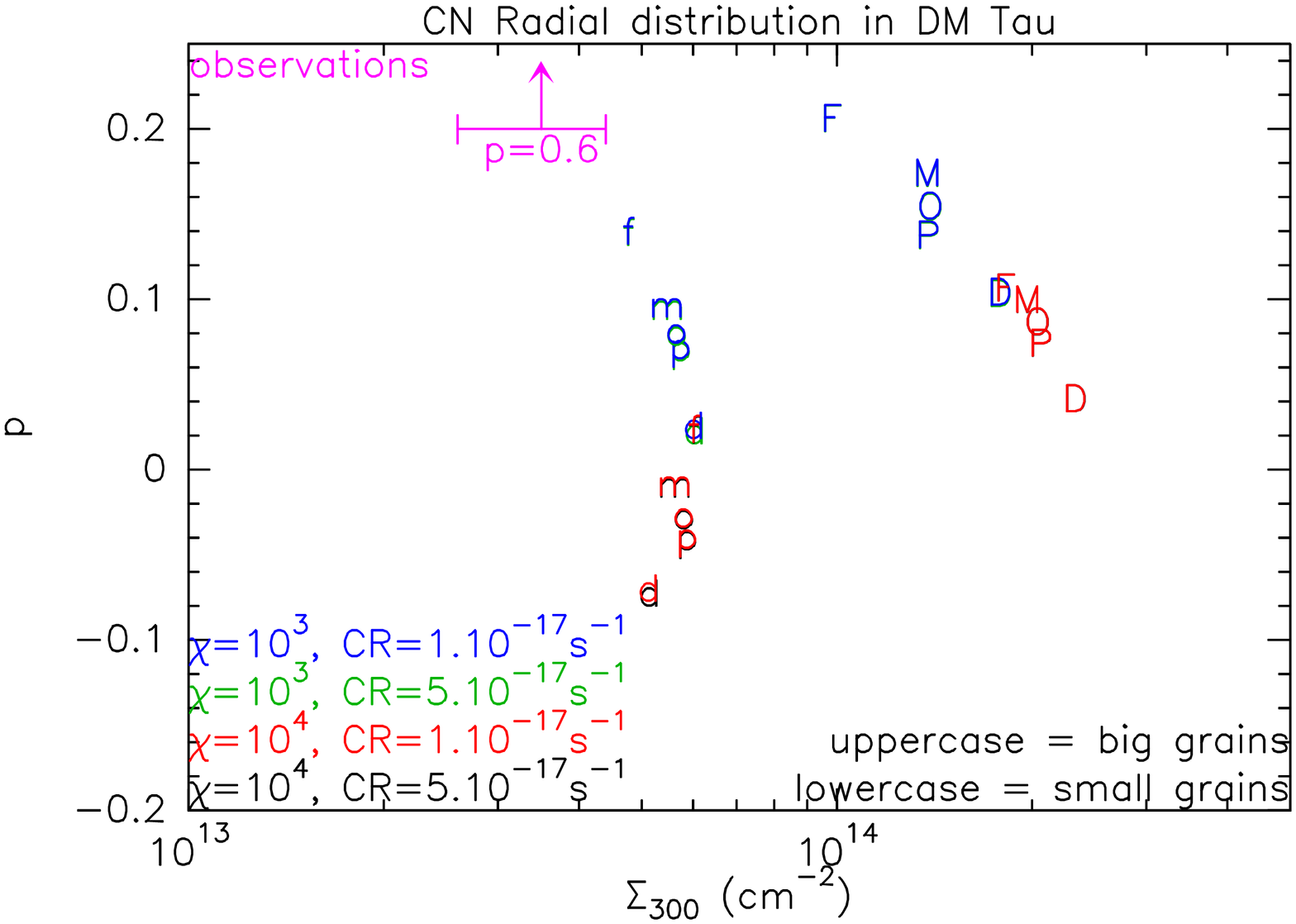}
      \includegraphics[angle=270,width=9cm]{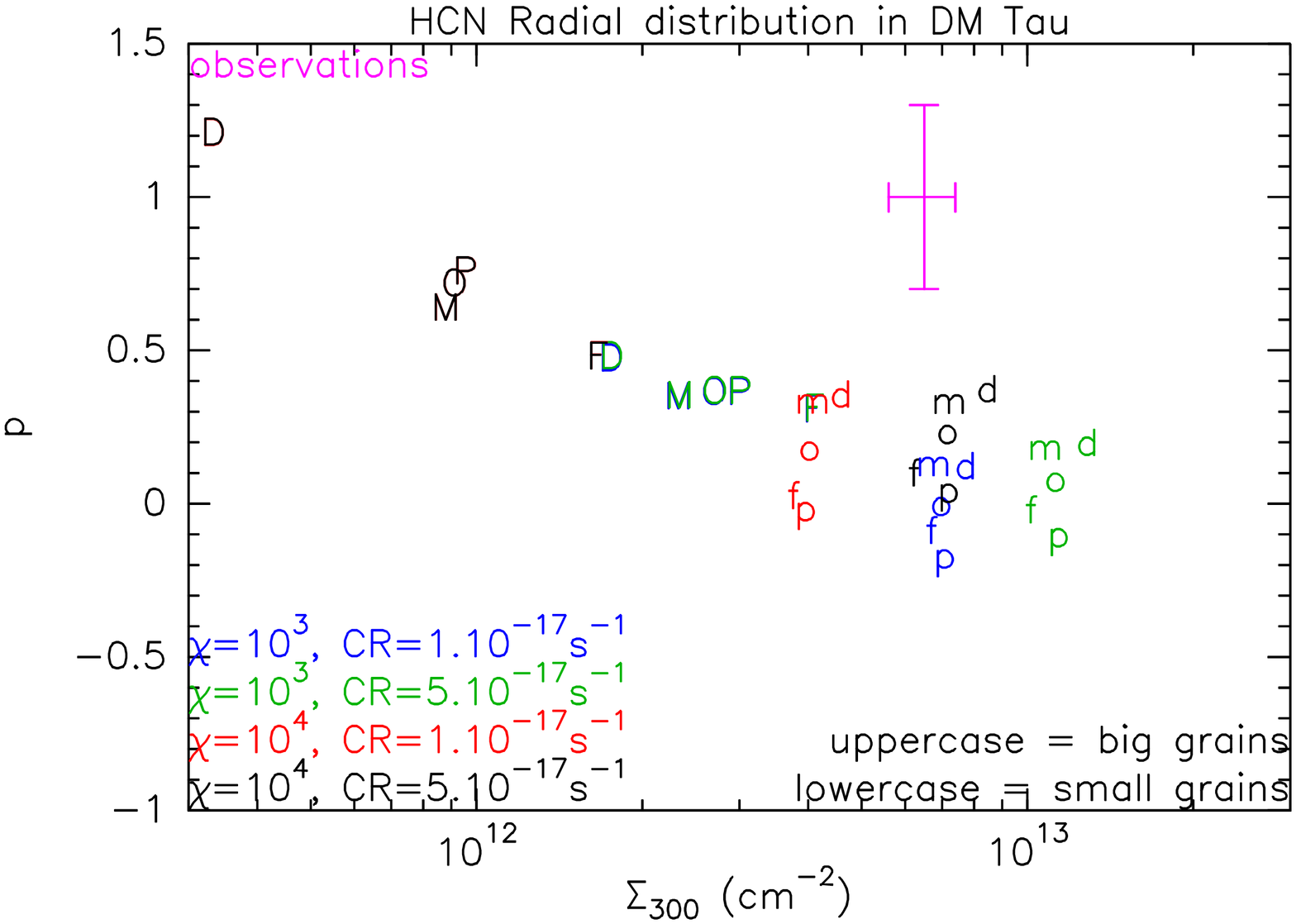}\\
 LkCa\,15\\
       \includegraphics[angle=270,width=9cm]{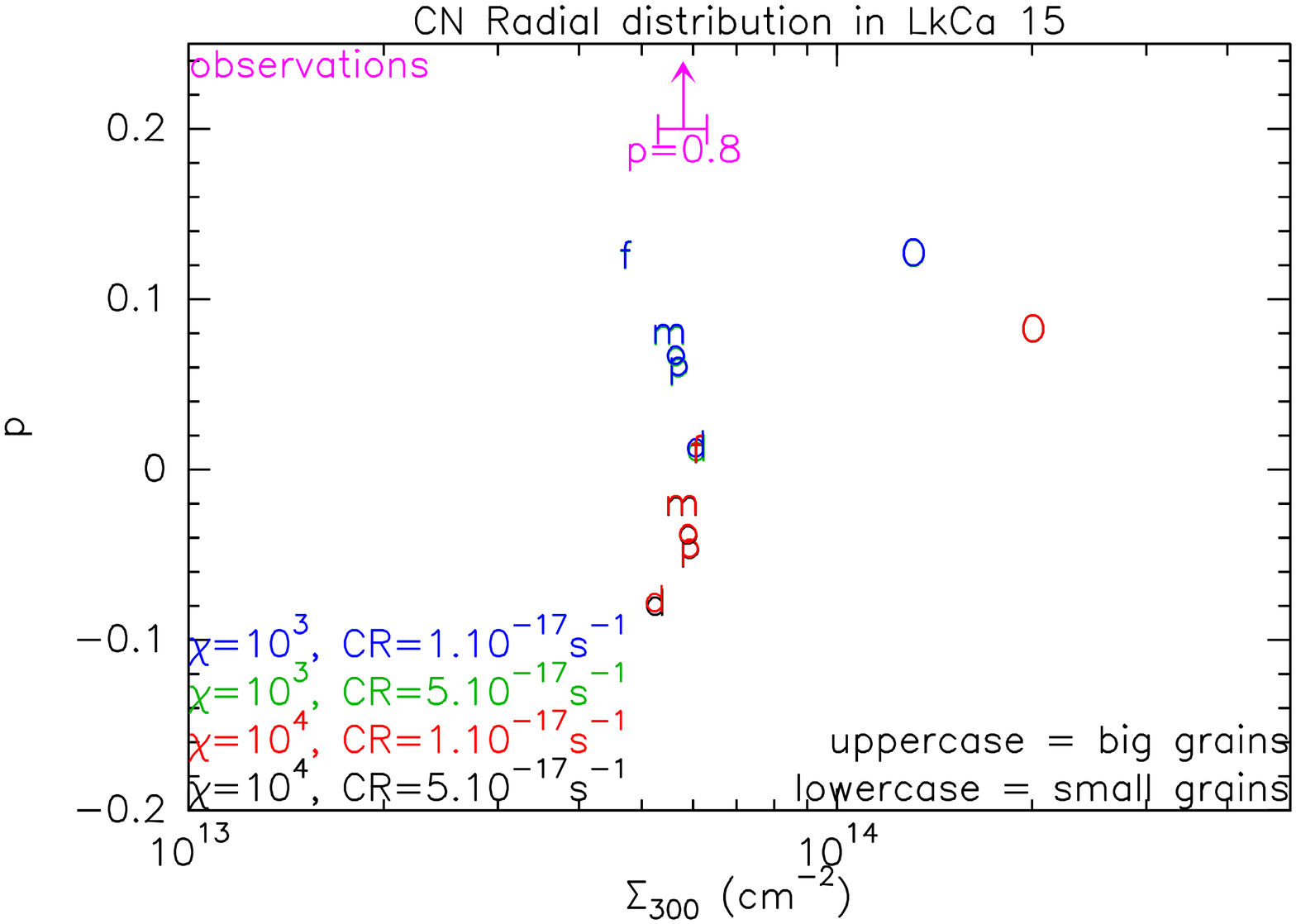}
      \includegraphics[angle=270,width=9cm]{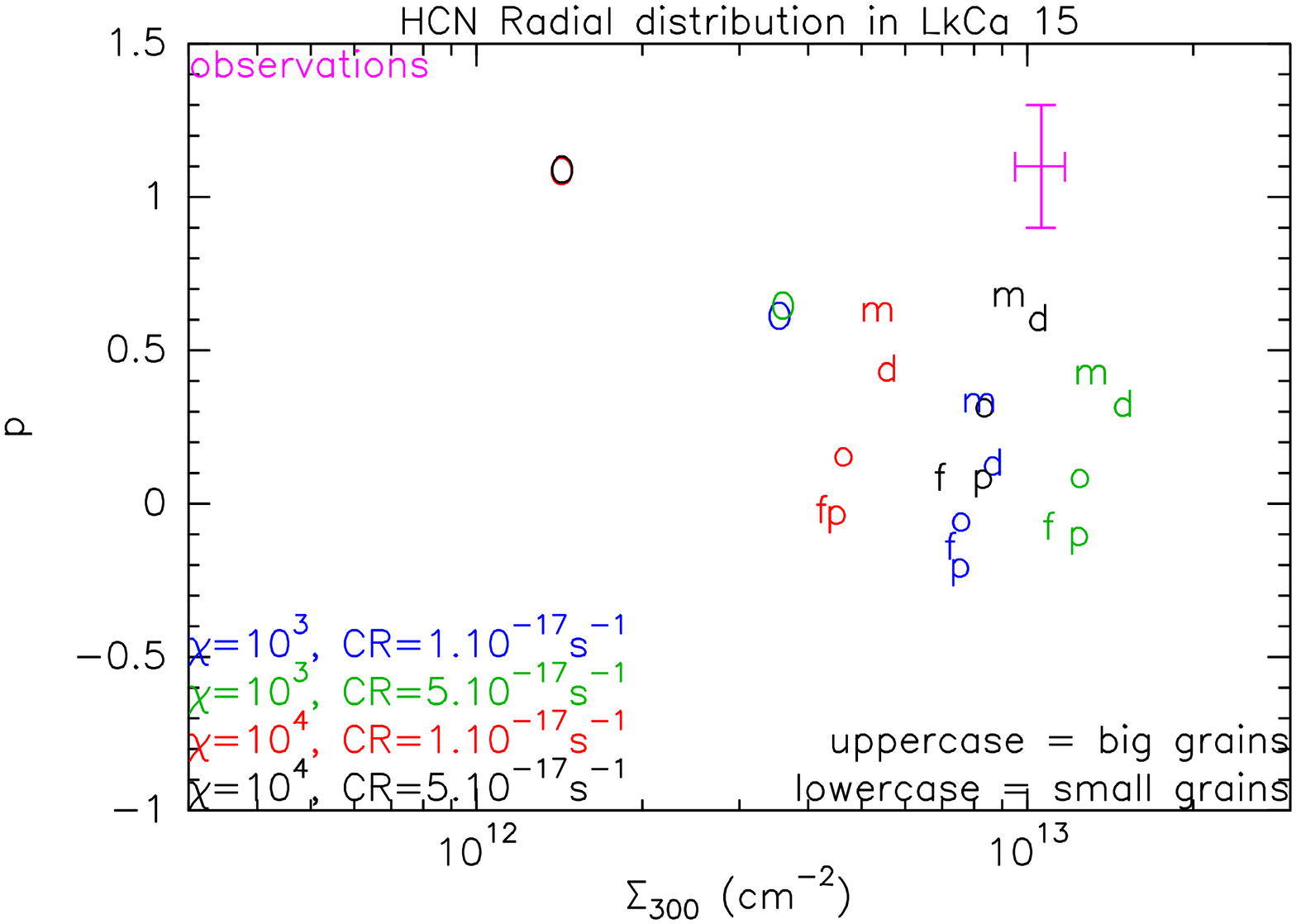}\\
 MWC\,480\\
       \includegraphics[angle=270,width=9cm]{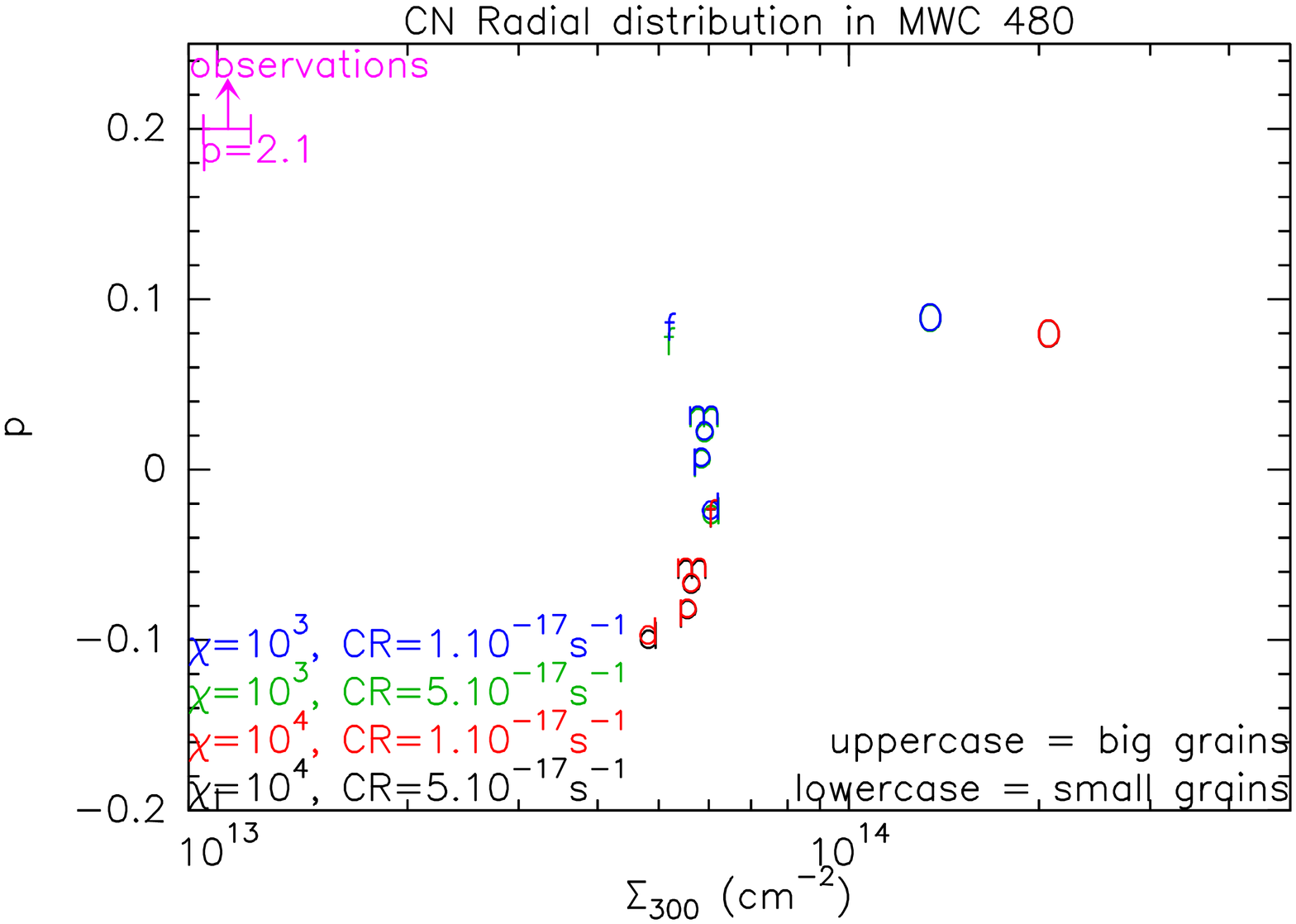}
      \includegraphics[angle=270,width=9cm]{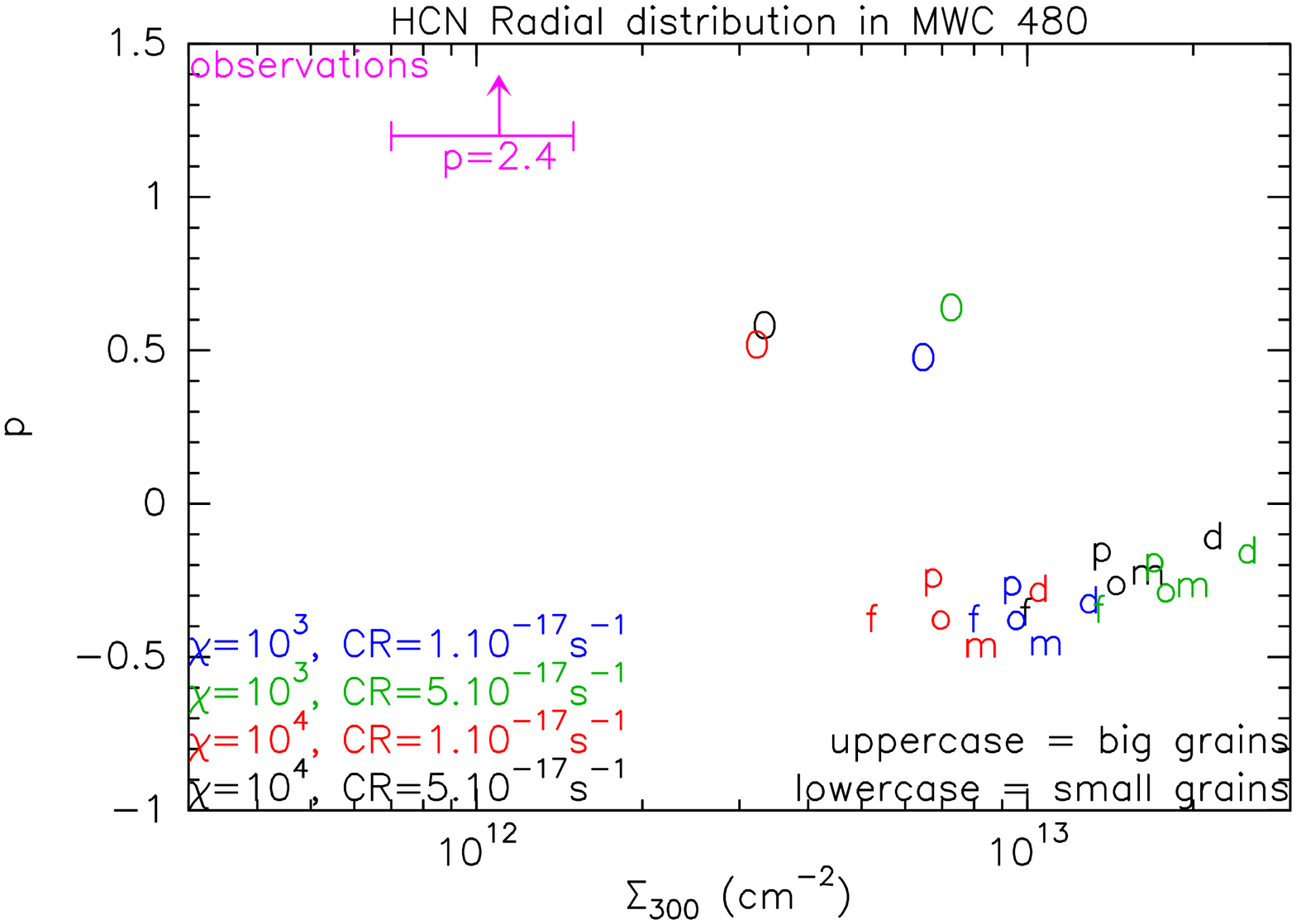}\\
\end{tabular}
\caption{Value of $\Sigma_{300}$ and p from our chemistry modeling for the three sources, with
our ``standard'' chemistry. \texttt{o}: density and temperature from theoretical models, \texttt{p}:
temperature increased by 30\%,  \texttt{m}: temperature decreased by 30\%,  \texttt{f}: density
multiplied by a factor 3 and \texttt{d}: density divided by a factor 3. In many models the blue and green marker overlap, as well as the red and black ones}
\label{fig:cn-hcn}
\end{figure*}

\begin{figure*}  
\begin{tabular}{lcc}
& small grains & big grains \\
$\chi=10^4$  &
\includegraphics[angle=270,width=8cm]{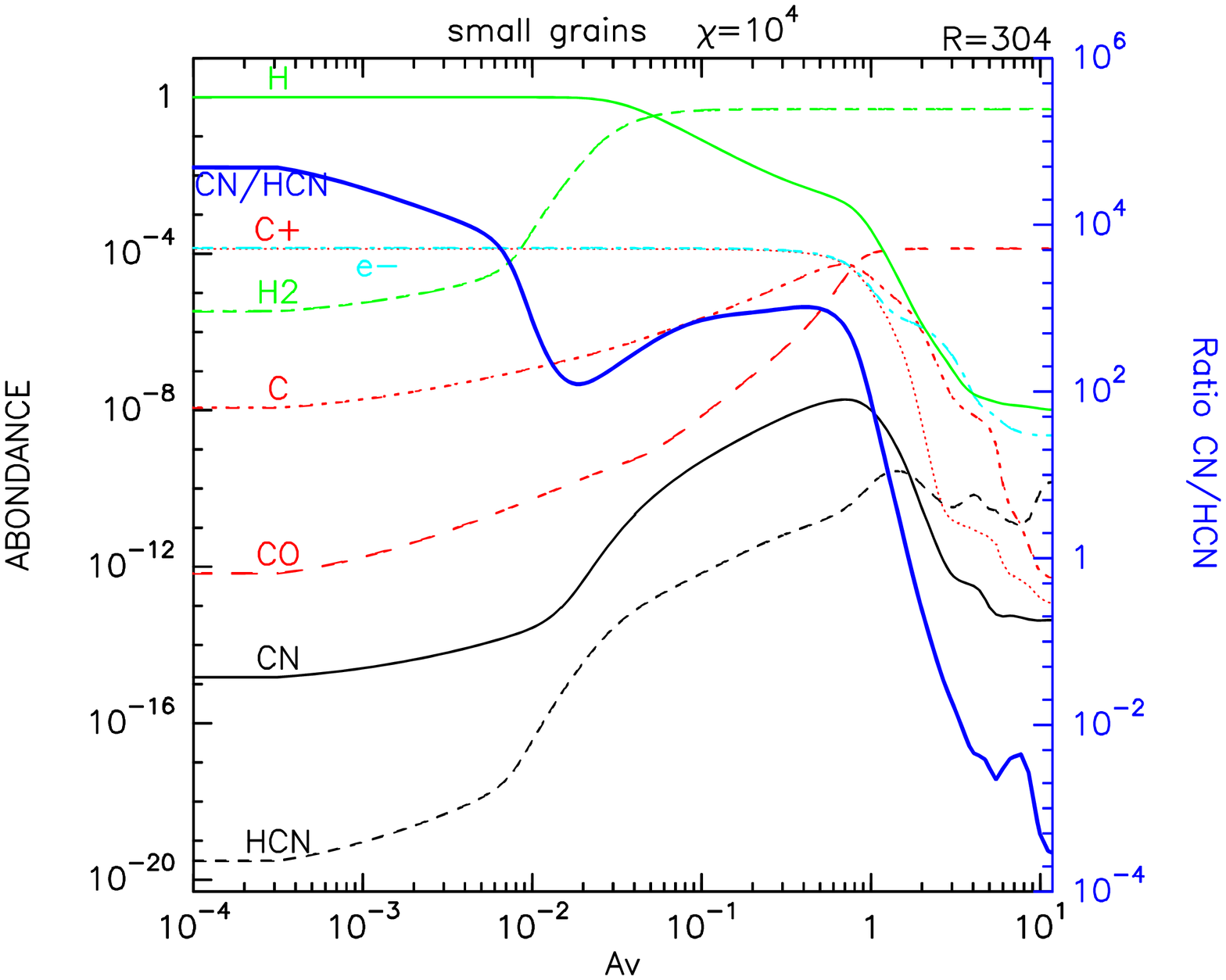}
&
\includegraphics[angle=270,width=8cm]{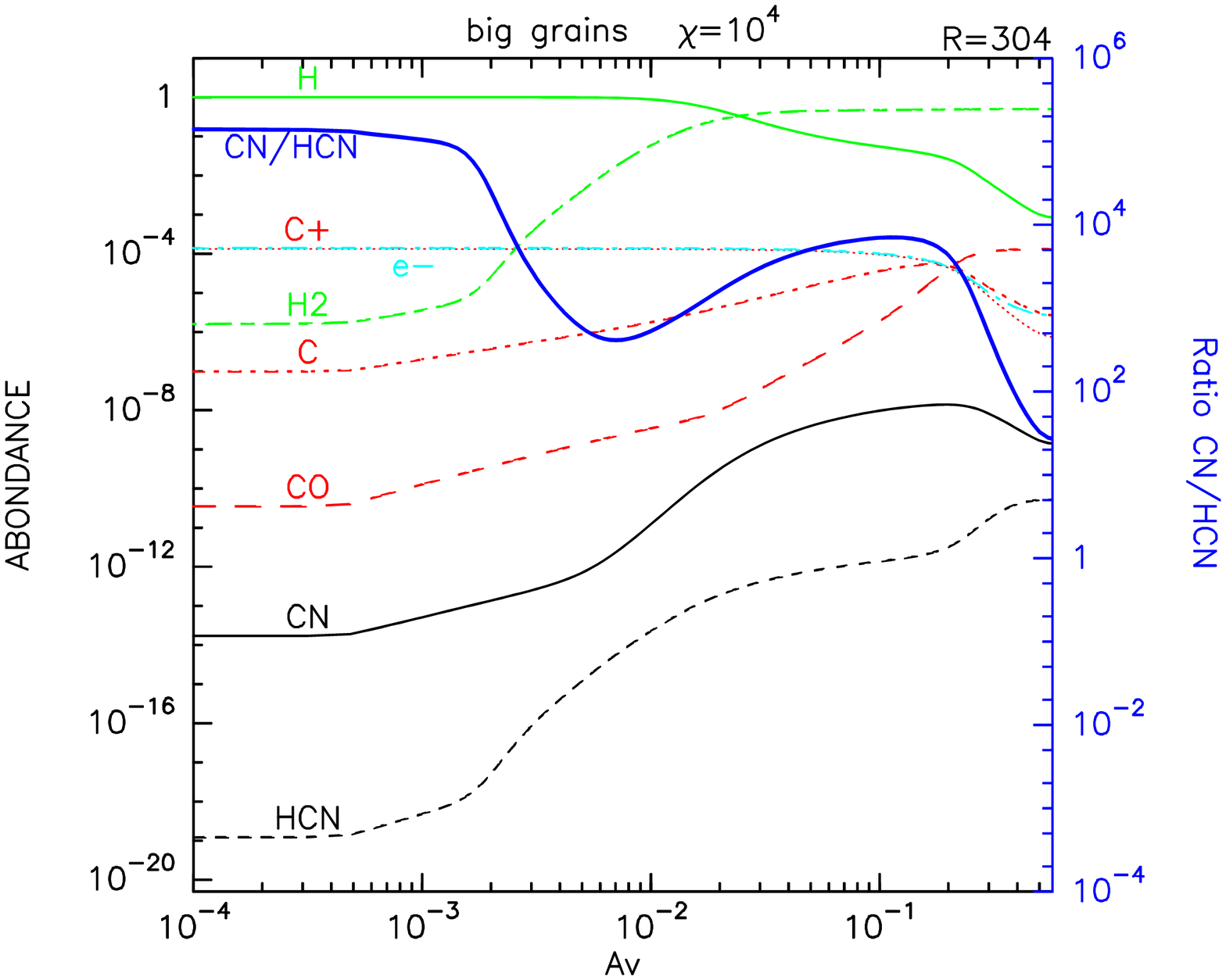}
\\
$\chi=10^3$  &
\includegraphics[angle=270,width=8cm]{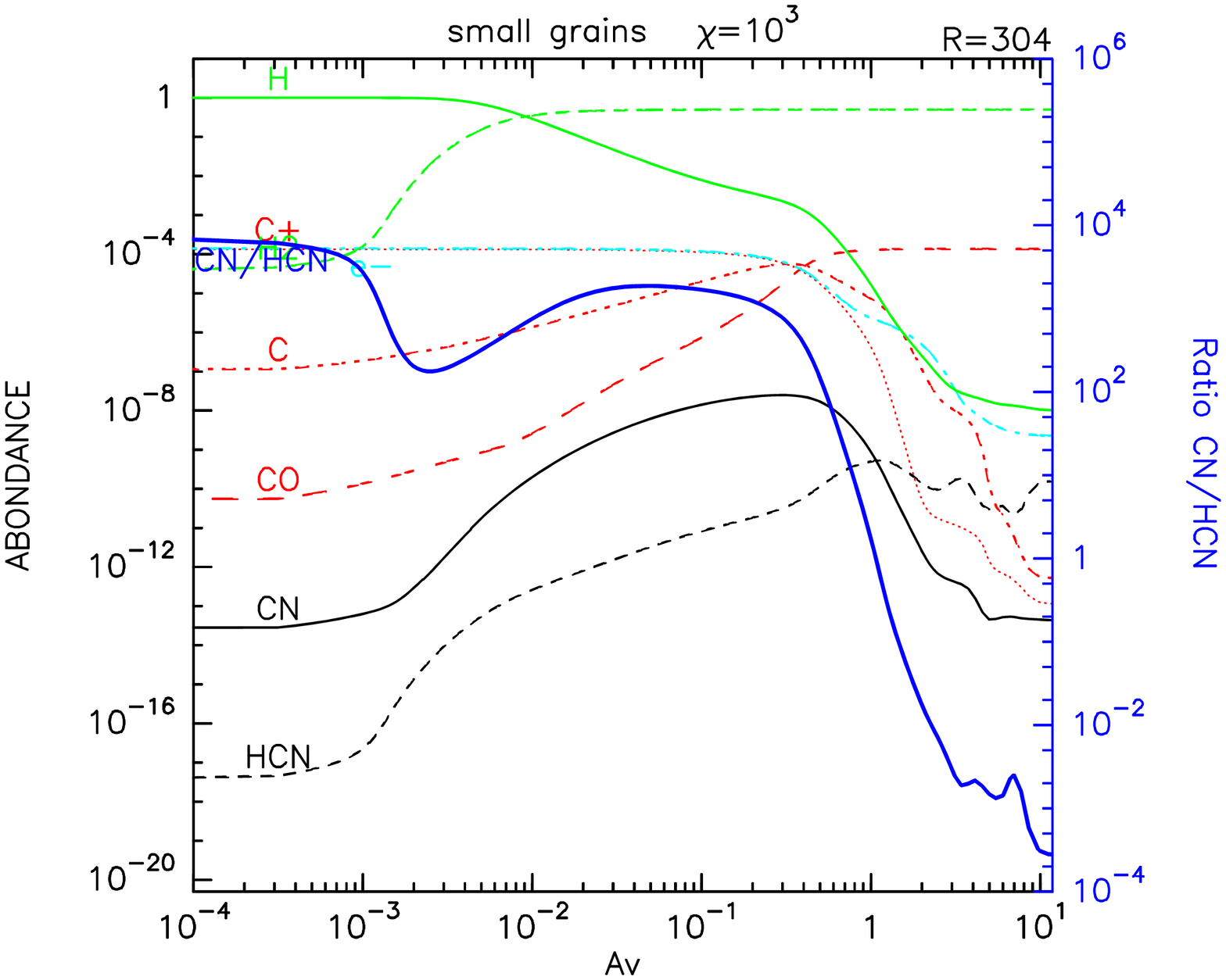}
&
\includegraphics[angle=270,width=8cm]{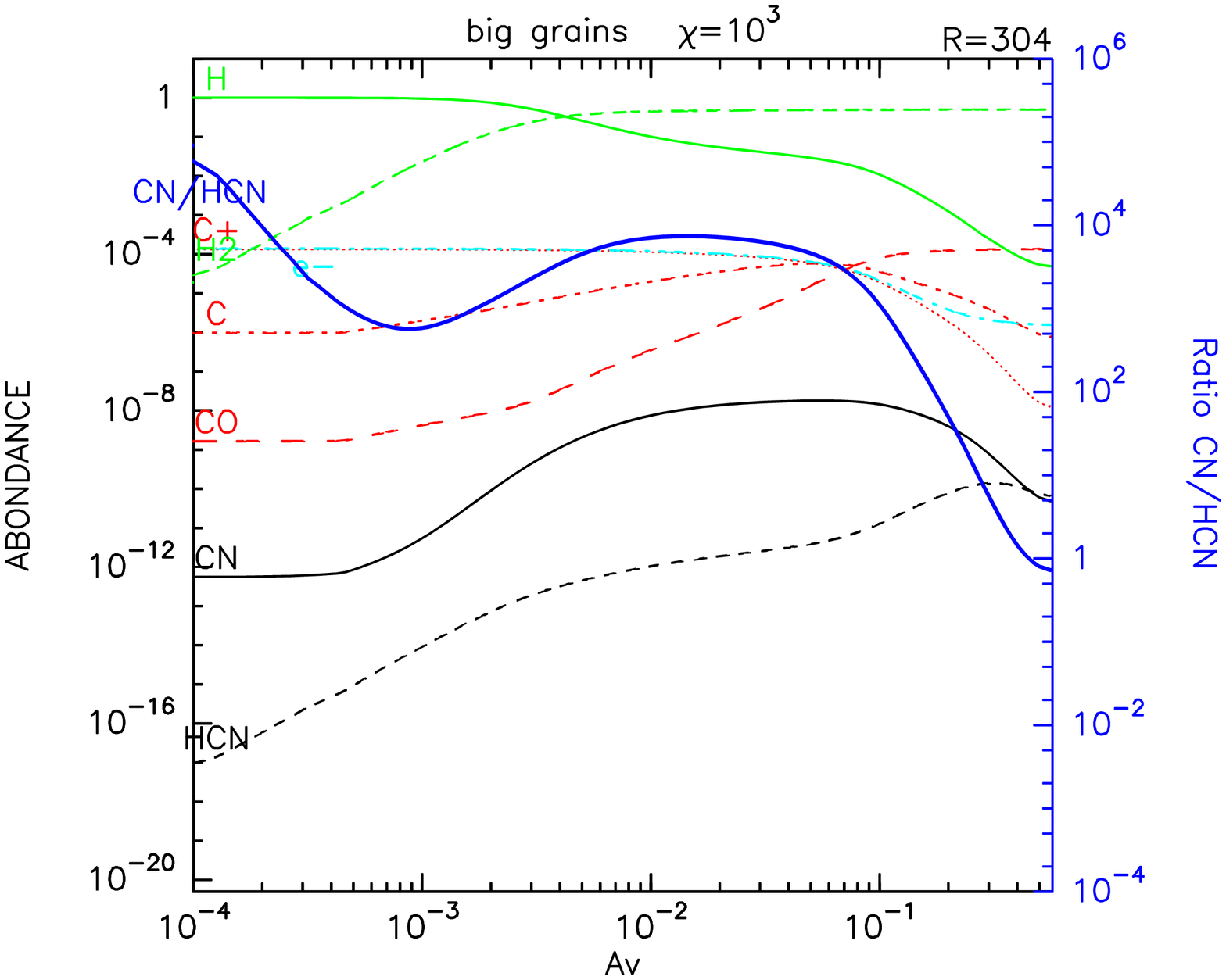}
\\
\end{tabular}
\caption{H, H$_2$, C$^+$, C, CO, CN, HCN abundance and CN/HCN ratio as a function of opacity from atmosphere (A$_v$=0) to mid-plane at \textbf{R=304\,AU} for DM\,Tau with CR=$10^{-17}$\,s$^{-1}$.}
\label{fig:abundance}
\end{figure*}

\end{appendix}

\end{document}